\documentclass[conference]{IEEEtran}
\def\thesection{\arabic{section}}                             

\usepackage{amsmath} 
\usepackage{amssymb} 
\usepackage{amsthm}
\usepackage{xspace}
\usepackage{pdfpages}

\usepackage{latexsym} 
\usepackage{colonequals} 
\usepackage{virginialake}

\usepackage{graphicx} 
\usepackage{tikz}
\usetikzlibrary{decorations.pathreplacing,patterns}
\usepackage[shortlabels]{enumitem}
\usepackage{multirow}

\usepackage{thmtools} 
\usepackage{thm-restate}

\newdimen\mydisplayskip
\newenvironment{smallequation*}
{\par\nobreak\vskip\mydisplayskip\noindent\bgroup\small\csname equation*\endcsname}{\csname endequation*\endcsname\egroup}

\theoremstyle{plain}
\newtheorem{theorem}{Theorem}[section]
\newtheorem{proposition}[theorem]{Proposition}
\newtheorem{lemma}[theorem]{Lemma}
\newtheorem{corollary}[theorem]{Corollary}
\theoremstyle{definition}

\newtheorem{notation}[theorem]{Notation}
\newtheorem{definition}[theorem]{Definition}
\newtheorem{construction}[theorem]{Construction}
\newtheorem{remark}[theorem]{Remark}
\newtheorem{example}[theorem]{Example}

\DeclareFontFamily{U} {MnSymbolC}{}
\DeclareFontShape{U}{MnSymbolC}{m}{n}{
	<-6>  MnSymbolC5
	<6-7>  MnSymbolC6
	<7-8>  MnSymbolC7
	<8-9>  MnSymbolC8
	<9-10> MnSymbolC9
	<10-12> MnSymbolC10
	<12->   MnSymbolC12}{}
\DeclareFontShape{U}{MnSymbolC}{b}{n}{
	<-6>  MnSymbolC-Bold5
	<6-7>  MnSymbolC-Bold6
	<7-8>  MnSymbolC-Bold7
	<8-9>  MnSymbolC-Bold8
	<9-10> MnSymbolC-Bold9
	<10-12> MnSymbolC-Bold10
	<12->   MnSymbolC-Bold12}{}
\DeclareSymbolFont{MnSyC}         {U}  {MnSymbolC}{m}{n}
\DeclareMathSymbol{\diamondplus}{\mathbin}{MnSyC}{124}
\DeclareMathSymbol{\boxtimes}{\mathbin}{MnSyC}{117}

\vlnostructuressyntax
\newcommand{\vlderivationauxnc}[1]{#1{\box\derboxone}\vlderivationterm}
\newcommand{\vlderivationnc}{\vlderivationinit\vlderivationauxnc}
\makeatletter
\newbox\@conclbox
\newdimen\@conclheight
\newcommand*{\A}{\mathcal{A}}

\newcommand{\qquand}{\qquad\mbox{and}\qquad}

\newcommand{\proviso}[1]{\mbox{\scriptsize #1}}
\newcommand{\defn}[1]{{\textit{\textbf{#1}}\/}}

\def\lef#1{#1^\bullet}
\def\rig#1{#1^\circ}
\def\srig#1{#1^\circ_{\star}}
\def\slef#1{#1^\bullet_{\star}}
\def\krig#1{#1^\circ_{\star}}
\def\klef#1{#1^\bullet_{\star}}
\def\wbox{\BOX}
\def\wdia{\DIA}

\newcommand{\w}{\rn{weak}}

\newcommand{\necr}{\rn{nec}}
\newcommand{\mpr}{\rn{mp}}
\newcommand{\cont}{\rn{cont}}

\newcommand{\monl}{\lef{\rn{mon}}}
\newcommand{\idr}{\rn{id}}

\newcommand*{\ax}[1]{\mathsf{#1}}
\newcommand*{\kax}[1][]{\ax{k_{#1}}}

\newcommand{\vax}{\ax{4}}

\newcommand{\tax}{\ax{t}}
\newcommand*{\IK}{\mathsf{IK}}
\newcommand*{\K}{\mathsf{K}}
\newcommand*{\IKfour}{\mathsf{IK4}}

\newcommand*{\ISfour}{\mathsf{IS4}}
\newcommand*{\ISfive}{\mathsf{IS5}}
\newcommand*{\Sfour}{\mathsf{S4}}
\newcommand*{\Sfive}{\mathsf{S5}}

\newcommand*{\labISf}{\lab\ISfour_{\le}}
\newcommand*{\labISfp}{\lab\ISfour_{\le}'}
\newcommand*{\labISfd}{\lab\ISfour_{\le}^{\star}}

\newcommand*{\NOT}{\neg}
\newcommand*{\AND}{\mathbin{\wedge}}
\newcommand*{\ANDk}{\mathbin{\wedge}_{\mathsf{k}}}
\newcommand*{\cand}{\mathbin{\wedge}}
\newcommand*{\TOP}{\mathord{\top}}
\newcommand*{\OR}{\mathbin{\vee}}
\newcommand*{\ORk}{\mathbin{\vee}_{\mathsf{k}}}

\newcommand*{\BOT}{\mathord{\bot}}
\newcommand*{\IMP}{\mathbin{\supset}}
\newcommand*{\IMPk}{\mathbin{\supset}_{\mathsf{k}}}

\newcommand*{\BOX}{\mathord{\Box}}
\newcommand*{\BOXk}{\mathord{\Box}}
\newcommand*{\DIA}{\mathord{\Diamond}}
\newcommand*{\DIAk}{\mathord{\Diamond}}
\newcommand{\lseq}[3]{#1 , #2 \SEQ #3}
\newcommand{\B}{\mathcal{R}}
\newcommand{\Left}{\Gamma} 
\newcommand{\Right}{\Delta} 
\newcommand*{\fm}[1]{{\color{notgreen}#1}}
\newcommand*{\lb}[1]{{\color{blue}#1}}
\newcommand*{\lbc}[1]{{\color{blue}#1}}
\newcommand*{\lbs}[1]{{\color{blue}#1}}

\newcommand{\fmb}[1]{\fm{#1^{{\color{black}\bullet}}}}
\newcommand{\fmw}[1]{\fm{#1^{{\color{black}\circ}}}}
\newcommand*{\rel}{R}

\newcommand{\lrel}[1]{R^{\leftrightarrow}_{\sq{#1}}}

\newcommand{\grel}[1]{R_{\sq #1}}
\newcommand*{\labels}[2]{\lb{#1}\mathord{:}\fm{#2}}
\newcommand*{\labelsb}[2]{\lb{#1}\mathord{:}\fmb{#2}}
\newcommand*{\labelsw}[2]{\lb{#1}\mathord{:}\fmw{#2}}
\newcommand*{\accs}[2]{\lb{#1}R\lb{#2}}

\newcommand{\sle}[1]{\mathbin{\le_{\sq{#1}}}}
\newcommand{\nsle}[1]{\mathbin{\not\le_{\sq{#1}}}}
\newcommand{\srel}[1]{\mathbin{\rel_{\sq{#1}}}}
\newcommand*{\accsq}[2]{\lb{#1}\mathord{\srel{G}}\lb{#2}}
\newcommand*{\accsqpr}[2]{\lb{#1}\mathord{\srel{G'}}\lb{#2}}
\newcommand*{\accsqar}[2]{\lb{#1}\mathord{\srel{\liftsat{\sq{G}}}}\lb{#2}}
\newcommand*{\accsqh}[2]{\lb{#1}\mathord{R^{\leftrightarrow}_{\sq{G}}}\lb{#2}}
\newcommand*{\accsqc}[2]{\lbc{#1}\mathord{\srel{G}}\lbc{#2}}
\newcommand*{\accshat}[2]{\lbc{#1}\mathord{\srel{\hG}}\lbc{#2}}

\newcommand*{\accsqp}[3]{\lb{#1}\mathord{\srel{#3}}\lb{#2}}
\newcommand*{\accslt}[4]{\lb{#1}\srel{#4}\lb{#2}\srel{#4}\lb{#3}}
\newcommand*{\accsl}[4]{\lb{#1}\srel{G}\lb{#2}\srel{G}\lb{#3}\srel{G}\lb{#4}}
\newcommand*{\futsq}[2]{\lb{#1}\mathord{\sle{G}}{\lb{#2}}}
\newcommand*{\futsqc}[2]{\lbc{#1}\mathord{\sle{G}}{\lbc{#2}}}

\newcommand*{\futsqpr}[2]{\lb{#1}\mathord{\sle{G'}}\lb{#2}}
\newcommand*{\futsqp}[3]{\lb{#1}\mathord{\sle{#3}}{\lb{#2}}}
\newcommand*{\nfutsq}[2]{\lb{#1}\mathord{\nsle{G}}{\lb{#2}}}
\newcommand*{\futs}[2]{\lb{#1}\mathord\le{\lb{#2}}}

\newcommand{\labelsof}[1]{\ell(\sq{#1})}
\newcommand{\labelsofx}[1]{\ell({#1})}
\newcommand{\forbiddenof}[1]{\mathcal{f}(\sq{#1})}
\newcommand{\forbiddenofx}[1]{\mathcal{f}({#1})}

\newcommand{\emb}{{\color{black}\mathsf{e}}}
\newcommand{\embof}[1]{{\color{black}\emb}(\lb{#1})}

\newcommand{\SEQ}{\Longrightarrow}

\newcommand{\unhs}[2]{\mathsf{unh}_{\DIA}(\sqset{#1}_{#2})}
\newcommand*{\rn}[1]  {\ensuremath{\mathsf{#1}}}
\newcommand*{\lab}{\mathsf{lab}}
\newcommand*{\rr}{\mathsf{r}}

\newcommand*{\labrn}[2][]  {\rn{#2}_{#1}}

\newcommand{\M}{\mathcal{M}}
\newcommand{\F}{\mathcal{F}}
\newcommand{\force}[3]{#1,\lb{#2}\Vdash\fm{#3}}
\newcommand{\nforce}[3]{#1,\lb{#2}\not\Vdash\fm{#3}}

\newcommand{\height}[1]{|#1|}
\newcommand{\Rref}{\rel\rn{rf}}

\newcommand{\Rtr}{\rel\rn{tr}}

\newcommand{\Lref}{\mathord\le\rn{rf}}

\newcommand{\Ltr}{\mathord\le\rn{tr}}

\newcommand{\fone}{\rn{F_1}}

\newcommand{\ftwo}{\rn{F_2}}

\newcommand{\fhat}{{\rn{F}}}

\def\tuple#1{\langle#1\rangle}

\newcommand{\der}{\mathcal{D}}
\newcommand{\deri}{\mathcal{D}}

\newcommand{\lbeq}{\sim}
\newcommand{\lbeql}[4]{{_{#2\mathstrut^{\mathstrut}}}\lb{#1}\sim{_{#4\mathstrut^{\mathstrut}}\lb{#3}}}

\newcommand{\sums}{ + }

\newcommand{\seqredsymb}{\rightsquigarrow}
\newcommand{\seqsetredsymb}{\mathbin{\lower.25ex\rlap{$\rightsquigarrow$}\raise.25ex\hbox{$\rightsquigarrow$}}}

\newcommand{\ssatred}{\seqsetredsymb_\mathsf{s}}
\newcommand{\tssatred}{\seqsetredsymb_\mathsf{s}^\ast}

\newcommand{\diared}{\mathbin{\seqredsymb_\Diamond}}
\newcommand{\sdiared}{\mathbin{\seqsetredsymb_\Diamond}}
\newcommand{\Diared}{\sdiared}
\newcommand{\tsdiared}{\mathbin{\seqsetredsymb^\ast_\Diamond}}
\newcommand{\tDiared}{\tsdiared}
\newcommand{\loopred}{\mathbin{\seqredsymb_\circlearrowright}}
\newcommand{\Loopred}{\mathbin{\seqsetredsymb_\circlearrowright}}
\newcommand{\tLoopred}{{\seqsetredsymb^\ast_\circlearrowright}}

\newcommand{\hp}{\hat p}
\newcommand{\hs}{\hat s}
\newcommand{\htt}{\hat t}
\newcommand{\hu}{\hat u}
\newcommand{\hU}{\hat U}
\newcommand{\hv}{\hat v}
\newcommand{\hx}{\hat x}
\newcommand{\hy}{\hat y}
\newcommand{\hz}{\hat z}
\newcommand{\hG}{{\hat G}}
\newcommand{\hH}{{\hat H}}

\newcommand{\hL}{{\hat L}}

\newcommand{\hw}{{\hat w}}

\newcommand{\hC}{{\hat C}}
\newcommand{\hCone}{{\hat C'_x}}
\newcommand{\hCtwo}{{\hat C''_x}}
\newcommand{\hS}{{\hat S}}

\newcommand{\Nat}{\mathbb{N}}
\newcommand{\Deri}{\mathcal{D}}
\newcommand{\cT}{\mathcal{T}}

\newcommand{\set}[1]{\{#1\}}

\newcommand{\sq}[1]{\mathfrak{#1}}

\newcommand{\sqset}[1]{\rlap{$\mathfrak{#1}$}\mkern.75mu\rlap{$\mathfrak{#1}$}\mkern0.75mu\mathfrak{#1}}

\newcommand{\coffsp}[2]{\childrenof{#2}}  

\makeatletter
\newcommand*{\overRightarrow}{\mathpalette{\overarrow@\Rightarrowfill@}}
\newcommand*{\overLeftarrow}{\mathpalette{\overarrow@\Leftarrowfill@}}
\makeatother

\newcommand{\liftsat}[1]{#1\mathord\uparrow}
\newcommand{\llift}[2]{\sq{#1}\mathord\uparrow^{\lb #2}}
\newcommand{\lllift}[2]{\sq{#1}\mathord\uparrow^{#2}}

\newcommand{\lliftf}[3]{\sq{#1}\mathord\uparrow^{\labels{#2}{#3}}}
\newcommand{\llniftf}[3]{{#1}\mathord\Uparrow^{\labels{#2}{#3}}}

\newcommand{\layerof}[1]{L_{\lb{#1}}}
\newcommand{\parentsof}[1]{\lb{\overleftarrow{#1}}}
\newcommand{\childrenof}[1]{\lb{\overrightarrow{#1}}}

\newcommand{\sizeof}[1]{\left|#1\right|}

\newcommand{\modelof}[1]{\M_{\sq{#1}}}

\newcommand{\blackf}[3]{\sq{#3},\labels{\!#1}{#2}^{\bullet}}
\newcommand{\whitef}[3]{\sq{#3},\labels{\!#1}{#2}^{\circ}}
\newcommand{\simul}{\rn{S}}
\newcommand{\unfold}{\rn{U}}

\newcommand{\wip}[1]{\check{#1}}
\newcommand{\nop}[1]{\bar{#1}}
\newcommand{\nopin}[3]{\bar{#1}_{#2/#3}}

\newcommand{\allfu}[3]{{#1}^{#3}_{#2}}

\newcommand{\weirdS}{\sqset S^\ast}

\usepackage{xcolor}
\usepackage{wasysym}
\usepackage{standalone}
\usepackage{dutchcal}
\definecolor{notgreen}{rgb}{.1,.6,.1}

\usepackage{ifthen}

\newboolean{arxiv}

\setboolean{arxiv}{true}

\usepackage{hyperref}

\ifCLASSOPTIONcompsoc
  \usepackage[nocompress]{cite}
\else
  \usepackage{cite}
\fi
\ifCLASSINFOpdf
\else
\fi
\begin{document}
%
\title{Intuitionistic S4 is decidable}

\author{\IEEEauthorblockN{Marianna Girlando}
	\IEEEauthorblockA{University of Amsterdam\\
	  Amsterdam, Netherlands
		}
	\and
	\IEEEauthorblockN{Roman Kuznets}
	\IEEEauthorblockA{TU Wien\\
	  Vienna, Austria
		}
	\and
	\IEEEauthorblockN{Sonia Marin}
	\IEEEauthorblockA{University of Birmingham\\
	  Birmingham, UK
		}
	\and
	\IEEEauthorblockN{Marianela Morales}
	\IEEEauthorblockA{Inria Saclay\\
	  Palaiseau, France
		}
	\and
	\IEEEauthorblockN{Lutz Stra{\ss}burger}
	\IEEEauthorblockA{Inria Saclay\\
	  Palaiseau, France
	}}



%


\ifthenelse{\boolean{arxiv}}{
  \IEEEoverridecommandlockouts
  \IEEEpubid{\makebox[\columnwidth]{This is the author version of the LICS'23 paper with detailed appendices. \hfill} \hspace{\columnsep}\makebox[\columnwidth]{ }}
}{
  \IEEEoverridecommandlockouts
  \IEEEpubid{\makebox[\columnwidth]{979-8-3503-3587-3/23/\$31.00~
      \copyright2023 IEEE \hfill} \hspace{\columnsep}\makebox[\columnwidth]{ }}
}

\maketitle

\begin{abstract}
  In this paper we demonstrate  decidability for the intuitionistic modal
  logic~$\Sfour$  first formulated by Fischer Servi. This solves a problem that has been open for almost thirty years since it had been posed in Simpson's PhD~thesis in~1994. We obtain this result by performing  proof search in a labelled deductive system that, instead of using only one binary relation on the  labels, employs two: one corresponding to the accessibility relation of modal logic and the other corresponding to the order relation of intuitionistic Kripke frames. Our search algorithm  outputs either a proof or a finite counter-model, thus, additionally  establishing the finite model property for  intuitionistic~$\Sfour$, which has been another long-standing open problem in the area.
\end{abstract}




%

\section{Introduction}

\label{sec:intro}

For a logic to be decidable, there must be a recursive procedure that determines, for each  formula, whether or not it is a theorem of the logic, ideally terminating with either a proof in a deductive system or a suitable countermodel.

\textbf{Decidability of intuitionistic propositional logic.}
Gentzen was the first to show decidability of intuitionistic propositional logic in~1935, using a sequent calculus he had designed~\cite{gentzen:35:I}.
His approach was to bound the number of consequences inferred from some given initial sequents.
It was later observed (independently by Ono~\cite{ono:1938}, Ketonen~\cite{ketonen:1945}, and Kleene~\cite{kleene:1952}) that, if structural rules were built in into the notation, it allowed for a root-first proof-search approach.
Namely, search for a sequent calculus proof until it either terminates with a proof or else reaches a sequent that has already occurred along the branch leading to it, in which case it is possible to reconstruct a Kripke countermodel from that branch~\cite{troel:schw:2001}.
For a detailed historical account, see Dyckhoff's survey~\cite{dyckhoff:2016}.

\textbf{Decidability for classical modal logics.}
Decidability for modal logics (on the classical propositional base) has been investigated early on.
Ladner provided decision procedures for some common modal logics, including~$\Sfour$, based purely on Kripke semantics, with no mention of sequent calculi~\cite{ladner:1977}. 
Otherwise, the procedure described above can similarly be applied to modal logic~$\Sfour$, as it is also  the logic of reflexive transitive Kripke frames.
Generally, the approach via proof search in a sequent calculus is available for those modal logics that have a cut-free sequent system~\cite{sato:1977} (or, in certain cases, a nested sequent system using a similar approach~\cite{brunnler:2009}).

\textbf{Decision procedures using labelled sequents.}
Labelled sequents internalize certain elements of Kripke semantics into the syntax of sequents, which turns out to have several interesting consequences for  decision procedures for  intuitionistic propositional logic~\cite{dyck:neg:2012}, as well as modal logics~\cite{negri:2005}.\looseness=-1

Firstly, in the setting of intuitionistic propositional logic, it is not necessary to restrict oneself to single-conclusion sequents (i.e.,~sequents with at most one formula in the succedent). 
This means that all the rules can be made invertible, thus, removing the need for backtracking and making the proof search procedure deterministic.
Secondly, in both settings, the loop-check for ensuring termination becomes local to the topmost sequent 
of a branch of a proof-search tree rather than applied between two sequents on this branch: it is sufficient to check whether two labels of the topmost sequent carry the same formulas.
Finally, in the case when a loop is found, a countermodel can be built directly from the topmost sequent.
Indeed, the syntax of labelled sequents makes it easy to read off a countermodel from a sequent by simply substituting labels or adding back edges, both constructions being available directly in the labelled syntax.

\textbf{Intuitionistic modal logic.}
Intuitionistic modal logic~$\ISfour$ is a way to combine intuitionistic propositional logic and modal logic~$\Sfour$.
It is  an extension of intuitionistic modal logic~$\IK$, which was 
first  studied in~\cite{fischer-servi:84,plotkin:stirling:86} and investigated in
detail in~\cite{simpson:phd}, where decidability was shown for
most logics in the intuitionistic modal $\Sfive$-cube. 
Decidability of~$\ISfive$ (aka~{\sf MIPQ}) had been shown earlier by Mints~\cite[\textsection11]{Mints:68}, but for~$\IKfour$~and~$\ISfour$
decidability  remained open. 
Indeed, the question of $\ISfour$'s~decidability hides a lot more complexity than either of its progenitors.
As $\ISfour$~can be interpreted on Kripke frames that combine the preorder relation of intuitionistic Kripke frames with the accessibility relation of modal Kripke frames, the  well-known problems of looping in decision procedures based on respective sequent calculi are exacerbated by the interactions between modalities and intuitionistic implication.

\textbf{Contribution.}
In this paper, we show that $\ISfour$~is decidable and  conjecture that the same method can also be applied to show  decidability of~$\IKfour$. 
We provide a constructive decision procedure, that, given a formula, produces either  a proof showing the formula to be valid or a finite countermodel falsifying the formula.
The procedure is based on a \emph{fully labelled proof system} for~$\ISfour$  presented in~\cite{mar:mor:str:2021}.
This system inherits the advantages of labelled systems for intuitionistic propositional logic and for classical modal logics, in particular, all inference rules are invertible 
and there is a direct correspondence between sequents and models.
This choice was crucial to the emergence of the solution, which also required an intricate organization of  proof search and loop-checking to reach a full decision procedure. 

\textbf{Related work.}
Several strategies have been explored for decidability of~$\ISfour$ and related logics.
The 
$\Box$-only 
 fragments of a wide range of intuitionistic multi-modal logics was shown to be decidable via a proof-search based procedure in~\cite{garg:genov:neg:2012}.\looseness=-1

Moreover, for $\Box$-only fragments of intuitionistic modal logics, decidability results were obtained through embedding into classical  bimodal logics, defined as \emph{fusions} of normal monomodal logics~\cite{wol:zak:1997:intuitionistic}. 
Logic~$\ISfour$ is closely related to a classical bimodal logic defined as the \emph{product} $\Sfour \times \Sfour$~\cite{fischer-servi:84} in that both are sound and complete with respect to Kripke models with two preorder relations forming a ``grid.'' Interestingly, logic~$\Sfour \times \Sfour$ was shown to be undecidable by Gabelaia~\emph{et~al.}~\cite{gabelaia:2005}.  This result shows how easily undecidability arises in modal logic; however, 
the proof strategy from~\cite{gabelaia:2005} does not scale to~$\ISfour$, whose models need to satisfy monotonicity over one of the preorder relations. 
%
Similarly, embeddings of intuitionistic modal logics into fragments of monadic second-order logic, which is decidable, were used to prove decidability~\cite{alech:shkat:2006}, but, as explained by the authors, the method could not be applied to~$\ISfour$.
Another common method of demonstrating decidability applied to some intuitionistic modal logics in~\cite{hasimoto:2001:finite} is
semantic filtration,  but again the  specificities of the logics from the intuitionistic $\Sfive$-cube have so far prevented its application to them.
On the other hand, constructive~$\Sfour$ (a different intuitionistic variant of modal logic~$\Sfour$) was shown to be decidable in~\cite{heilala:pientka:07},
and the finite model property for it was proved in~\cite{bal:die:fer:2021}. 

\textbf{Organization.}
In Sect.~\ref{sec:models}, we recall the syntax and semantics of~$\ISfour$.  In  Sect.~\ref{sec:problem}, we explain the difficulties of devising a  decision procedure for it.
In Sect.~\ref{sec:labelled_seq}, we present the first ingredient of our solution:   proof system~$\labISf$ for~$\ISfour$ based on fully labelled sequents.
In  Sect.~\ref{sec:sequents_models}, we describe how to read off a model from such a sequent. In Sect.~\ref{sec:clusters}, we establish useful properties of sequents occurring during a proof search  in~$\labISf$.
In Sect.~\ref{sec:algorithm}, we develop  concepts necessary to present our decision procedure based on a modified proof-search algorithm.
In Sect.~\ref{sec:countermodel}, we demonstrate  how to retrieve a countermodel from a failed proof search. In Sect.~\ref{sec:unfolding}, we show how to reconstruct a real proof when the algorithm succeeds.
In Sect.~\ref{sec:termination}, we prove that the algorithm always terminates. Finally,  Sect.~\ref{sec:conclusion}, contains our conclusions.
\ifthenelse{\boolean{arxiv}}{
  All the proofs and some detailed examples can be found in the Appendix.
}{}

\section{Preliminaries: Logic $\ISfour$}
\label{sec:models}

In this paper, formulas are denoted 
by capital letters~$\fm A$, $\fm B$, $\fm C$,~\ldots\ and are constructed from a countable set~$\mathcal{A}$ of \emph{atomic propositions} (denoted by lowercase~$\fm a$, $\fm b$, $\fm c$,~\ldots) as
\[
\scalebox{.95}{$
	\fm A \coloncolonequals
	\fm \BOT \mid \fm a \mid \fm{(A \AND A)} \mid \fm{(A \OR A)} \mid  \fm{(A \IMP A)} \mid \fm{\BOX A} \mid \fm{\DIA A}$}
\]

Intuitionistic modal logic~$\K$, or $\IK$~for short, is obtained from
intuitionistic propositional logic by extending the syntax with 
two modalities~$\wbox$~and~$\wdia$, standing most generally for \emph{necessity} and \emph{possibility} respectively, and by adding \emph{$\kax{}$-axioms}
\begin{equation}
  \label{eq:kax}
  \hskip-.95em%
  \scalebox{.89}{$
	\begin{array}{r@{\,}c}
		\kax[1]\colon&\fm{\BOX(A\IMP B)\IMP(\BOX A\IMP\BOX B)}\\
		\kax[2]\colon&\fm{\BOX(A\IMP B)\IMP(\DIA A\IMP\DIA B)}
        \end{array}
	\begin{array}{r@{\,}c}
		\kax[3]\colon&\fm{\DIA(A\OR B)\IMP(\DIA A\OR\DIA B)}\\
		\kax[4]\colon&\fm{(\DIA A\IMP \BOX B)\IMP\BOX(A\IMP B)}\\
		\kax[5]\colon&\fm{\DIA\BOT\IMP\BOT}\\
	\end{array}
        $}
\end{equation}

A formula is a theorem of~$\IK$ if{f} it is derivable from this set via the rules of \emph{necessitation} and \emph{modus ponens}:
\[
	\vlinf{\necr}{}{\fm{\wbox A}}{\fm A}
	\qquand
	\vliiinf{\mpr}{}{\fm B}{\fm A}{}{\fm{A\IMP B}}
	\quad
\]
Intuitionistic modal logic~$\Sfour$, or $\ISfour$~for short, is obtained from~$\IK$ by
adding  axioms
\begin{equation}
	\label{eq:vax}
	\begin{array}{r@{\;}l}
		\vax\colon&
		\fm{(\wdia\wdia A\IMP\wdia A)\cand(\wbox A\IMP\wbox\wbox A)}
		\\ 
		\tax\colon&
		\fm{(A\IMP\wdia A)\cand(\wbox A\IMP A)}
	\end{array}
\end{equation}

Note that in the classical case  axioms~$\kax[2]$--$\kax[5]$
in~\eqref{eq:kax} would follow from~$\kax[1]$, but due to the lack of
De~Morgan duality, this is not the case in intuitionistic
logic. Similarly, in~\eqref{eq:vax} both conjuncts are
needed because they do not follow from each other as in the classical
case.

Let us now recall
the \emph{birelational models}~\cite{plotkin:stirling:86,ewald:86} for
intuitionistic modal logics, which combine the Kripke
semantics for intuitionistic propositional logic and
classical modal logics.

\begin{definition}
	A \defn{birelational frame}~$\F$ is a triple~$\langle W, \rel, \le \rangle$
	of a nonempty set~$W$ of \defn{worlds}  equipped with an \defn{accessibility relation}~$\rel$ and a preorder~$\le$ (i.e.,~a reflexive and transitive relation) satisfying:
	\begin{enumerate}
		\item[($\rn{F_1}$)] For all~$\lb x, \lb y, \lb z \in W$, if $\accs xy$ and $\futs yz$, there exists~$\lb u \in W$ such that $\futs xu$ and $\accs uz$ (see figure below left);
		\item[($\rn{F_2}$)] For all~$\lb x, \lb y, \lb z \in W$, if $\futs xz$ and $\accs xy$, there exists~$\lb u \in W$ such that $\accs zu$ and $\futs yu$ (see figure below right).
	\end{enumerate}	
	
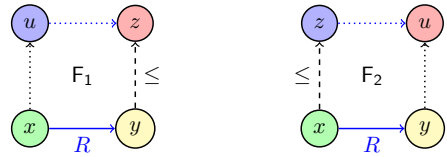
\begin{center}
			
	\begin{tikzpicture}[thick, every node/.style={scale=1.2}]

		\tikzstyle{node}=[circle,draw]
		\tikzstyle{s-node}=[circle,draw,inner sep=2pt]
		\tikzstyle{nonode}=[inner sep=0pt]
		\tikzstyle{access}=[->,blue]
		\tikzstyle{future}=[->,dashed]
		\tikzstyle{deleted}=[->,gray]
		\tikzstyle{loop}=[->,red]
		
		\node[] (3) at (5,3) [] {$\mathsf{F_1}$};
		
		\node[label=below :{}] (3) at (4,2) [node, fill=green!30] {$x$};
		\node[label=below :{}] (4) at (6,2) [node, fill=yellow!30] {$y$};
		\node[label=above :{}] (10) at (4,4) [node, fill=blue!30] {$u$};
		\node[label=above :{}] (11) at (6,4) [node, fill=red!30] {$z$};
		
		\draw[access] (3) edge[below] node {$R$} (4);
		\draw[access,dotted] (10) -- (11);

		\draw[future,dotted] (3) -- (10);
		\draw[future] (4) edge[right] node {$\le$}  (11);
	\end{tikzpicture}

\hspace*{2cm}

\begin{tikzpicture}[thick, every node/.style={scale=1.2}]

	\tikzstyle{node}=[circle,draw]
	\tikzstyle{s-node}=[circle,draw,inner sep=2pt]
	\tikzstyle{nonode}=[inner sep=0pt]
	\tikzstyle{access}=[->,blue]
	\tikzstyle{future}=[->,dashed]
	\tikzstyle{deleted}=[->,gray]
	\tikzstyle{loop}=[->,red]
	
	\node[] (3) at (5,3) [] {$\mathsf{F_2}$};
	
	\node[label=below :{}] (3) at (4,2) [node, fill=green!30] {$x$};
	\node[label=below :{}] (4) at (6,2) [node, fill=yellow!30] {$y$};
	\node[label=above :{}] (10) at (4,4) [node, fill=blue!30] {$z$};
	\node[label=above :{}] (11) at (6,4) [node, fill=red!30] {$u$};
	
	\draw[access] (3) edge[below] node {$R$} (4);
	\draw[access,dotted] (10) -- (11);
	
	\draw[future] (3) edge[left] node {$\le$} (10);
	\draw[future,dotted] (4) -- (11);
\end{tikzpicture}
	\end{center}
\end{definition}

\begin{definition}
	\label{model}
	A \defn{birelational model}~$\M$ is a quadruple~$\langle W, \rel,\le,V \rangle$ with~$\langle W, \rel, \le \rangle$ a birelational frame and $V\colon W \to 2^\mathcal{A}$ a \defn{valuation function}, that is, a function mapping each world~$\lb w$ to the subset of propositional atoms that are true at~$\lb w$, additionally subject to the \defn{monotonicity condition}:
	if $\futs w{w'}$, then $V(\lb w)\subseteq V(\lb{w'})$.
 
We write $\force\M wa$ if{f} $\fm a \in V(\lb w)$ and recursively extend relation~$\Vdash$ to all formulas following the rules for both intuitionistic and modal Kripke models: $\nforce\M w\BOT$;
\[
\begin{array}{@{}r@{\;}c@{\;\;}l}
\force\M w{A \AND B} & \mbox{if{f}} & \force\M wA \mbox{ and } \force\M wB;\\

\force\M w{A \OR B} & \mbox{if{f}} & \force\M wA \mbox{ or } \force\M wB;\\

\force\M w{A \IMP B} & \mbox{if{f}} & \mbox{for all } \lb{w'} \mbox{ with } \futs w{w'},
\\ &&\mbox{if $\force\M{w'}A$, then $\force\M{w'}B$};\\

\force\M w{\BOX A} & \mbox{if{f}} & \mbox{for all } \lb{w'} \mbox{ and } \lb u \mbox{ with } \futs w{w'}\\
&&\mbox{and } \accs{w'}u, \mbox{ we have } \force\M uA; \hfill \\ 

\force\M w{\DIA A} & \mbox{if{f}} & \mbox{there exists } \lb u \mbox{ such that }\\
&&\accs wu \mbox{ and } \force\M uA.
\end{array}
\]
\end{definition}

From the monotonicity of valuation function~$V$, we get the monotonicity property for relation~$\Vdash$:

\begin{proposition}(Monotonicity) 
	For any formula~$\fm A$ and for any~$\lb w, \lb{w'} \in W$, if $\futs w{w'}$ and $\force\M wA$, then $\force\M{w'}A$.
\end{proposition}

\begin{definition}[Validity]
	A formula~$\fm A$ is \defn{valid in a model}~$\M = \langle W, \rel, \le, V \rangle$ if{f}  $\force\M wA$ for all~$\lb w \in W$.
	A formula~$\fm A$ is \defn{valid in a frame}~$\F = \langle W, \rel, \le \rangle$ if{f} it is valid in~$\langle W, R, \le, V \rangle$ for all valuations~$V$.
\end{definition}

The correspondence between syntax and semantics for~$\ISfour$ can be stated as follows:

\begin{theorem}[Completeness~\cite{fischer-servi:84,plotkin:stirling:86}]\label{thm:plotkin}
  A formula~$\fm A$ is a theorem of\/~$\ISfour$ if and only if $\fm A$~is valid in every birelational frame~$\langle W, \rel, \le \rangle$ where $\rel$~is reflexive and transitive.
\end{theorem}

\section{Why Is $\ISfour$'s Decidability Hard?}
\label{sec:problem}


In this section we highlight the main difficulties that we encountered in tackling the decidability problem for  $\ISfour$ and give a hint at the key ingredients of our method in the process.

One way to prove decidability for a logic is to perform proof search in a sound and complete deductive system with the intention of either finding a proof or constructing a countermodel from a failed proof search. %

For~$\ISfour$ several such deductive systems exist, the first being Simpson's labelled systems~\cite{simpson:phd}. 
Moreover, there are
 various kinds of nested sequents systems: single-conclusion~\cite{str:fossacs13}, multiple conclusion~\cite{kuz:str:2017maehara}, and also focused variants~\cite{cha:mar:str:fscd16}.
A natural question to ask is why none of these systems has been used to prove decidability of~$\ISfour$.  

\textbf{The need for more labelling.}
The aforementioned systems rely on what we could call a mixed approach: they internalize the modal accessibility relation $R$ within the sequent syntax, using either  labels and relational atoms or nesting, but they rely on a traditional structural approach for the intuitionistic aspect of the logic, e.g., single-conclusion sequents (at least in certain rules). 
One might think that combining the traditional loop-check for the intuitionistic part with the label-based loop-check for the modal part would be a way to a decision procedure.
But the situation is more complicated because
\begin{enumerate}
	\item 
	the classical $\Sfour$-loop test, which is looking for a
	repetition along the $\rel$-relation, cannot be applied to the right-hand-side of a sequent, as the conclusion formula can sometimes  be replaced by a new one;
	
	\item the structural approach to the intuitionistic system also means the rules are not all invertible and the procedure requires backtracking, so the
	modality loop-check also needs to be combined with the necessary backtracking.
\end{enumerate}

Both of these problems can be overcome by using a fully labelled proof system that incorporates both relations~$\rel$~and~$\le$~\cite{mar:mor:str:2021, maffezioli:naibo:negri:13}.
This has the same advantages as moving from a structural to a labelled approach for intuitionistic propositional logic, mentioned in the introduction.
Not only does this system  re-establish the close relationship between a sequent and a model, as is known from classical modal logic, it also enables us to make all rules in the system invertible.
Moreover, explicit relational atoms in the sequent syntax  make it easy to implement the loop-checks and to represent the back edges explicitly when constructing a countermodel.\looseness=-1

\textbf{The backtracking/termination trade-off.}
Naive proof search is not terminating, with two possible
sources of non-termination: the first inherited from the classical
modal logic~$\Sfour$, and the second from intuitionistic
propositional logic. 
In both those cases independently, the problem would be solved by a
simple loop-check. However, it is not straightforward to combine the two, as the following example shows:

	\begin{example}
		Consider the following formula, which is not provable in~$\ISfour$:
	\begin{equation}
		\label{eq:back-example}
		\fm{\wbox\Bigl((\wbox
			a\IMP\BOT)\cand\bigl((a\IMP\BOT)\IMP\BOT\bigr)\Bigr)\IMP
			\BOT}
		\text{.}
	\end{equation}
	Let $\fm A=\fm{\wbox a\IMP\BOT}$ and $\fm B=\fm{(a\IMP\BOT)\IMP\BOT}$.  In
	order to construct a countermodel~$\M$, we need a world~$\lb {w_1}$ that forces
	$\fm{\wbox(A\cand B)}$ and, therefore, also  $\fm A$~and~$\fm B$.
	Consequently, every world~$\lb{w'}$ such that $\futs {w_1}{w}$ and $\accs w{w'}$ for some~$\lb w$ should force these formulas. 
	Then all these worlds must
	force neither $\fm{\wbox a}$ nor $\fm{a\IMP\BOT}$. The latter means that for
	each such world~$\lb{w'}$, there must be a world~$\lb v$ with $\futs{w'}v$ and
	$\force\M va$. Of course, in turn, $\lb v$~must not force~$\fm{\wbox a}$, so there must be  worlds~$\lb u$ and $\lb{u'}$ with $\futs v{u'}$, $\accs{u'}u$, and
	$\nforce\M ua$. But this~$\lb u$ must not force~$\fm{a\IMP\BOT}$ so there must be a world~$\lb{v_1}$ with $\futs u{v_1}$ and $\force\M{v_1}a$,  and so on.  Thus, a naive implementation of a
	countermodel construction via proof search will keep adding worlds
	\emph{ad~infinitum} because neither of the two loop-checks will detect the repetition (see Fig.~\ref{fig:inf-model}, Left).
	\begin{figure}
		\begin{center}
			\begin{tikzpicture}[thick, every node/.style={scale=1.2}]
		
		\tikzstyle{node}=[circle,draw]
		\tikzstyle{s-node}=[circle,draw,inner sep=2pt]
		\tikzstyle{nonode}=[inner sep=0pt]
		\tikzstyle{access}=[->,blue]
		\tikzstyle{future}=[dashed,->]
		\tikzstyle{suricata}=[dashed,red]
		\tikzstyle{deleted}=[->,gray]
		\tikzstyle{loop}=[->,red]		
		
		\node[label={[align=center] right :{$\Vdash \BOX(A \AND B), \Vdash A, \Vdash B$\\$\not\Vdash \BOX a, \not\Vdash a \IMP\BOT$}}] (1) at (2,0) [node] {$w_1$}; 
		
		\node[] (2) at (2,2) [node] {$w_2$};
		\node[label= right :{$\not\Vdash a$}] (3) at (4,2) [node, fill=green!30] {$w_3$};
		%
		
		\node[label= right :{$\Vdash a$}] (4) at (4,4) [node, fill=purple!30] {$w_4$};
		\node[] (5) at (2,4) [node] {$w_5$};
		
		\node[label= right:{}] (6) at (4,6) [node] {$w_6$};
		\node[label= right :{$\not\Vdash a$}] (7) at (6,6) [node,fill=green!30] {$w_7$};
		\node[] (8) at (2,6) [node] {$w_8$};
		
		\node[label= right :{$\Vdash a$}] (9) at (6,8) [node,fill=purple!30] {$w_9$};
		\node[label= right :{}] (10) at (4,8) [s-node] {$w_{10}$};
		\node[] (11) at (2,8) [s-node] {$w_{11}$};

		\draw[access] (2) -- (3);
		
		\draw[access] (5) -- (4);
		
		\draw[access] (8) -- (6);
		\draw[access] (6) -- (7);
		
		\draw[access] (11) -- (10);
		\draw[access] (10) -- (9);
		
		\draw[future] (1) -- (2);
		\draw[future] (2) -- (5);
		\draw[future] (5) -- (8);
		\draw[future] (8) -- (11);
		\draw[future] (11) -- (2,9);
		\draw[future] (3) -- (4);
		\draw[future] (4) -- (6);
		\draw[future] (6) -- (10);
		\draw[future] (10) -- (4,9);
		\draw[future] (7) -- (9);
		\draw[future] (9) -- (6,9);

	\end{tikzpicture}
\hspace*{2cm}
\begin{tikzpicture}[thick, every node/.style={scale=1.2}]

\tikzstyle{node}=[circle,draw]
\tikzstyle{s-node}=[circle,draw,inner sep=2pt]
\tikzstyle{nonode}=[inner sep=0pt]
\tikzstyle{access}=[->,blue]
\tikzstyle{future}=[dashed,->]
\tikzstyle{suricata}=[dashed,red]
\tikzstyle{deleted}=[->,gray]
\tikzstyle{loop}=[->,red]		

\node[] (1) at (2,0) [node] {$w_1$}; 

\node[] (2) at (2,2) [node] {$w_2$};
\node[label= right :{$\not\Vdash a$}] (3) at (4,2) [node, fill=green!30] {$w_3$};
%

\node[label= right :{$\Vdash a$}] (4) at (4,4) [node, fill=purple!30] {$w_4$};
\node[] (5) at (2,4) [node] {$w_5$};

\node[label= right:{}] (6) at (4,6) [node] {$w_6$};
\node[] (8) at (2,6) [node] {$w_8$};

\node[label= right :{}] (10) at (4,8) [s-node] {$w_{10}$};
\node[] (11) at (2,8) [s-node] {$w_{11}$};

\draw[access] (2) -- (3);

\draw[access] (5) -- (4);

\draw[access] (8) -- (6);
\draw[access] (6) edge[bend right] (3);

\draw[access] (11) -- (10);
\draw[access] (10) edge[bend left] (4);

\draw[future] (1) -- (2);
\draw[future] (2) -- (5);
\draw[future] (5) -- (8);
\draw[future] (8) -- (11);
%
\draw[future] (3) -- (4);
\draw[future] (4) -- (6);
\draw[future] (6) -- (10);
%
%

\end{tikzpicture}
	
		\end{center}
		\caption{Left: Illustration of the potential non-termination issue. \quad Right: Illustration of a break of condition $\rn {F_1}$ when identifying nodes unrestrictedly.}
		\label{fig:inf-model}
	\end{figure}
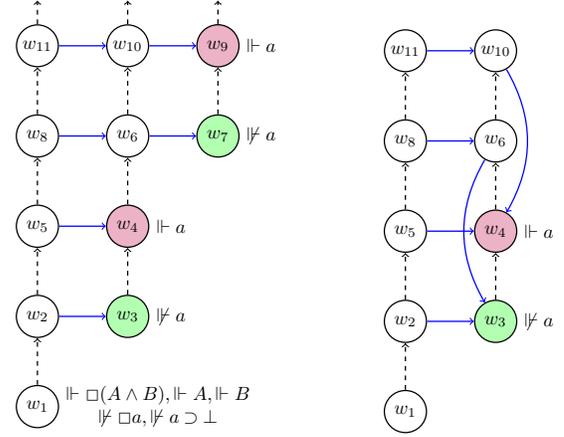
        In other words, 
\begin{enumerate}\setcounter{enumi}{2}
	\item to account for the interaction of modalities and intuitionistic implications, the two loop-checks must get along well with each other. 
\end{enumerate}

	\end{example}

\begin{figure*}[!t]
	\scalebox{1}{\fbox{
		\begin{minipage}{.97\textwidth}
			\begin{center}
				\begin{tabular}{c c}
				$\vlinf{\labrn{id}}{}{\B, \futs xy, \Left, \labels xa \SEQ \Right, \labels ya }{}$
				&
				$\vlinf{\lef\BOT}{}{\B, \Left, \labels x\BOT \SEQ \Right}{}$\\[0.5cm]
				$ 	\vlinf{\lef\AND}{}{\B,\Left, \labels x{A \AND B} \SEQ \Right}{\B, \Left, \labels xA, \labels xB \SEQ \Right}$
				&
				$\vliinf{\rig\AND}{}{\B,\Left \SEQ \Right, \labels x{A \AND B}}{\B, \Left \SEQ \Right, \labels xA}{\B, \Left \SEQ \Right,  \labels xB} $\\[0.5cm]
				$\vliinf{\lef\OR}{}{\B, \Left, \labels x{A \OR B} \SEQ \Right}{\B, \Left, \labels xA \SEQ \Right}{\B, \Left, \labels xB \SEQ \Right}$
				&
				$	\vlinf{\rig\OR}{}{\B, \Left \SEQ \Right, \labels x{A \OR B}}{\B, \Left \SEQ \Right,  \labels xA, \labels xB}$ \\[0.5cm]
				$ 	\vliinf{\lef\IMP}{}{\B, \futs xy, \Left, \labels x{A \IMP B} \SEQ \Right}{\B, \futs xy, \labels x{A \IMP B}, \Left \SEQ \Right, \labels yA}{\B, \futs xy, \Left, \labels yB \SEQ \Right}$
				&
				$\vlinf{\rig\IMP}{\proviso{$\lb z$ fresh}}{\B, \Left \SEQ \Right, \labels x{A \IMP B}}{\B, \futs xz, \Left, \labels zA \SEQ \Right,  \labels zB} $
				\\[0.5cm]
				$ 	\vlinf{\lef\BOX}{}{\B, \futs xy, \accs yz, \Left, \labels x{\BOX A} \SEQ \Right}{\B, \futs xy, \accs yz, \Left, \labels x{\BOX A}, \labels zA \SEQ \Right}$ 
				&
				$\vlinf{\rig\BOX}{\proviso{$\lb u,\lb z$ fresh}}{\B, \Left \SEQ \Right, \labels x{\BOX A}}{\B, \futs xu, \accs uz, \Left \SEQ \Right,  \labels zA} $
				\\[0.5cm]
				$\vlinf{\lef\DIA}{\proviso{$\lb y$  fresh}}{\B, \Left, \labels x{\DIA A} \SEQ \Right}{\B, \accs xy, \Left, \labels yA \SEQ \Right}$
				&
				$		\vlinf{\rig\DIA}{}{\B, \accs xy, \Left \SEQ \Right, \labels x{\DIA A}}{\B, \accs xy, \Left \SEQ \Right, \labels x{\DIA A}, \labels yA}$ \\[0.5cm]
				\multicolumn{2}{l}{\hbox to .97\linewidth{\dotfill}}\\[0.3cm]
				$ 	\vlinf{\Lref}{}{\B, \Left \SEQ \Right}{\B, \futs xx, \Left \SEQ \Right}$
				&
				$\vlinf{\Ltr}{}{\B, \futs xy, \futs yz, \Left \SEQ \Right}{\B, \futs xy, \futs yz, \futs xz, \Left \SEQ \Right}
				 $\\[0.5cm]
				 $ 	\vlinf{\Rref}{}{\B, \Left \SEQ \Right}{\B, \accs xx, \Left \SEQ \Right}$
				 &
				 $	 \vlinf{\Rtr}{}{\B, \accs xy, \accs yz, \Left \SEQ \Right}{\B, \accs xy, \accs yz, \accs {x}z ,\Left \SEQ \Right} $\\[0.5cm]
				 $ \vlinf{\fone}{\proviso{$\lb u$ fresh}
				 }{\B, \accs xy, \futs yz, \Left \SEQ \Right}{\B, \accs xy, \futs yz, \futs xu, \accs uz, \Left \SEQ \Right}$
				 &
				 $  \vlinf{\ftwo}{\proviso{$\lb u$ fresh}}{\B, \accs xy, \futs xz, \Left \SEQ \Right}{\B, \accs xy, \futs xz, \futs yu, \accs zu, \Left \SEQ \Right } $\\[0.3cm]
			\end{tabular}
				\end{center}
			\end{minipage}
		}		
	}
	\caption{System~$\labISf$}
	\label{fig:labIKp}
\end{figure*}

\begin{figure*}[!t]
	\scalebox{1}{\fbox{
			\begin{minipage}{.97\textwidth}
				\[
				\vlinf{\monl}{}{\lseq{\B, \futs xy}{\Left, \labels xA}\Right}{
					\lseq{\B, \futs xy}{\Left, \labels xA, \labels yA}\Right}
				\qquad
				\vlinf{\lef\vax}{}{\B, \accs xy, \labels{x}{\BOX A}, \Left \SEQ \Right}{\B, \accs xy, \labels{x}{\BOX A}, \labels{y }{\BOX A},  \Left \SEQ \Right }
				\qquad
				\vlinf{\lef{\cont}}{}{\B, \Left, \labels{x}{A} \SEQ \Right}{\B, \Left, \labels{x}{A}, \labels{x}{A} \SEQ \Right}
				$$
				$$
				\quad
				\vlinf{\w}{}{\lseq{\B,\B'}{\Left,\Left'}{\Right,\Right'}}{\lseq{\B}{\Left}{\Right}}
				\qquad\quad
				\vlinf{\rig\vax}{}{\B, \accs xy, \Left \SEQ \Right, \labels{x}{\DIA A}}{\B, \accs xy,  \Left \SEQ \Right,  \labels{x}{\DIA A}, \labels{y }{\DIA A} }
				\qquad
				\vlinf{\rig{\cont}}{}{\B, \Left \SEQ \Right, \labels{x}{A}}{\B, \Left \SEQ \Right, \labels{x}{A}, \labels{x}{A}}
				\]
				
				%
			\end{minipage}
	}}		
	\caption{Admissible rules}
	\label{fig:AdmRules}
\end{figure*}

\textbf{The interleaving of proof search and loops.}
For solving this third problem, we have to implement a more sophisticated loop-check involving both relations.
This directly leads us to the fourth problem.

\begin{example}
  Assume, for the sake of example, that we designed a suitable loop-check such that we could stop proof search at the stage of the structure presented on Fig.~\ref{fig:inf-model} (Left) and that we identified $\lb{w_7}$ to $\lb{w_3}$ and $\lb{w_{9}}$ to $\lb{w_4}$. 
		This would create ``backlinks'' between $\lb{w_{10}}$ and $\lb{w_4}$ as well as between~$\lb{w_6}$~and~$\lb{w_3}$.
		However, this would lead to a violation of~$\fone$  as now $\lb{w_{10}} R \lb{w_4} \mathord{\le} \lb{w_6}$ but there is no $\lb{w'}$ such that $\lb{w_{10}} \mathord{\le} \lb{w'} R \lb{w_6}$. (see Fig.~\ref{fig:inf-model}, Right).
                This means that %
\begin{enumerate}\setcounter{enumi}{3}
	\item the standard method of constructing a (finite) countermodel from a failed proof search via identifying labels/worlds that create a loop fails in the setting of birelational models. We break the \rn{F_1}/\rn{F_2} properties, which in turn would force us to add new worlds, which would mean we have to continue proof search.
\end{enumerate}
\end{example}

We solve this problem by identifying (substituting) labels not only after the completion of but also during   proof search.  
This preserves unprovability, but could, \emph{a priori}, be unsound. 
This means that, when terminating a branch on a non-axiom\-at\-ic sequent, it is still possible to extract a countermodel from it.
However, when reaching only axiomatic leaves, it remains to be shown that a sound proof can be obtained from the proof attempt (potentially containing identification of labels).
So, instead of doing naive proof search and then constructing a countermodel from a failed proof search by ``folding'' the failed sequent, we perform the  folding already during the proof search, which is now a countermodel search, and then construct a proper proof from a failed countermodel search by ``unfolding'' the search tree.
The loop-check ensuring termination has to be subtly calibrated for this final step of  unfolding the proof attempt into a real proof.


\begin{figure*}
	\scalebox{.9}{
		$
		\vlderivationnc{
			\vlin{\rig\IMP  }{}{\SEQ\labels{1}{\BOX (\DIA A \AND \DIA b) \IMP \BOT}}{
				\vlin{\Rref+\lef\BOX+\lef\AND}{}{\futs{1}{2},\labels{2}{\Box (\DIA A \AND \DIA b) }\SEQ\labels{2}{\BOT}}{
				\vlin{\lef\DIA}{}{\futs{1}{2},\accs{2}{2},\labels{2}{\Box (\DIA A \AND \DIA b) }, \labels{2}{\Diamond A}, \labels{2}{\DIA b}\SEQ\labels{2}{\BOT}}{
					\vliin{\Lref+\lef\IMP}{}{\futs{1}{2},\accs{2}{2}, \accs{2}{3},\labels{2}{\Gamma},\labels{3}{A}\SEQ\labels{2}{\BOT}}{
						\vlin{\rig\IMP}{}{\futs{1}{2},\accs{2}{2}, \accs{2}{3},\futs{3}{3},\labels{2}{\Gamma},\labels{3}{A}\SEQ\labels{2}{\BOT},\labels{3}{c \IMP \DIA b}}{
							\vlin{\lef\BOX+\lef\AND}{}{\futs{1}{2},\accs{2}{2}, \accs{2}{3},\futs{3}{3},\futs{3}{5},\labels{2}{\Gamma},\labels{3}{A},\labels{5}{c}\SEQ\labels{2}{\BOT},\labels{5}{\DIA b}}{
							\vlin{\lef\DIA}{}{\futs{1}{2},\accs{2}{2}, \accs{2}{3},\futs{3}{3},\futs{3}{5},\labels{2}{\Gamma},\labels{3}{A}, \labels{5}{c},\labels{3}{\DIA A}, \labels{3}{\DIA b}\SEQ\labels{2}{\BOT},\labels{5}{\DIA b}}{
								\vlin{\rn{F_2}+\lef{\rn{mon}}}{}{\futs{1}{2},\accs{2}{2}, \accs{2}{3},\futs{3}{3},\futs{3}{5},\accs{3}{4},\labels{2}{\Gamma},\labels{3}{A},\labels{5}{c},\labels{3}{\DIA A}, \labels{4}{b}\SEQ\labels{2}{\BOT},\labels{5}{\DIA b}}{
									\vlin{\rig\DIA}{}{\futs{1}{2},\accs{2}{2}, \accs{2}{3},\futs{3}{3},\futs{3}{5},\accs{3}{4}, \futs{4}{6}, \accs{5}{6}, \labels{2}{\Gamma},\labels{3}{A},\labels{5}{c},\labels{3}{\DIA A}, \labels{4}{b}, \labels{6}{b} \SEQ\labels{2}{\BOT},\labels{5}{\DIA b}}{
										\vlin{\idr}{}{\futs{1}{2},\accs{2}{2}, \accs{2}{3},\futs{3}{3},\futs{3}{5},\accs{3}{4}, \futs{4}{6}, \accs{5}{6},\labels{2}{\Gamma},\labels{3}{A},\labels{5}{c},\labels{3}{\DIA A}, \labels{4}{b}, \labels{6}{b} \SEQ\labels{2}{\BOT},\labels{5}{\DIA b}, \labels{6}{b}}{
											\vlhy{}
										}
									}
								}
							}
						}
					}
					}{
					\vlin{\hspace*{-1.5cm}\lef\BOT}{}{\sq G}{\vlhy{}}
					}
				}
			}
		}
		}
		$
		\hspace*{1cm}
			
	\begin{tikzpicture}[thick, every node/.style={scale=1.2}]

		\tikzstyle{node}=[circle,draw]
		\tikzstyle{s-node}=[circle,draw,inner sep=2pt]
		\tikzstyle{nonode}=[inner sep=0pt]
		\tikzstyle{access}=[->,blue]
		\tikzstyle{future}=[->,dashed]
		\tikzstyle{deleted}=[->,gray]
		\tikzstyle{loop}=[->,red]
		
		
		\node (1) at (2,0) [node] {$1$};
		
		\node[label={[align=center]above :{$\lef{\Box (\DIA A \AND \DIA b)},$\\$\lef\DIA b,\rig\bot$}}] (2) at (2,2) [node,fill=purple!30] {$2$};
		\node[label=below :{$\lef{\DIA A},\lef A$}] (3) at (4,2) [node, fill=green!30] {$3$};
		\node[label=below :{$\lef b$}] (4) at (6,2) [node, fill=yellow!30] {$4$};
		%
		
		\node[label=above :{$ \lef c, \rig{\DIA b}$}] (10) at (4,4) [node, fill=blue!30] {$5$};
		\node[label=above :{$\lef b, \rig b$}] (11) at (6,4) [node, fill=red!30] {$6$};
		
		
		\draw[access] (2) -- (3);
		\draw[access] (3) -- (4);
		%
		
		\draw[access] (10) -- (11);
		%
		
		\draw[future] (1) -- (2);
		\draw[future] (3) -- (10);
		\draw[future] (4) -- (11);
		%
	\end{tikzpicture}

	}
	\caption{Left: Proof in $\labISf$ of  $ \fm{\BOX (\DIA A \AND \DIA b) \IMP \BOT}$ with $ \fm A = \fm{(c \IMP \DIA b ) \IMP \BOT} $, $\fm\Gamma = \{\fm{\Box (\DIA A \AND \DIA b) }, \fm{\DIA b}\}$ and
	$ \sq G = \futs{1}{2},\accs{2}{2}, \accs{2}{3}, \futs{3}{3},\labels{2}{\Gamma},\labels{3}{\BOT}\SEQ\labels{2}{\BOT}$.
	%
	 Right: Diagrammatic representation of the top left sequent in the derivation.
	}
	\label{fig:ex-deriv}
\end{figure*}

\section{Preliminaries: Fully Labelled Sequents}
\label{sec:labelled_seq}

In this section we present the fully labelled sequent proof system~$\labISf$~\cite{mar:mor:str:2021} for intuitionistic modal logics we are using in order to prove decidability of~$\ISfour$. The starting point is the notion of a \emph{labelled formula} which is a pair~$\labels xA$ of a label~$\lb x$ and a formula~$\fm A$. 
A \emph{relational atom} is either an expression $\accs xy$ or $\futs xy$ where $\lb x$~and~$\lb y$~are labels. 
A \emph{(labelled) sequent}~$\sq G$ is a triple~$\lseq\B\Left\Right$
where $\B$~is a set of relational atoms and $\Left$~and~$\Right$~are multisets of labelled formulas, all written as comma-separated lists. 
Figure~\ref{fig:labIKp} shows the inference rules of~$ \labISf$.\footnote{Note that $\labISf$~has no cut rule, meaning that cut is admissible. For a syntactic proof of cut elimination, see~\cite{mar:mor:str:2021}.}
Figure~\ref{fig:AdmRules} shows some additional rules, and we define $\labISfp \colonequals \labISf \cup\set{\monl, \w, \lef{\vax}, \rig{\vax}, \lef{\cont}, \rig{\cont}}$.

A \defn{derivation} is a finite tree where each edge is a sequent and each internal node is a rule.
A derivation is a \defn{proof} of the sequent at the root if{f} each leaf is a rule with no premises, i.e.,~$\lef\BOT$~or~$\idr$.
As shown in~\cite{mar:mor:str:2021}, $\labISf$ is sound and complete for~$\ISfour$.\looseness=-1

\begin{theorem}[\cite{mar:mor:str:2021}]
  \label{thm:labIS4}
	A formula~$\fm A$~is a theorem of\/~$\ISfour$ if{f} for every $\lb x$, the sequent  $\SEQ\labels xA$ is derivable in\/~$\labISf$.
\end{theorem}

\begin{example}
	\label{ex:ex-deriv}
	On the left of Fig.~\ref{fig:ex-deriv} we have a proof in $ \labISf $ for the valid formula  $\fm F = \fm{\BOX (\DIA A \AND \DIA b) \IMP \BOT}$, with $ \fm A = \fm{(c \IMP \DIA b ) \IMP \BOT} $, where, for reasons of space,  several rules are sometimes  condensed into a  single step. 
\end{example}

\begin{corollary}\label{cor:adm}
	Rules\/ $\monl$,\/ $\w$,\/ $\lef{\vax}$,\/ $\rig{\vax}$, $\lef{\cont}$, and\/~$\rig{\cont}$ are admissible for\/~$\labISf$. Therefore, a formula~$\fm A$ is derivable in\/~$\labISfp$ if{f} it is derivable in\/~$\labISf$.
\end{corollary}

\section{From Sequents to Models}
\label{sec:sequents_models}

In this section we establish the promised correspondence between sequents and models.

\begin{notation}\label{not:fully-labelled}
	Let $\sq G$~be a sequent $\lseq\B\Left\Right$. 
	We write

	\begin{itemize}
		\item  $\futsq xy$ if{f} $\futs xy$ occurs in~$\B$;
		\item  $\accsq xy$ if{f} $\accs xy$ occurs in~$\B$;
		\item  $\blackf xAG$ 
		if{f} $\labels xA$ occurs in~$\Left$\\ (in this case we also say that $\fmb A$~\defn{occurs} at~$\lb x$ in~$\sq G$);
		\item  $\whitef xAG$ 
		if{f} $\labels xA$ occurs in~$\Right$\\ (in this case we also say that $\fmw A$~\defn{occurs} at~$\lb x$ in~$\sq G$);
              \item $\labelsof G$ for the set of labels occurring in $\sq G$.
	\end{itemize}
\end{notation}

This notation is used extensively throughout the paper.\footnote{The use of $\bullet$ and $\circ$ in this way goes back to Lamarche~\cite{lamarche:2003}.}

\begin{example}
	\label{ex:model_notation}
	This notation also enables us to represent sequents graphically, as shown on the right of Fig.~\ref{fig:ex-deriv}, which
        depicts the top-left sequent of the proof showcased on the left of the figure.  The sequent is represented by means of a directed graph whose nodes  are the labels of the sequent (for convenience we shall use natural numbers), dashed edges are the $\leq$-relations, and solid edges are the $R$-relations. We can now simply write a formula  $\fmb{A}$~($\fmw{A}$) next to a label $ \lb w $, to indicate that the labelled formula $\labels{w}{A}$ occurs on the left-hand side (right-hand side) of~$\SEQ$.
\end{example}

\begin{definition}[Happy labelled formula]\label{def:happyformula}
	A formula~$\fmb A$~($\fmw A$) is \defn{happy} at a label~$\lb x$ in a sequent~$\sq G$, or $\blackf xAG$ ($\whitef xAG$) is \defn{happy} for short, if{f} the following conditions hold:
	\begin{itemize}
		\item $\blackf xaG$ is always happy; 
		\item $\whitef xaG$ is happy if{f} we do not have $\blackf xaG$;
		\item $\blackf x\BOT G$ is never happy;
		\item $\whitef x\BOT G$ is always happy;
		\item $\blackf x{A \AND B}G$ is happy if{f} $\blackf xAG$ and $\blackf xBG$; 
		\item
		$\whitef x{A \AND B}G$ is happy if{f} $ \whitef xAG$ or $\whitef xBG$;
		\item
		$\blackf x{A \OR B}G$ is happy if{f} $\blackf xAG$ or $\blackf xBG$;
		\item
		$\whitef x{A \OR B}G$ is happy if{f} $\whitef xAG$ and $\whitef xBG$;
		\item
		$\blackf x{A \IMP B}G$ is happy if{f} 
                  $\whitef xAG$ or $\blackf xBG$;
		\item
		$\whitef x{A \IMP B}G$ is happy if{f}\\
		\strut\hfill 
		$\blackf yAG$ and $\whitef yBG$ for some~$ \lb y $ s.t.\ $\futsq xy $;
		\item
		$\blackf x{\BOX A}G$ is happy if{f}\\
		\strut\hfill  
		$\blackf zAG$ and $\blackf z{\BOX A}G$ for  all~$ \lb z$ with $ \accsq xz $;
		\item
		$\whitef x{\BOX A}G$ is happy if{f}\\
		\strut\hfill  
		$\whitef zAG$ for some $ \lb y$, $\lb z $ s.t.\ $ \futsq xy$ and $ \accsq yz $;
		\item
		$\blackf x{\DIA A}G$ is happy if{f} 
		$\blackf yAG$ for some~$ \lb y $ s.t.\ $ \accsq xy $; 
		\item 
		$\whitef x{\DIA A}G$ is happy if{f}\\
		\strut\hfill  
		$\whitef yA G$ and $\whitef y{\DIA A}G$ for all~$\lb y$ s.t.\ $ \accsq xy $.
	\end{itemize}
	Otherwise, the formula is \defn{unhappy}.	
\end{definition}

\begin{definition}[Happy label]\label{def:happy-label}
	A label~$ \lb x $ occurring in a sequent~$\sq{G}$ is \defn{happy} if{f} all formulas occurring at~$ \lb x$ in~$\sq G$ are happy. Label~$ \lb x $ is \defn{almost happy} if{f} all formulas occurring at~$\lb x$ are happy except, possibly, those of the shapes $\fmb\BOT$, $\fmw a$, $\fmw{A \IMP B}$, and~$\fmw{\BOX A}$.
	Label~$\lb x$ is \defn{naively happy} if{f} all formulas occurring at~$\lb x$ are happy except, possibly, those
 of the shapes~$\fmb\BOT$, $\fmb{\DIA A}$, $\fmw a$, $\fmw{A \IMP B}$,~and~$\fmw{\BOX A}$. 
	
\end{definition}
\begin{definition}[Structurally saturated sequent] 
	\label{def:struct-sound-seq}
	A sequent~$\sq{G}$ is \defn{structurally saturated} if{f} the following holds:
	\begin{enumerate}[leftmargin=3.5em]
		\item[($ \monl $)] if $\futsq xy$  and $\blackf xCG$, then $\blackf yCG$;
		\item[($\fone$)] if $ \accsq xy $ and  $ \futsq yz $, then there is~$\lb u$ such that $ \futsq xu $ and $ \accsq uz$;		
		\item[($ \ftwo $)] if $\accsq xy$ and  $\futsq xz$, then there is~$\lb u$ such that $\futsq yu $ and $ \accsq zu$;
		\item[($ \Ltr $)] if $ \futsq xy $ and $\futsq yz $, then $\futsq xz $;	
		\item[($ \Lref $)] $\futsq xx$ for all $ \lb x $ occurring in $\sq{G}$; 
		\item [($\Rtr$)] if  $ \accsq xy $ and $\accsq yz $, then  $\accsq xz$; 
		\item [($\Rref$)]  $\accsq xx$ for all $ \lb x $ occurring in $\sq{G}$.
		
	\end{enumerate}
\end{definition}

\begin{definition}[Happy sequent]\label{def:happysequent}
	A sequent~$\sq{G}$ is \defn{happy} if{f} it is structurally saturated and all labels in the sequent are happy.
\end{definition}

\begin{definition}[Model of a sequent]
	\label{dfn:modelofsequent}
	Let $\sq{G}$~be a sequent. We define the \defn{model~$\modelof{G}$ of}~$\sq{G}$ to be the quadruple
	$\modelof{G}=\tuple{\labelsof G,\sle G,\srel G,V}$ where $\labelsof G,\sle G,\srel G$ are as defined in Notation~\ref{not:fully-labelled} and $V$~is the
	 function $V\colon \labelsof G\to 2^\A$ defined such that for all atoms $\fm a \in \A$ we have 
	$\fm a \in V(\lb w)$ if{f} $\blackf waG$.
\end{definition}

\begin{theorem}[restate = completeness, name = Completeness] \label{thm:completeness}
	For a happy sequent\/~$\sq{G}$, its model~$\modelof{G}=\tuple{\labelsof G,\sle G,\srel G,V}$ is a transitive and reflexive birelational model with the following two properties:
	\begin{itemize}
		\item if\/ $\blackf xAG$, then $\force{\modelof{G}}xA$;
		\item if\/ $\whitef xAG$, then $\nforce{\modelof{G}}xA$.
	\end{itemize} 
\end{theorem}

\begin{definition}[Axiomatic sequent]
	A sequent~$\sq{G}$ is \defn{axiomatic} if{f} there is a label~$\lb x$ such that either $\blackf xaG$ and $\whitef xaG$ for some~$\fm a$, or $\blackf x\BOT G$.  Otherwise, $\sq{G}$~is called \defn{non-axiomatic}.
\end{definition}

\begin{remark}
	An axiomatic sequent~$\sq G$ is never happy, because either $\whitef xaG$ or $\blackf x\BOT G$ is unhappy.
\end{remark}

\section{Layers, Clusters, and Saturated Sequents}
\label{sec:clusters}

All derivations of $\SEQ\labels xA$ obtained by a proof search in $\labISf$  have a particular structure: their labels are partitioned into \emph{layers}, with each layer having a tree structure. This plays an important role in our proof search algorithm. 

\begin{definition}[Layer]
	For a sequent~$\sq G$, we define the relation~$\lrel G$ to be the transitive and reflexive closure of $\grel G \cup \grel{G}^{-1}$. Since this is an equivalence relation, we can define
	a \defn{layer}~$L$ in~$\sq G$ to be an equivalence class of~$\lrel G$.
\end{definition}

\begin{definition}[Layered sequent]\label{def:lay-sq}
	We say that a sequent~$\sq{G}$ is \defn{layered} if{f} for any 
	labels~$\lb x$, $\lb{x'}$, $\lb y$,~and~$\lb{y'}$ occurring in~$\sq G$: 
	\begin{enumerate}
		\item if   $\accsqh xy$ for $\lb x \neq \lb y$, then $\nfutsq xy$ and $\nfutsq yx$; and
		\item 
		\label{def:lay-sq-it2}
		 if  $\accsqh xy$, $\accsqh{x'}{y'}$, and $\futsq{x}{x'}$ for $\lb x \neq \lb {x'}$,  then  $\nfutsq{y'}{y}$.
	\end{enumerate}
        %
	For layers~$L_1$~and~$L_2$, we define $L_1 \le L_2$ whenever there are labels $\lb x \in L_1$ and $\lb y \in L_2$ such that $\futsq xy$. We write $L_1<L_2$ if{f} $L_1 \le L_2$ and $L_1 \neq L_2$. 
\end{definition}

\begin{proposition}[restate = OrderLayers, name = ] 
\label{prop:order}
	For a layered structurally saturated sequent\/~$\sq G$, relation\/~$\le$ is an order relation on its layers.%
\end{proposition}

The second condition in Def.~\ref{def:lay-sq} is only needed to establish antisymmetry in Prop.~\ref{prop:order}. If $\sle G$ on labels is already antisymmetric then the second condition follows from the first. 
In all sequents that we discuss here, the order relation $\le$ on layers defines a tree structure.

\begin{definition}[Tree-layered sequent]
  A layered sequent is \defn{tree-layered} if{f} (i)~there is a layer $L_0$ such that $L_0\le L$ for all layers $L$, and (ii)~for all layers~$L$, $L'$,~and~$L''$, whenever $L'\le L$ and $L''\le L$, then either $L'\le L''$ or $L''\le L'$.
\end{definition}

\begin{definition}[Topmost and inner layer]
	\label{def:topmostlayer}
	A layer~$L$ in a layered sequent~$\sq G$ is called a \defn{topmost layer} if{f} it is maximal with respect to~$\le$, i.e.,~whenever $L\le L'$ for some layer~$L'$, then $L=L'$. Otherwise, the layer~$L$ is called an \defn{inner layer}.
\end{definition}

Note that in a tree-layered structurally saturated sequent, topmost layers are  the leaves of the tree structure w.r.t.~$\le$.\looseness=-1

In a sequent $\sq G$  constructed by $\labISf$,  each layer also has a tree structure with respect to~$\grel G$. But, as said before, in order to search for a proof and a countermodel at the same time, we need to weaken this tree structure on the layers.%

\begin{definition}[Cluster]
\label{def:cluster}
  If~$\sq G$ is structurally saturated, then $\grel G \cap \grel{G}^{-1}$~is an equivalence relation, and we can define a \defn{cluster}~$\lbc C$ in~$\sq G$ to be an equivalence class $\lbc C=\{\lb{x_1},...,\lb{x_n}\}$ of~$\grel G \cap \grel{G}^{-1}$. A cluster~$\lbc C=\{\lb{x_1}\}$ containing only one label is called \defn{singleton}. On clusters, we define two binary relations:
  \begin{itemize}
  \item $\futsqc{C_1}{C_2}$ if{f} for all $\lb y\in \lbc{C_2}$ there is $\lb x\in \lbc{C_1}$ with $\futsq xy$.
  \item $\accsqc{C_1}{C_2}$ if{f} there are $\lb x\in\lbc C_1$ and $\lb y\in\lbc C_2$ with $\accsq xy$. 
  \end{itemize}
  We sometimes abuse the notation and replace one of the clusters by a label $\lb x$ even when $\{\lb x\}$ is not a cluster. Nevertheless, the definitions are then applied verbatim to $\{\lb x\}$.
\end{definition}

\begin{proposition}[restate = OrderClusters, name = ] 
	\label{prop:order_clusters}
	For a structurally saturated sequent\/~$\sq G$, relation $\grel G$~is an order and\/ $\sle G$~is a preorder on its clusters.  If\/ $\sq G$~is layered, then\/ $\sle G$~is also an order.%
\end{proposition}

\begin{definition}[Tree-clustered sequent]
  A structurally saturated sequent $\sq G$ is called \defn{tree-clustered} if{f} (i)~for any clusters~$\lbc {C'}$~and~$\lbc{C''}$ there is a cluster $\lbc C$, such that $\accsq{C}{C'}$ and $\accsq{C}{C''}$, and (ii)~for all clusters~$\lbc C$, $\lbc{C'}$,~and~$\lbc{C''}$, whenever $\accsq{C'}{C}$ and $\accsq{C''}{C}$, then either  $\accsq{C'}{C''}$ or $\accsq{C''}{C'}$.
\end{definition}

\begin{definition}[Happy layer]
	A layer~$L$ is \defn{happy/almost happy/naively happy} if{f} all labels of~$L$ are happy/almost happy/naively happy respectively. 
\end{definition}

\begin{remark}
\label{rem:tobereferred}
Observe that the definitions of happy labelled formulas for~$\fmb{A\IMP B}$~and~$\fmb{\BOX A}$ in Def.~\ref{def:happyformula} are different from the forcing conditions of the corresponding connectives.
This mismatch is intentional  with the missing conditions on $\le$ outsourced to $(\monl)$-structural saturation (Def.~\ref{def:struct-sound-seq}) instead. As a result, happiness becomes almost a local property for layers: if a sequent is modified (by adding labelled formulas or relational atoms) outside of a given happy layer, the only formulas whose happiness needs to be reinspected are~$\fmw{A\IMP B}$~and ~$\fmw{\BOX A}$. We use this observation in the notion of  stable sequents below, where inner layers are never modified and always remain happy.
\end{remark}

\begin{definition}[Stable and saturated sequent]
  A sequent~$\sq G$ is \defn{stable} if{f} it is tree-layered and tree-clustered and, additionally, all its inner layers are happy.  A sequent~$\sq G$ is \defn{saturated} if{f} it is stable and all its topmost layers are almost happy. It is \defn{semi-saturated} if{f} it is stable and all its topmost layers are naively happy. A set~$\sqset S$ of sequents is called \defn{stable/saturated/semi-saturated} if{f} it is finite and all elements of~$\sqset S$ are stable/saturated/semi-saturated respectively.
\end{definition}

Since tree-layered sequents are layered and tree-clustered sequents are structurally saturated (in fact, the definition of clusters does not make sense otherwise), all stable sequents are layered and structurally saturated.


\section{Search Algorithm}
\label{sec:algorithm}

The algorithm that we will present here works on stable sequents. Roughly speaking, we work on the topmost layers of such sequents until these layers are almost happy. This is achieved by first semi-saturating (which essentially means applying all inference rules to a layer that do not add new labels to the sequent), then picking an unhappy $\labelsb x{\DIA A}$ in the layer and making it happy (either  by applying the $\lef\DIA$-rule or, in case of loop detection, by creating a cluster). Then we semi-saturate again, and continue until all $\labelsb x{\DIA A}$ in the layer are happy and, therefore, the layer is almost happy.

After all layers are almost happy,  the only unhappy formulas remaining in the resulting saturated sequent are  in topmost layers and of the shape $\labelsb x\BOT$, $\labelsw xa$, $\labelsw x{A\IMP B}$ or $\labelsw x{\BOX A}$. For the first two shapes, the sequent is axiomatic, and we can stop working on it.

If all unhappy formulas are in topmost layers and of the shape $\labelsw x{A\IMP B}$ or $\labelsw x{\BOX A}$, we make them happy by creating new layers that become new topmost layers to be saturated while  formerly topmost layers turn into happy inner layers. This process is repeated, until we see a repetition of layers. In order to achieve termination of this process, we implement two other loop detection mechanisms (different from the one used for $\labelsb x{\DIA A}$ formulas).

In the remainder of this section, we formally introduce the concepts needed to understand the details of this algorithm, which is then shown in Fig.~\ref{fig:algorithm}.

\subsection{Semi-saturation}

We begin by making happy  all formulas  that are not of the form $\labelsb x\BOT$, $\labelsw xa$, $\labelsb x{\DIA A}$, $\labelsw x{A\IMP B}$, or $\labelsw x{\BOX A}$.

\begin{definition}[Semi-saturation]\label{def:sat-sqset}
	We define a rewrite relation~$\ssatred$ 
	on sets~$\sqset S$ of sequents:  $\sqset S\ssatred\sqset S'$ if{f} there is  a sequent~$\sq G\in\sqset S$ such that  some  labelled formula~$\fm C$ of one of the below shapes occurs at~$\lb x$ in~$\sq G$ and is unhappy, and $\sqset S'$ is obtained by replacing sequent~$\sq G$ in set~$\sqset S$ either with sequent~$\sq G'$  according to cases~1)--4) below, or with sequents~$\sq G'$~and~$\sq{G''}$ according to cases~5)--7) below, 
	 as follows:\looseness=-1
		\begin{enumerate}
			\item for $\fm C = \fmb{A \AND B}$, obtain~$\sq{G'}$ by adding to~$\sq{G}$ all of~$\labelsb xA$ and $\labelsb xB$  that are missing;
			\item for $\fm C = \fmw{A \OR B}$, obtain~$\sq{G'}$ by adding  to~$\sq{G}$ all  of~$\labelsw xA$~and~$\labelsw xB$ that are missing;
			\item for $\fm C = \fmb{\BOX A}$, obtain~$\sq{G'}$ by adding to~$\sq G$ all~$\labelsb zA$ and~$\labelsb z{\BOX A}$ that are missing whenever $\accsq xz$; 
			\item for $\fm C = \fmw{\DIA A}$, obtain~$\sq{G'}$ by adding to~$\sq G$ all~$\labelsw yA$ and~$\labelsw y{\DIA A}$ that are missing whenever $\accsq xy$; 
			\item  for $\fm C = \fmb{A \OR B}$, obtain~$\sq{G'}$ (respectively~$\sq{G''}$) by adding to~$\sq G$ the labelled formula $\labelsb xA$ (respectively~$\labelsb xB$);
			\item for $\fm C = \fmw{A \AND B}$, obtain~$\sq{G'}$ (respectively~$\sq{G''}$) by adding to~$\sq G$ the labelled formula  $\labelsw xA$ (respectively~$\labelsw xB$);
			\item  for $\fm C = \fmb{A \IMP B}$,  obtain~$\sq{G'}$ (respectively~$\sq{G''}$) by adding to~$\sq G$ the labelled formula $\labelsw xA$ (respectively~$\labelsb xB$). 
		\end{enumerate}
	We write~$\tssatred$ for the transitive and reflexive closure of~$\ssatred$. If $\sqset{S}\tssatred\sqset{S}'$ and $\sqset S'$~is in normal form w.r.t.~$\ssatred$, i.e., if  all labels occurring in all sequents from~$\sqset S'$ are naively happy, then $\sqset{S}'$~is called the \defn{semi-saturation} of~$\sqset S$.
\end{definition}

The use of this term is justified because, the semi-saturation of a stable set of sequents is semi-saturated.  

\begin{lemma}[restate = SaturationStable, name = ] 
	\label{lemma:ssat}
	All the following statements hold:
	\begin{enumerate}[a)]
		\item 
		\label{lemma:ssat:stable}
		If a set\/~$\sqset S$ is stable and\/ $\sqset{S} \ssatred\sqset{S}'$, then\/ 
		 $\sqset{S}'$~is stable. 
		\item 
		\label{lemma:ssat:term}
		On finite sets\/~$\sqset S$, the rewrite relation\/~$\ssatred$ is terminating.
		\item 
		\label{lemma:ssat:ssat}
		A set\/~$\sqset S$ is semi-saturated if{f} it is stable and in normal form w.r.t.\/~$\ssatred$.
	\end{enumerate}
\end{lemma}
%

\subsection{Saturation}

In the next step we show how to deal with unhappy~$\labelsb x{\DIA A}$. This is the first source of non-termination, and for this we employ the same method that is commonly used for classical~$\Sfour$. However, whereas in the case of classical~$\Sfour$, loop detection completes the proof search in the current branch, here we have  to continue proof search. For this reason we realize  loops by clusters.\looseness=-1

\begin{definition}[Equivalent labels/clusters]\label{def:eq-labels}
  Labels~$\lb x$~and~$\lb y$  occurring in sequents~$\sq{G}$~and~$\sq H$ respectively are \defn{equivalent}, in symbols $\lbeql{x}{\sq{G}}{y}{\sq{H}}$, if{f}  $\blackf xA G \Longleftrightarrow \blackf yA H$ and also $\whitef xA G \Longleftrightarrow \whitef yA H$ for all formulas~$\fm A$.
   If $\sq G$~and~$\sq H$~are clear from the context, we simply write $\lb x \lbeq \lb y$.
	For structurally saturated sequents~$\sq{G}$~and~$\sq H$, we can generalize this to clusters:  $\lbeql{C_1}{\sq G}{C_2}{\sq H}$ if{f} there is a bijection $f\colon \lbc{C_1}\to \lbc{C_2}$, such that $\lb{f x}\lbeq\lb x$ for all $\lb x\in \lbc{C_1}$.
\end{definition}

\begin{definition}[Having no past]
  A label~$\lb x$ in a sequent~$\sq G$ \defn{has no past} if{f} $\futsq yx$~implies $\lb y =  \lb x$ for all~$\lb y$.
  %
\end{definition}

\begin{definition}[Saturation]
	\label{def:dia-sat}
	 We define a binary relation~$\diared$ 
         on sequents as follows: $\sq G\diared\sq G'$ if{f} there is a label~$\lb y$ and a formula~$\fm A$,
          such that $\blackf y{\DIA A} G$ is unhappy, and all labels $\lb u \ne \lb y$ with $\accsq uy$ but not $\accsq yu$ are almost happy, and either
	\begin{enumerate}
	\item
Option~1: 
		\label{def:dia-sat:one}
		 there is 
		 a label $\lb x\neq\lb y$ such that $\accsq xy$ and \hbox{$\lb x\lbeq\lb y$}, and $\blackf x{\DIA A} G$ is happy, and every label $\lb u$ with~$\accsq xu$ has no past in~$\sq G$. Then $\sq G'$~is obtained from~$\sq G$ by first substituting~$\lb x$ for each  occurrence of~$\lb y$  and then closing the resulting~$\grel G$ under transitivity;
                         or
		\item
		\label{def:dia-sat:two}
	 	Option 2: There is no such label~$\lb x$. Then $\sq G'$~is obtained from~$\sq G$ by first adding $\futs zz$, $\accs yz$, $\labelsb z{A}$ for some fresh label~$\lb z$, and then  closing the resulting~$\grel G$ under transitivity and reflexivity.
         	\end{enumerate}
	For sets $\sqset S$ and $\sqset S'$ of sequents, we write $\sqset S\Diared\sqset S'$
	if{f} $\sqset S$~is semi-saturated, $\sq G\diared\sq G'$  for some $\sq G\in\sqset S$ and $\sqset S'$ is a semi-saturation of $\bigl(\sqset S\setminus\set{\sq G}\bigr) \cup \set{\sq G'}$.\footnote{Strictly speaking, in the first case $\sq G'$ is already semi-saturated, but to simplify the presentation, we semi-saturate  in both cases.}
	We write~$\tDiared$ for the transitive and reflexive closure of~$\Diared$.
	If $\sqset S\tDiared\sqset S'$ and $\sqset S'$ is in normal form w.r.t.~$\Diared$, then $\sqset S'$~is called a \defn{saturation} of~$\sqset S$.  
\end{definition}

As before, this term is justified because, a saturation of a semi-saturated set of sequents is saturated.

\begin{lemma}[restate = DiaSaturationStable, name = ] 
	\label{lemma:sat}
		All the following statements hold:
	        \begin{enumerate}[a)]
	        \item   
		\label{lemma:sat:stable}
		If a set\/~$\sqset S$ is 
		semi-saturated
		and\/ $\sqset S\Diared\sqset S'$, then\/ $\sqset S'$~is 
		semi-saturated.
		\item \label{lemma:sat:term}
		The rewrite relation\/~$\Diared$ is terminating. 
		
		\item \label{lemma:sat:sat} A set\/~$\sqset S$  is saturated if{f} it is semi-saturated and in normal form w.r.t.\/~$\Diared$.
	\end{enumerate}
\end{lemma}

\subsection{Lifting saturation}

After having discussed how to saturate sets of sequents, we are now going to show how to deal with unhappy $\rig{\fm{A \IMP B}}$ and $\rig{\fm{\BOX A}}$. For this, we have to create new layers. The naive way would be to just  copy the $R$-structure of the layer that contains the unhappy $\labelsw x{A \IMP B}$ or $\labelsw x{\BOX A}$. However, due to the presence of clusters, we sometimes need to create two copies of the cluster that contains  the unhappy formula.

\begin{definition}[Sequent sum]
	Let $\sq G$ and $\sq G'$ be  sequents $\lseq\B\Left\Right$ and $\lseq{\B'}{\Left'}{\Right'}$ respectively. We define their \defn{sum} ${\sq G} \sums {\sq G'}$ to be the sequent $\lseq{\B,\B'}{\Left,\Left'}{\Right,\Right'}$.
\end{definition}

\begin{construction}[Layer Lifting I]
	\label{def:lifting}
	Let $\sq G$ be a saturated sequent, $L$ be one of its topmost layers, and let $\lb x \in L$. 
	Then  $\lb x$ belongs to a cluster $\lbc{C_x}=\set{\lb{x_1},\ldots,\lb{x_h}}$ with $h\geq 1$ and $\lb x=\lb {x_m}$ for some $1 \leq m \leq h$, which we often abbreviate as $m = 1..h$. In particular, $\accsq{x_i}{x_{i'}}$ for any $i,i' = 1..h$. 
	Let $\set{\lb{y_1},\ldots,\lb{y_l}}=L\setminus \lbc{C_x}$ where $l\geq 0$. 
	Then, if $h=1$, let  $\hL$ be the set of fresh labels $\set{\lb{\hy_1},\ldots, \lb{\hy_l},\lb{\hx}}$. Otherwise, if $h>1$, let 
	$\hL \colonequals \set{\lb{\hy_1},\ldots, \lb{\hy_l},\lb{\hx},\lb{\hx_1'},\ldots, \lb{\hx_h'},\lb{\hx_1''},\ldots, \lb{\hx_h''}}$ 
	where again all labels are fresh. We  define $\llift Gx$ to consist of:%
	\begin{enumerate}
		\item
		relational atoms $\futs {v}{v}$ and $\accs vv$ for all $\lb v\in\hL$; 
		\item
		for each $i=1..l$, each $j=1..h$, and each  $\lb w \in \labelsof{G}$: 
		\begin{enumerate}
			\item relational atom $\futs {w}{\hy_i}$ whenever $\futsq {w}{y_i}$,  
			\item
			relational atom $\futs {w}{\hx}$ whenever $\futsq {w}{x}$,
			\item	 only for $h>1$: relational atoms $\futs {w}{\hx_j'}$ and $\futs
							{w}{\hx_j''}$ whenever $\futsq {w}{x_j}$;
		\end{enumerate}
		\item
		\label{def:lifting:rels}
		 	for all $i,i'=1..l$ and $j,j'=1..h$,
		\begin{enumerate}
			\item 
			\label{def:lifting:ys}
			relational atom $\accs{\hy_i}{\hy_{i'}}$ whenever $\accsq{y_i}{y_{i'}}$,
			\item 
			\label{def:lifting:yxs}
			 only for $h>1$:  relational atoms $\accs{\hy_i}{\hx_{j}'}$ and
			$\accs{\hy_i}{\hx''_{j}}$ whenever $\accsq{y_i}{x_{j}}$,
			\item 
			\label{def:lifting:yx}
			relational atom $\accs{\hy_i}{\hx}$ whenever $\accsq{y_i}{x}$,
			\item 
			\label{def:lifting:xsy}
			 only for $h>1$: relational atoms $\accs{\hx'_j}{\hy_i}$ and
			$\accs{\hx''_j}{\hy_i}$ whenever $\accsq{x_j}{y_i}$,
			\item 
			relational atom $\accs{\hx}{\hy_i}$ whenever $\accsq{x}{y_i}$,	 
			\item 
			\label{def:lifting:xc}
			 only for $h>1$: relational atoms  $\accs{\hx_j'}{\hx_{j'}'}$, $\accs{\hx'_j}{\hx}$, 
			   $\accs{\hx}{\hx''_{j}}$, and $\accs{\hx''_j}{\hx''_{j'}}$. 
		\end{enumerate}
		\item 
		\label{def:lifting:itemize:mon}
		For each $i=1..l$, each $j=1..h$,  and  each~$\fm C$, add:
		\begin{enumerate}
			\item
			labelled formulas $\labelsb{\hy_i}{C}$ whenever $\blackf
			{y_i}{C} G$,
			\item	 only for $h>1$: labelled formulas $\labelsb{\hx'_j}{C} $ and $\labelsb{\hx''_j}{C}$ whenever $\blackf {x_j}{C} G$,
			\item labelled formulas $\labelsb{\hx}{C}$ whenever $\blackf {x}{C} G$. 
		\end{enumerate}
	\end{enumerate}
\end{construction}

\begin{figure}[!t]
	\begin{center}
			\begin{tikzpicture}[thick, every node/.style={scale=1.2},font = {\large}]

		\tikzstyle{node}=[circle,fill=black,inner sep=1.5pt]
		\tikzstyle{nonode}=[inner sep=0pt]
		\tikzstyle{access}=[blue]
		\tikzstyle{future}=[dashed,->]
		
		\node[] (Lh) at (-1,6) {$\hat L$};
		\node[] (L) at (-1,2) {$L$};
		
		\node[label=below:{$x_2$}] (x2) at (5.8,2.95) [node] {};
		\node[label=below:{$x_1$}] (x1) at (5.5,1.15) [node] {};
		\node[label=below:{$x_3$}] (x3) at (6.8,1.4) [node] {};
		\node[label=below:{$y_1$}] (y1) at (2,2) [node] {};
		\node[label=below:{$y_2$}] (y2) at (3,2) [node] {};
		\node[label=left:{$y_3$}] (y3) at (4,2.7) [node] {};
		\node[label=below:{$y_4$}] (y4) at (8,2) [node] {};
		\node[label=below:{$y_5$}] (y5) at (9,2.7) [node] {};
		\node[label=below:{$y_6$}] (y6) at (9.5,1.3) [node] {};
		%

		\node[label=above:{$\hat x_1'$}] (x1') at (3.5,5.15) [node] {};
		\node[label=above:{$\hat x_1''$}] (x1'') at (7.5,5.15) [node] {};
		\node[label=above:{$\hat x_2'$}] (x2') at (3.8,6.95) [node] {};
		\node[label=above:{$\hat x$}] (xhat) at (5.8,6) [node] {};
		\node[label=above:{$\hat x_2''$}](x2'') at (7.8,6.95) [node] {};
		\node[label=above:{$\hat x_3'\:$}] (x3') at (4.8,5.4) [node] {};
		\node[label=above:{$\hat x_3''\:$}] (x3'') at (8.8,5.4) [node] {};
		\node[label={$z$}] (z) at (6.5,6.5) [node] {};
		\node[label=above:{$\hat y_1$}] (y1h) at (0,6) [node] {};
		\node[label=above:{$\hat y_2$}] (y2h) at (1,6) [node] {};
		\node[label=above:{$\hat y_3$}] (y3h) at (2,6.7) [node] {};
		\node[label=above:{$\hat y_4$}] (y4h) at (10,6) [node] {};
		\node[label=above:{$\hat y_5$}] (y5h) at (11,6.7) [node] {};
		\node[label=above:{$\hat y_6$}] (y6h) at (11.5,5.3) [node] {};
		%
			
		\draw[access] (y1) -- (5,2);
		\draw[access] (y2) -- (y3);
		\draw[access] (6,2) circle [radius=1cm];
		\draw[access] (7,2) -- (y4);
		\draw[access] (y4) -- (y5);
		\draw[access] (y4) -- (y6);
		
		\draw[access] (y1h) -- (3,6);
		\draw[access] (y2h) -- (y3h);
		\draw[access] (4,6) circle [radius=1cm];
		\draw[access] (5,6) -- (7,6);
		\draw[access] (xhat) -- (z);
		\draw[access] (8,6) circle [radius=1cm];
		\draw[access] (9,6) -- (y4h);
		\draw[access] (y4h) -- (y5h);
		\draw[access] (y4h) -- (y6h);
		
		\draw[future] (x1) -- (x1');
		\draw[future] (x1) -- (x1'');
		\draw[future] (x2) -- (x2');
		\draw[future] (x2) -- (xhat);
		\draw[future] (x2) -- (x2'');
		\draw[future] (x3) -- (x3');
		\draw[future] (x3) -- (x3'');
		\draw[future] (y1) -- (y1h);
		\draw[future] (y2) -- (y2h);
		\draw[future] (y3) -- (y3h);
		\draw[future] (y4) -- (y4h);
		\draw[future] (y5) -- (y5h);
		\draw[future] (y6) -- (y6h);

	\end{tikzpicture}
	\end{center}
	\caption{Layer lifting as defined in Constructions~\ref{def:lifting} and~\ref{def:liftingII}}
	\label{fig:mm-lifting}
\end{figure}
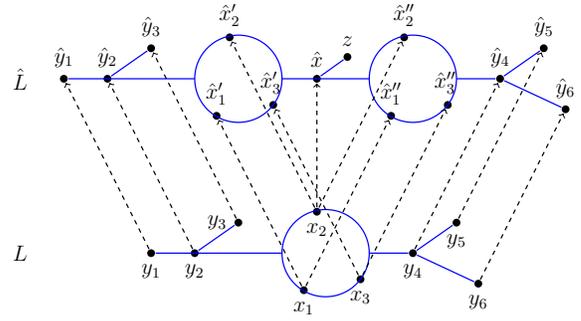

This construction lifts a layer $L$ of $\sq G$ w.r.t.~a label~$\lb x \in L$. If $\lb x$ is a singleton cluster, this is a simple lifting as one might expect. Otherwise, if $\lb x$ is in a non-singleton cluster, this cluster is duplicated and the lifting of~$\lb x$ is put in between the two copies, as indicated in Fig.~\ref{fig:mm-lifting} (ignore label~$\lb z$ for the moment). 

In Construction~\ref{def:lifting}, Points~1)--2) ensure that in $\sq G\sums\llift Gx$, the new layer is indeed above $L$ as intended and ($\Ltr$), ($\Lref$), and ($\Rref$) are satisfied; Point~3) ensures ($\Rtr$), ($\fone$), and ($\ftwo$); Point~4) ensures $(\monl)$. Thus, \mbox{$\sq G\sums\llift Gx$} is tree-clustered, tree-layered, and structurally saturated. However, it may not be stable because $L$ is now an inner layer. It is almost happy due to the saturation of~$\sq G$ but may contain unhappy $\rig{\fm{A \IMP B}}$ and $\rig{\fm{\BOX B}}$ formulas.

\begin{construction}[Layer Lifting II] 
  \label{def:liftingII}
	Let $\sq G$ be a saturated sequent with $\whitef xF G$ being unhappy
	for some label $\lb x$ and some formula $\fm F$ of shape
	$\fm{A\IMP B}$ or $\fm{\BOX B}$. We reuse the notation from Construction~\ref{def:lifting} (note that $L$ must be a topmost layer of $\sq G$ due to it being saturated).
	We define $\lliftf Gx{F}$ as follows: 
	\begin{itemize}
		\item If $\fm F = \fm{A\IMP B}$, then 
		$\lliftf Gx{A\IMP B}$ 
		is $\llift Gx$ to which we add
		labelled 
		formulas $\labelsb \hx {A}$ and $\labelsw \hx {B}$ and call $\lb\hx$  a \defn{suricata label of $\lb x$}
		\item If $\fm A = \fm{\Box B}$, then 
		$\lliftf Gx{\Box  B}$ is $\llift Gx$ to which we add 
                    $\accs zz$, $\futs zz$,
                  and $\labelsw z {B}$ for a fresh label~$\lb z$, and additionally add relational atoms $\accs{v}{z}$ whenever $\lb v\in\hL$ and  $\accs{v}{\hx}$ is in~$\llift{G}{x}$. Here $\lb z$ is called a \defn{suricata label of $\lb x$}. 
	\end{itemize}
\end{construction}

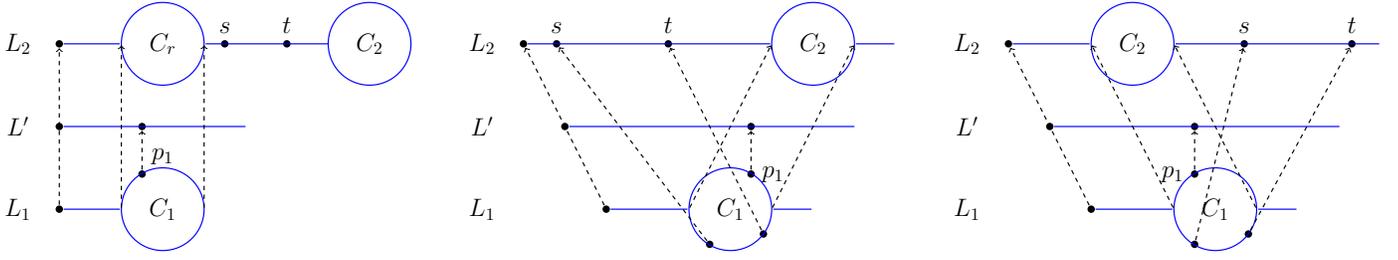
\begin{figure*}[t!]
		\begin{tikzpicture}[thick, every node/.style={scale=1.2},font = {\Large}]

		\tikzstyle{node}=[circle,fill=black,inner sep=1.5pt]
		\tikzstyle{nonode}=[inner sep=0pt]
		\tikzstyle{access}=[blue]
		\tikzstyle{future}=[dashed,->]
		
		\node[] (L1) at (2.5,2) {$L_1$};
		\node[] (L') at (2.5,4) {$L'$};
		\node[] (L2) at (2.5,6) {$L_2$};
		
		\node[label=above right:{$p_1$}] (p1) at (5.5,2.85) [node] {};
		\node[label=below:{}] (C1) at (6,2) {$C_1$};
		\node (l1) at (3.5,2) [node] {};
		\node (x1) at (5,2) [nonode] {};
		\node (y1) at (7,2) [nonode] {};


		\node[label=above:{}] (sur') at (5.5,4) [node] {};
		\node (l') at (3.5,4) [node] {};
		
		\node[label=below:{}] (Cr) at (6,6) {$C_r$};
		\node[label=below:{}] (C2) at (11,6) {$C_2$};
		\node[label=above:{$s$}] (s) at (7.5,6) [node] {};
		\node[label=above:{$t$}] (t) at (9,6) [node] {};
		%
		\node (l2) at (3.5,6) [node] {};
		\node (x2) at (5,6) [nonode] {};
		\node (y2) at (7,6) [nonode] {};

		\draw[access] (l1) -- (x1);
		\draw[access] (C1) circle [radius=1cm];
		
		\draw[access] (l') -- (8,4);

		\draw[access] (l2) -- (x2);
		\draw[access] (Cr) circle [radius=1cm];
		\draw[access] (y2) -- (10,6);
		\draw[access] (C2) circle [radius=1cm];

		\draw[future] (l1) -- (l2);
		\draw[future] (p1) -- (sur');
		\draw[future] (x1) -- (x2);
		\draw[future] (y1) -- (y2);
	\end{tikzpicture}

\hspace*{1cm}

\begin{tikzpicture}[thick, every node/.style={scale=1.2},font = {\Large}]
	
	\tikzstyle{node}=[circle,fill=black,inner sep=1.5pt]
	\tikzstyle{nonode}=[inner sep=0pt]
	\tikzstyle{access}=[blue]
	\tikzstyle{future}=[dashed,->]
	
	\node[] (L2) at (0,6) {$L_2$};
	\node[] (L') at (0,4) {$L'$};
	\node[] (L1) at (0,2) {$L_1$};
	
	\node[label=below :{}] (C1) at (6,2) {$C_1$};
	\node[label=right:{$p_1$}] (p1) at (6.5,2.85) [node] {};
	\node[] (x1) at (5.5,1.15) [node] {};
	\node[] (x3) at (6.8,1.4) [node] {};
	\node[] (y1) at (3,2) [node] {};
	\node[] (y2) at (5,2) [nonode] {};
	\node[] (y3) at (7,2) [nonode] {};
	\node[] (y4) at (8,2) [nonode] {};
	%
	
	\node[] (y1') at (2,4) [node] {};
	\node[] (sur') at (6.5,4) [node] {};
	
	\node[label=above :{}] (C2) at (8,6) {$C_2$};
	%
	\node[label=above:{$s$}] (x1') at (1.8,6) [node] {};
	\node[label=above:{$t$}] (x3') at (4.5,6) [node] {};
	%
	\node[] (y1h) at (1,6) [node] {};
	\node[] (y2h) at (7,6) [nonode] {};
	\node[] (y3h) at (9,6) [nonode] {};
	\node[] (y4h) at (10,6) [nonode] {};
	%
	
	\draw[access] (y1) -- (y2);
	\draw[access] (C1) circle [radius=1cm];
	\draw[access] (y3) -- (y4);
	
	\draw[thick,blue] (y1') -- (9,4);
	
	\draw[access] (y1h) -- (y2h);
	\draw[access] (C2) circle [radius=1cm];
	\draw[access] (y3h) -- (y4h);
	
	\draw[future] (x1) -- (x1');
	%
	\draw[future] (p1) -- (sur');
	%
	\draw[future] (x3) -- (x3');
	%
	\draw[future] (y1) -- (y1h);
	\draw[future] (y2) -- (y2h);
	\draw[future] (y3) -- (y3h);
	%
	
\end{tikzpicture}

\hspace*{1cm}

\begin{tikzpicture}[thick, every node/.style={scale=1.2},font = {\Large}]
	
	\tikzstyle{node}=[circle,fill=black,inner sep=1.5pt]
	\tikzstyle{nonode}=[inner sep=0pt]
	\tikzstyle{access}=[blue]
	\tikzstyle{future}=[dashed,->]
	
	\node[] (L2) at (0,6) {$L_2$};
	\node[] (L') at (0,4) {$L'$};
	\node[] (L1) at (0,2) {$L_1$};
	
	\node[label=below :{}] (C1) at (6,2) {$C_1$};
	\node[label=left:{$p_1$}] (p1) at (5.5,2.85) [node] {};
	\node[] (x1) at (5.5,1.15) [node] {};
	\node[] (x3) at (6.8,1.4) [node] {};
	\node[] (y1) at (3,2) [node] {};
	\node[] (y2) at (5,2) [nonode] {};
	\node[] (y3) at (7,2) [nonode] {};
	\node[] (y4) at (8,2) [nonode] {};
	%
	
	\node[] (y1') at (2,4) [node] {};
	\node[] (sur') at (5.5,4) [node] {};
	
	\node[] (C2) at (4,6) {$C_2$};
	%
	\node[label=above:{$s$}] (x1') at (6.7,6) [node] {};
	\node[label=above:{$t$}] (x3') at (9.3,6) [node] {};

	\node[] (y1h) at (1,6) [node] {};
	\node[] (y2h) at (3,6) [nonode] {};
	\node[] (y3h) at (5,6) [nonode] {};
	\node[] (y4h) at (10,6) [nonode] {};
	%
	
	\draw[access] (y1) -- (y2);
	\draw[access] (6,2) circle [radius=1cm];
	\draw[access] (y3) -- (y4);
	
	\draw[access] (y1') -- (9,4);
	
	\draw[access] (y1h) -- (y2h);
	\draw[access] (C2) circle [radius=1cm];
	\draw[access] (y3h) -- (y4h);
	
	\draw[future] (x1) -- (x1');
	\draw[future] (p1) -- (sur');
	\draw[future] (x3) -- (x3');
	\draw[future] (y1) -- (y1h);
	\draw[future] (y2) -- (y2h);
	\draw[future] (y3) -- (y3h);
	
\end{tikzpicture}
	\caption{Left: Structure of an unhappy R-triangle loop defined in Def.~\ref{def:triangle} \quad Middle and Right: Structure of an unhappy U-triangle loop defined in Def.~\ref{def:u-triangle}}
	\label{fig:R-triangle}
	\label{fig:U-triangle}
\end{figure*}

Informally speaking, sequent $\sq G \sums \lliftf GxF$ contains precisely the relational atoms and labeled formulas that need to be added to $\sq G \sums \llift{G}{x}$ so that (i)~the unhappy $\whitef xFG$ becomes happy (the suricata label contains the white formula responsible for this happiness) and (ii)~the result is still structurally saturated (and tree-layered and cluster-layered). Figure~\ref{fig:mm-lifting} shows the case of~$\labelsw x{\BOX B}$ with the additional fresh label~$\lb z$ (that has no past). 
In order to also preserve the property of being stable, we need to add such a layer for all unhappy $\whitef xFG$ in the same  layer~$L$.

\begin{construction}[Layer Lifting III] 
\label{def:liftingIII}
	Let $L$ be a topmost layer in a  saturated non-axiomatic sequent $\sq G$ where  $\labelsw {x_1}{F_1}$, \ldots, $\labelsw {x_m}{F_m}$ (of the form $\fm{A\IMP B}$ or $\fm{\BOX B}$) are all the unhappy formulas in~$L$. Then  
	$\lllift GL\!\colonequals\!  \lliftf G{x_1}{F_1} \!\sums\! \cdots \!\sums\! \lliftf G{x_m}{F_m}$ where we assume the sets of fresh labels introduced in each of $\lliftf G{x_1}{F_1} , \ldots , \lliftf G{x_m}{F_m}$ to be pairwise disjoint. 
\end{construction}

If we do this addition of layers naively, we will not terminate. Therefore, we need a loop-check, which could be viewed as a generalized version of a loop-check in sequent calculi for intuitionistic logic.

\begin{definition}[Simulation of layers]\label{def:simul-layers}
  Let $L'$ and $L$ be layers in a layered sequent~$\sq G$.
  A \defn{layer simulation} between~$L'$~and~$L$ is a non-empty binary relation $\simul \subseteq (L' \times L)\;\cap\sim$ such that for all $\lb{x'} \in L'$, $\lb x,\lb y \in L$,
  \begin{enumerate}
  \item [($\rn{S1}$)]   whenever $\lb {x'} \simul \lb x$ and $\accsq{x}{y}$, there exists $\lb{y'} \in L'$ such that 	$\accsq{x'}{y'}$ and $\lb {y'} \simul \lb y$, and 
  \item [($\rn{S2}$)]   whenever $\lb {x'} \simul \lb x$ and $\accsq{y}{x}$, there exists $\lb{y'} \in L'$ such that 	$\accsq{y'}{x'}$ and $\lb {y'} \simul \lb y$. 
  \end{enumerate}
  We say $L'$ \defn{simulates} $L$ if{f} there is a layer simulation  between~$L'$~and~$L$.
\end{definition}

\begin{proposition}
\label{prop:sim_bij}
  Let $L'$ and $L$ be layers in a layered sequent\/~$\sq G$. If\/ $\simul$ is a layer simulation between $L'$ and $L$ then, for all $\lb x\in L $, there is a $\lb {x'}\in L'$ such that $\lb{x'}\simul\lb x$. %
\end{proposition}

\begin{definition}\label{def:simul-layersII}
	Let $L$ be a topmost layer in a layered sequent~$\sq G$. We say $L$  \defn{is simulated} if{f} there is a layer~$L'$ in~$\sq G$, such that $L'< L$ and $L'$ simulates~$L$.
\end{definition}

\begin{definition}[Lifting Saturation]
  \label{def:lsat}
    Let $\sq G$ be a saturated non-axiomatic sequent and  $L_1,\ldots,L_n$ be all topmost layers in~$\sq G$ that are not simulated.
    We define the \defn{lifting saturation} of $\sq G$ as
    $\liftsat {\sq G}\colonequals \sq G \sums \lllift G{L_1} \sums \cdots \sums \lllift G{L_n}$
    where the sets of fresh labels introduced in each of $\lllift G{L_1} , \ldots , \lllift G{L_n}$ are pairwise disjoint. 
\end{definition}

\begin{lemma}[restate = LiftingSaturationStable, name = ] 
\label{lemma:lifting_stable}
	If\/ $\sq G$ is saturated, then\/ $\liftsat{\sq{G}}$ is stable. 
\end{lemma}

\subsection{Loop saturation}
We have presented all the steps that make the sequent 
grow,  by adding new labelled formulas or relational atoms. 
As a result,  unhappy formulas become happy, or the sequent becomes structurally saturated. 
Definition~\ref{def:lsat} exhibits a condition under which we stop 
applying the growing steps: 
namely, when a topmost layer is simulated by a layer below.

However, observe that, in general, new layers are larger than previous ones and there are two sources for the growth. First, in the case of an unhappy $\labelsw x{\BOX B}$, a fresh label $\lb z$ is added, and second, if  label $\lb x$ is in a non-singleton cluster, this cluster is duplicated. Both effects can be seen in Fig.~\ref{fig:mm-lifting}.

For this reason, we need to find a way to \emph{shrink} a layer. This will be done by creating clusters  similar to how Def.~\ref{def:dia-sat} does for~$\lef\DIA$. The difference is that this time the potential repetition will involve several layers rather than being  local to a single layer. The difficulty is that some part of a layer $L_1$ will be repeated in a layer $L_2$ 
occurring above it, 
but in order to keep the sequent tree-layered, we cannot create clusters across several layers. The solution we implement here is to create a cluster inside layer $L_2$ for a part that would be repeated in a future layer, provided that we can repeat   in layer $L_2$ what happened in layer $L_1$.
We call  such loops \emph{triangle loops}. 
There are two kinds of such loops. The first kind occurs if the cluster to be created in $L_2$ is in a part of $L_2$ that has no past in $L_1$. This could be caused for example by repetitions of $\rig\BOX$. We call these loops \emph{R-triange loops}. The second kind occurs when the repetition is caused by a repeated duplication of clusters in the layer lifting (see Fig.~\ref{fig:mm-lifting}). We call these loops \emph{U-triangle loops}.\looseness=-1

Before we give the formal definitions, observe that all new layers that are created in our algorithm are of the shape  $\lliftf GxF$, as defined in Construction~\ref{def:liftingII}, and each such layer contains exactly one suricata label. This can, therefore, be called \emph{the suricata label of the layer}, and it is (immediately after the lifting) the only label that contains a white formula.\footnote{On the other hand,  observe that a label $\lb x$ can have several suricata labels in other layers if there is more than one $\rig\IMP$- or $\rig\BOX$-formula in $\lb x$.}





\begin{figure*}[!t]
	$$
	\scalebox{1}{\fbox{
			\begin{minipage}{.85\textwidth}
				\begin{enumerate}[leftmargin=1.5em]\setcounter{enumi}{-1}
					\item Given a formula $\fm F$, define $\sq G_0(\fm F)$ to be the sequent ~$\futs rr, \accs rr \SEQ\labels rF$~ and let $\sqset S_0' \colonequals\set{\sq G_0(\fm F)}$. 
					\item For  set $\sqset S_i'$, calculate a full saturation $\sqset S_i$. 
					\item If all sequents in $\sqset S_i$ are axiomatic, then  terminate.\\
					The formula $\fm F$ is provable and we can give a proof of $\SEQ{\labels rF}$ in $\labISf$ (see Sect.~\ref{sec:unfolding}).
					\item Otherwise, pick a non-axiomatic sequent $\sq G_i \in \sqset S_i$ and compute its lifting saturation $\liftsat {\sq G_i}$ (see Def.~\ref{def:lsat}). 
					\item If $\liftsat {\sq G_i}=\sq G_i$, then terminate.\\ The formula $\fm F$ is not provable, and sequent $\sq G_i$ defines a countermodel (see Sect.~\ref{sec:countermodel}).
					
					\item Otherwise, set $\sqset S_{i+1}'\colonequals\bigl(\sqset S_i\setminus\set{\sq G_i}\bigr)\cup\set{\liftsat{\sq G_i}}$.
					\item Go to Step 1).
					
				\end{enumerate}
				
			\end{minipage}
	}}	
	$$
	\caption{Proof search algorithm}	
	\label{fig:algorithm}
\end{figure*}

\begin{definition}[R-triangle loop]\label{def:triangle}
  Let $\sq G$ be a saturated sequent with two layers $L_1 < L_2$. 
  We say that  clusters  $\lbc{C_1}\subseteq L_1$ and  $\lbc{C_2}\subseteq L_2$  form an \defn{R-triangle loop} if{f}\looseness=-1
  \begin{enumerate}
  \item $\lb{C_1}\lbeq\lb{C_2}$;
  \item there is a label $\lb{p_1}\in\lb{C_1}$ such that there is a layer~$L'$ with $L_1<L'\le L_2$ that contains a suricata label of $\lb {p_1}$;
  \item there is a cluster $\lbc{C_r}$ such that $\futsq {C_1}{C_r}$, and $\accsq{C_r}{C_2}$, and
   no label $\lb v\in L_2\setminus \lbc {C_r}$ with $\lbc{C_r}\srel G\lb v\srel G\lbc{C_2}$ has a past in $L_1$, i.e., $\nfutsq uv$ for any $\lb u \in L_1$.
  \end{enumerate}
  This R-triangle loop is \defn{unhappy} if{f} additionally 
  \begin{enumerate}\setcounter{enumi}3
  \item $L_2$ is a topmost layer;
  \item\label{i:st} there are labels $\lb s,\lb t\in L_2\setminus\lbc{C_2}$
    such that $\lb s\lbeq\lb t$, and $\lb s \neq \lb t$, and $\accsl{C_r}{s}{t}{C_2}$;
  \item\label{i:suricata}
    there is no suricata label $\lb u$ with $\lb{C_r}\srel{G}\lb{u}\srel{G}\lb{t}$;
  \item\label{i:p2}
    we do \emph{not} have $\lb{C_2}\srel{G}\lb{t}$.
  \end{enumerate}
  (Note that  $\lb s$ and $\lb t$ may be in a common cluster.)
\end{definition}

Figure~\ref{fig:R-triangle} (Left) illustrates this definition.

\begin{definition}[U-triangle loop]\label{def:u-triangle}
  Let $\sq G$ be a saturated sequent with two layers  $L_1< L_2$.
  We say that clusters $\lbc{C_1}\subseteq L_1$ and $\lbc{C_2}\subseteq L_2$ form a \defn{U-triangle loop} if{f} 
  \begin{enumerate}
  \item $\lb{C_1}\lbeq\lb{C_2}$;
  \item there is a label $\lb{p_1}\in\lb{C_1}$ such that there is a layer~$L'$ with $L_1<L'\le L_2$ that contains a suricata label of $\lb {p_1}$;
  \item $\futsq {C_1}{C_2}$.
  \end{enumerate}
  The U-triangle loop is \defn{unhappy} if{f}  additionally  
  \begin{enumerate}\setcounter{enumi}3
  \item $L_2$ is a topmost layer;
  \item\label{i:v-st} there are labels $\lb s,\lb t\in L_2\setminus\lbc{C_2}$ such that $\futsq{C_1}s$, and $\futsq{C_1}t$, and $\lb s\lbeq\lb t$, and $\lb s \neq \lb t$, and either $\accslt{C_2}{s}{t}G$ or $\accslt{s}{t}{C_2}G$;
  \item\label{i:v-suricata}
    there is no suricata label $\lb u$ with $\lb{s}\srel{G}\lb{u}\srel{G}\lb{t}$;
  \item\label{i:v-p2}
    we do \emph{not} have $\lb{s}\srel{G}\lb{C_2}\srel{G}\lb{t}$.
  \end{enumerate}
  (Again, $\lb s$ and $\lb t$ may be in the same cluster.)
\end{definition}

The two  possibilities envisioned by this definition, depending on whether $\accslt{C_2}{s}{t}G$ or $\accslt{s}{t}{C_2}G$, are illustrated in Fig.~\ref{fig:U-triangle} (Middle and Right).

Informally, we speak of a \emph{triangle loop} when we can reproduce the steps that started with the creation of $L'$ and led to $L_2$. The loop is unhappy if we can observe some repetition in the new part of $L_2$. If this is the case, we can collapse this repetition by creating a cluster, or shrinking an existing cluster, as follows:

\begin{definition}[Loop saturation]\label{def:loop-sat}
	Let $\sq G$ and $\sq{G'}$ be saturated sequents. We write $\sq{G} \loopred \sq{G'}$ if{f} there is an unhappy R-triangle or U-triangle loop in $\sq{G}$ where the labels $\lb s$ and $\lb t$ are as in  Def.~\ref{def:triangle} or Def.~\ref{def:u-triangle} respectively  and $\sq G'$ is obtained from~$\sq G$ by substituting  $\lb s$ for all occurrences of $\lb t$ and closing the resulting $\grel G$ under transitivity. 
	For sets $\sqset{S}$ and $\sqset S'$ of sequents, we write $\sqset{S} \Loopred \sqset S' $ if{f} $\sqset S$ is saturated, $\sq{G} \loopred \sq{G'}$ for some $\sq G\in \sqset S$, and $\sqset{S'}=\bigl(\sqset S\setminus\set{\sq G}\bigr)\cup\set{\sq G'}$. 
	We write~$\tLoopred$ for the transitive and reflexive closure of~$\Loopred$.
	If $\sqset S\tLoopred\sqset S'$ and $\sqset S'$ is in normal form w.r.t.~$\Loopred$, then $\sqset S'$ is a \defn{loop saturation} of $\sqset S$.%
\end{definition}

As before,  this term is justified because a loop saturation of a saturated set of sequents is saturated.

\begin{lemma}[restate = LoopSaturationStable, name = ] 
	\label{lemma:loop-sat}
	It holds that: 
	\begin{enumerate}[a)]
		\item   
		\label{lemma:loop-sat:sat}
		If\/ $\sqset S$  is saturated and\/ $\sqset S\Loopred\sqset S'$, then\/ $\sqset S'$ is saturated.\looseness=-1
		\item \label{lemma:loop-sat:term}
		The rewrite relation\/ $\Loopred$ is terminating. 
	\end{enumerate}
\end{lemma}
%

\begin{definition}[Full saturation]
  Let $\sqset S$ be a set of stable sequents. Let $\sqset S'$ be a semi-saturation of~$\sqset S$, and $\sqset S''$ be a saturation of~$\sqset S' $, and $\sqset S''' $ be a loop saturation of $\sqset S''$. Then  $\sqset S'''$ is a \defn{full saturation} of~$\sqset S$.%
\end{definition}

\begin{corollary}[restate = FullSaturationStable, name = ] 
\label{cor:full_saturation_stable}
	If a set\/ $\sqset S$ of sequents is stable and\/ $\sqset S''' $ is its full saturation, then\/ $\sqset S''' $ is saturated. 
\end{corollary}

\subsection{Search algorithm}

We have now all the ingredients for our proof/countermodel search algorithm, which is presented in Fig.~\ref{fig:algorithm}. It produces a sequence $\sqset S_0 ,\sqset S_1 ,\sqset S_2 ,\ldots$ of sets of  saturated sequents using the transformations discussed in this section. We terminate at step $i$ if either all sequents in $\sqset S_i $ are axiomatic (in that case we can produce a proof in $\labISf$, see Sect.~\ref{sec:unfolding}) or there is a sequent $\sq G_i\in\sqset S_i $ that does not grow anymore (in that case we can construct a countermodel, see Sect.~\ref{sec:countermodel}).

\begin{figure}[t!]
	\begin{center}
		\begin{tikzpicture}[thick, every node/.style={scale=1.2}]

		\tikzstyle{node}=[circle,draw]
		\tikzstyle{s-node}=[circle,draw,inner sep=2pt]
		\tikzstyle{nonode}=[inner sep=0pt]
		\tikzstyle{access}=[->,blue]
		\tikzstyle{future}=[->,dashed]
		\tikzstyle{deleted}=[->,gray]
		\tikzstyle{loop}=[->,red]
		
		\node[font = {\Large}] (L0) at (0,0) {$L_0$};
		\node[font = {\Large}] (L1) at (0,2) {$L_1$};
		\node[font = {\Large}] (L2) at (0,6) {$L_2$};
		
		\node[label=right :{$ \rig{\Box (\DIA A \AND \DIA b) \IMP \bot}$}] (1) at (2,0) [node] {$1$};
		
		\node[label=left :{$\rig\bot$}] (2) at (2,2) [node] {$2$};
		\node[label=below :{$\lef A, \rig{c\IMP\DIA b}$}] (3) at (4,2) [node, fill=green!30] {$3$};
		\node[label=below :{$\lef b$}] (4) at (6,2) [node, fill=yellow!30] {$4$};
		\node[] (5) at (8,2) [node,gray,fill=green!10] {$5$};
		\node[label=below :{$\lef b$}] (6) at (4.5,4) [node, fill=yellow!30] {$6$};
		\node[label=below :{$\lef A,\rig{c\IMP\DIA b} $}] (7) at (6.5,4) [node,fill=green!30] {$7$};
		\node[] (8) at (8.5,4) [node, gray,fill=yellow!10] {$8$};

		\node[] (9) at (2,6) [node] {$9$};
		\node[label=below :{$\lef A, \rig{c \IMP \DIA b}$}] (10) at (4,6) [s-node, fill=green!30] {$10$};
		\node[label=below :{$\lef b$}] (11) at (6,6) [s-node, fill=yellow!30] {$11$};
		
		\node[label=above :{$\lef b$}] (12) at (3.5,8) [s-node, fill=yellow!30] {$12$};
		\node[label=below :{$\lef A,\rig{c\IMP\DIA b}$}] (13) at (5,8) [s-node, fill=green!30] {$13$};
		\node[label={[align=center]above :{$\lef A,\rig{c\IMP\DIA b}$\\$\lef c,\rig{\DIA b} ,\rig b$}}] (14) at (6.5,8) [s-node,fill=blue!30] {$14$};
		\node[label=below :{$\lef b,\rig{\DIA b},\rig b$}] (15) at (8,8) [s-node,fill=red!50] {$15$};
		\node[label={[align=center]above :{$\lef A,\rig{c\IMP\DIA b}$\\$\rig{\DIA b},\rig b$}}] (16) at (9.5,8) [s-node, fill=purple!30] {$16$};
		
		\draw[access] (2) -- (3);
		\draw[access] (3) edge[bend right] (4);
		\draw[deleted] (4) -- (5);
		\draw[access] (4) edge[bend right] (3);
		\draw[access] (2) -- (6);
		\draw[access] (6) edge[bend right] (7);
		\draw[deleted] (7) -- (8);
		\draw[access] (7) edge[bend right] (6);
		
		\draw[access] (9) -- (10);
		\draw[access] (10) edge[bend right] (11);
		\draw[access] (11) edge[bend right] (10);
		\draw[access] (9) -- (12);
		\draw[access] (12) edge[bend right] (13);
		\draw[access] (13) edge[bend right] (12);
		\draw[access] (13) -- (14);
		\draw[access] (14) -- (15);
		\draw[access] (15) edge[bend right] (16);
		\draw[access] (16) edge[bend right] (15);
		
		\draw[future] (1) -- (2);
		\draw[future] (2) -- (9);
		\draw[future] (7) -- (14);
		
	\end{tikzpicture}	
	\end{center}
	\caption{Formula $\fm{\BOX (\DIA A \AND \DIA b) \IMP \BOT}$, with $ \fm A = \fm{(c \IMP \DIA b ) \IMP \BOT} $. The set  $\fmb{\Gamma} = \{\fmb{\Box (\DIA A \AND \DIA b) }, \fmb{\DIA A \AND \DIA b}, \fmb{\Diamond A}, \fmb{\DIA b}\}$ is in all nodes, except $ \lb 1 $. 
	}
	\label{fig:fml_provable:maintext}
\end{figure}

\begin{example}
	Figure \ref{fig:fml_provable:maintext} contains a {(partial)} diagrammatic representation of one sequent generated by the algorithm, when run on the valid $ \ISfour $ formula from Example~\ref{ex:ex-deriv}. 
	At layer $ L_1 $ clusters $ \{\lb 3, \lb 4\}  $ and $ \{\lb 6, \lb 7 \} $  are generated by saturation (Option~\ref{def:dia-sat:one} of Def.~\ref{def:dia-sat}). 
	Layer $ L_2 $ is created by lifting saturation (Def.~\ref{def:lsat}), where $\lb{14}$ is a suricata label and both clusters $\set{\lb{12}, \lb{13}}$ and $\set{\lb{15}, \lb{16}}$  come from $\set{\lb{6}, \lb{7}}$. The sequent is axiomatic because of $\lb{15}$.  
\end{example}

\section{Countermodel Construction}
\label{sec:countermodel}

Assume we initiate the algorithm with a formula~$\fm F$. If we terminate at Step~4, we have found a saturated sequent $\sq G_i$ such that $\whitef rF{G_\mathit{i}}$ and $\liftsat {\sq G}_i=\sq G_i$. This means that each topmost layer 
of $\sq G_i$  is either happy or  simulated. 
Thus,  for each  unhappy layer $L$ there is some happy inner layer $L'$ that  simulates $L$ via a simulation $\simul_L$. We now define $\sq G_i^\star$ to be obtained from $\sq G_i$ by adding a relational atom $\futs x{x'}$  whenever $\lb {x'}\simul_L \lb x$ for some unhappy layer $L$, and by closing the result under transitivity of $\le$. 
This makes all unhappy $\fmw{A\IMP B}$ and $\fmw{\BOX A}$ in the 
topmost layers
happy and preserves structural saturation. Hence, $\sq G_i^\star$ is a happy sequent, which allows to apply Theorem~\ref{thm:completeness} to obtain a finite countermodel for~$\fm F$.

\begin{theorem}[restate = countermodel, name = ] 
	\label{thm:step4}
  If the algorithm shown in Fig.~\ref{fig:algorithm} terminates in Step~4, then
   formula~$\fm F$ is not a theorem of\/~$\ISfour$.
\end{theorem}

\begin{example}
	Figure~\ref{fig:fml_R_triangle_main} represents (a $\leq$-branch of) a sequent, generated by the algorithm when run on formula~\eqref{eq:back-example} from Sect.~\ref{sec:problem}, that is on $\fm{\BOX ( A \AND B ) \IMP \BOT }$ with $\fm{A} = \fm{\BOX a \IMP \BOT}$ and $\fm{B} = \fm{(a \IMP \BOT) \IMP \BOT}$. 
	If we include label $ \lb{12} $ (but remove  atom $\lb 9\le \lb {14}$), the figure depicts the moment when the algorithm finds an unhappy R-triangle loop: employing the terminology from Def.~\ref{def:triangle}, we take  $\lb{C_1} = \{\lb{3}\}$, $\lb p_1=\lb{3}$, $\lb{C_2} = \{\lb{13}\}$, $\lb{s} = \lb{14} $, $\lb{C_r}= \{\lb s\}$, and  $\lb{t} = \lb{12} $. Then $\lb {12}$ is replaced with $ \lb{14} $, and after the loop saturation is no longer present in the sequent. 
	Then the search on the  depicted $\leq$-branch stops because the topmost layer $L_6$ can be simulated by layer $L_{4}$. By adding relational atoms $\futs{15}{8}$, $\futs{14}{6}$, and $\futs{13}{7}$ to the sequent (Theorem~\ref{thm:step4}), we obtain (a part of) the countermodel for our formula.  
\end{example}

\begin{figure}[t!]
	\begin{center}
		\begin{tikzpicture}[thick, every node/.style={scale=1.2}]
	
	\tikzstyle{node}=[circle,draw]
	\tikzstyle{s-node}=[circle,draw,inner sep=2pt]
	\tikzstyle{nonode}=[inner sep=0pt]
	\tikzstyle{access}=[->,blue]
	\tikzstyle{future}=[dashed,->]
	\tikzstyle{suricata}=[dashed,red]
	\tikzstyle{deleted}=[->,gray]
	\tikzstyle{loop}=[->,red]
	
	\node[font = {\Large}] (L0) at (0,-2) {$L_0$};
	\node[font = {\Large}] (L1) at (0,0) {$L_1$};
	\node[font = {\Large}] (L2) at (0,2) {$L_2$};
	\node[font = {\Large}] (L3) at (0,4) {$L_3$};
	\node[font = {\Large}] (L4) at (0,6) {$L_4$};
	\node[font = {\Large}] (L5) at (0,8) {$L_5$};
	\node[font = {\Large}] (L6) at (0,10) {$L_6$};
	
	\node[label=right :{$\rig{\BOX(A\AND B) \IMP\BOT}$}] (0) at (2,-2) [node] {$0$}; 
	
	\node[label=right :{$\rig{\BOT}$}] (1) at (2,0) [node] {$1$}; 
	
	\node[] (2) at (2,2) [node, fill=blue!30] {$2$};
	\node[label= right :{$\rig a$}] (3) at (4,2) [node, fill=green!30] {$3$};
	%
	
	\node[label= right :{$\lef a,\rig\bot$}] (4) at (4,4) [node, fill=purple!30] {$4$};
	\node[] (5) at (2,4) [node, fill=blue!30] {$5$};
	
	\node[label=below left :{$\lef a$}]  (6) at (4,6) [node, fill=yellow!30] {$6$};
	\node[label= right :{$\rig a$}] (7) at (6,6) [node,fill=green!30] {$7$};
	\node[] (8) at (2,6) [node, fill=blue!30] {$8$};
	
	\node[label= right :{$\lef a,\rig\bot$}] (9) at (6,8) [node,fill=purple!30] {$9$};
	\node[label=below left :{$\lef a$}] (10) at (4,8) [s-node, fill=yellow!30] {$10$};
	\node[] (11) at (2,8) [s-node, fill=blue!30] {$11$};
	
	\node[label={[text=gray]below left:{$\lef a$}}] (12) at (6,10) [s-node, gray, fill=yellow!10] {$12$};
	%
	\node[label= right:{$\rig a$}] (13) at (8,10) [s-node, fill=green!30] {$13$};
	%
	\node[label=below left :{$\lef a$}] (14) at (4,10) [s-node,fill=yellow!30] {$14$};
	\node[] (15) at (2,10) [s-node, fill=blue!30] {$15$};
	
	\draw[access] (2) -- (3);
	
	\draw[access] (5) -- (4);
	
	\draw[access] (8) -- (6);
	\draw[access] (6) -- (7);
	
	\draw[access] (11) -- (10);
	\draw[access] (10) -- (9);
	
	\draw[access] (15) -- (14);
	\draw[deleted] (14) -- (12);
	\draw[deleted] (12) -- (13);
	\draw[access] (14) edge[bend left] (13);
	
	\draw[future] (0) -- (1);
	\draw[future] (1) -- (2);
	\draw[future] (2) -- (5);
	\draw[future] (5) -- (8);
	\draw[future] (8) -- (11);
	\draw[future] (11) -- (15);
	\draw[future] (3) -- (4);
	\draw[future] (4) -- (6);
	\draw[future] (6) -- (10);
	\draw[future] (10) -- (14);
	\draw[future] (7) -- (9);
	\draw[future,gray] (9) -- (12);
	\draw[future] (9) -- (14);
	\draw[future,red,->] (15) edge[bend left] (8);
	\draw[future,red,->] (14)  edge[bend left] (6);
	\draw[future,red,->] (13) edge[bend left] (7);
	
\end{tikzpicture}
	
	\end{center}

	\caption{Formula $\fm{\BOX ( A \AND B ) \IMP \BOT }$ with $\fm{A} = \fm{\BOX a \IMP \BOT}$ and $\fm{B} = \fm{(a \IMP \BOT) \IMP \BOT}$.  $\fmb{\Gamma} = \{ \fmb{\BOX (A \AND B)},\fmb{A \AND B}, \fmb{A}, \fmb{B}, \fmw{\Box a}, \fmw{a \IMP \BOT}\}$ is in all nodes but~$\lb 0$.
	}
	\label{fig:fml_R_triangle_main}
\end{figure}

\section{Sequent Proof Unfolding}
\label{sec:unfolding}

Let us now turn to the case when the algorithm terminates in Step~2. Then all sequents in~$\sqset{S}_i$ are axiomatic, and we want to construct a proof of $\SEQ{\labels rF}$ in~$\labISf$. For this we are going to simulate the steps of the algorithm by applying the inference rules of~$\labISfp$, starting with rules~$\Rref$~and~$\Lref$ to obtain the sequent $\futs rr, \accs rr \SEQ\labels rF$ that is the input of the algorithm. On the one hand, this unfolding seems easy because our search algorithm was designed as organized proof search in~$\labISfp$ and most steps can indeed be executed by applying the rules of~$\labISfp$.
However, the difficulties come from
the fact that  sequents produced by~$\labISfp$ do not have non-singleton clusters. We call such sequents \emph{proper}.


\begin{definition}[Vertical sequent]
  A layered sequent~$\sq G$ is \defn{vertical} if{f} for all labels~$\lb u$, $\lb {u'}$, $\lb v$,~and~$\lb {v'}$ in~$\sq G$, (a)~if $\futsq uv$, and $\futsq u{v'}$, and $\lb v\lrel G\lb{v'}$, then $\lb v=\lb{v'}$, and (b)~if  $\futsq uv$, and $\futsq{u'}v$, and $\lb u\lrel G\lb{u'}$, then $\lb u=\lb{u'}$, i.e., for each label there is at most one future per layer and at most one past per layer.
\end{definition}

\begin{definition}[Proper layer/sequent]
	A layer~$L$ in a tree-clustered sequent~$\sq{G}$ is called \defn{proper} if{f} all clusters $\lbc C\subseteq L$ are singletons. A tree-clustered sequent~$\sq{G}$ is \defn{proper} if{f} it is tree-layered and vertical and all its layers are proper. 
\end{definition}

The basic idea of constructing our proof is to mimic the algorithm with a derivation that works ``layer by layer.''
The key observation is that whenever we create a cluster in the algorithm,  we can \emph{unfold} it, i.e.,~repeat this cluster arbitrarily often in an actual proof with only proper sequents. And this property is preserved when we lift the cluster to the next layer.\looseness=-1

\begin{definition}[Unfolding]
  \label{def:unfolding}
  For $n\in\Nat$, 
  a proper sequent~$\sq\hG$~is called an \defn{$n$-unfolding} of a stable sequent~$\sq G$ if{f} there is a binary \defn{$n$-unfolding relation} $\unfold\subseteq \labelsof G\times\labelsof\hG$ that has the following properties, where we assume $\lb x\unfold\lb\hx$ and $\lb y\unfold\lb\hy$ and write~$\layerof u$ to denote the layer of~$\lb u$:
  \begin{enumerate}[($\unfold$1),leftmargin=2.5em]
  \item\label{unf:formula}  
  	\emph{formula invariance}: 
		$\lbeql{x}{}{\hx}{}$;
  \item\label{unf:R} 
  	\emph{$\rel$-invariance}: 
		if $\lb x$ and~$\lb y$ are not in the same cluster in~$\sq G$, 
		then $\accsq xy$ if{f} $\accshat \hx\hy$; 
  \item\label{unf:layer} 
  	\emph{layer invariance}:
    		$\layerof x\sle G\layerof y$  if{f} $\layerof \hx\sle{\hG}\layerof \hy$;
  \item\label{unf:nopast}  
  	\emph{``no-past'' invariance}: 
		$\lb x$~has no past in~$\sq G$ if{f} $\lb\hx$~has no past in~$\sq\hG$;
  \item\label{unf:singleton}  
  	\emph{singleton-cluster invariance}: 
		if $\lbc C=\set{\lb z}$ is a singleton cluster in~$\sq G$, 
		then there is a unique $\lb\hz\in\labelsof{\sq \hG}$ with $\lb z\unfold\lb\hz$;
  \item\label{unf:nonsingleton} 
  	\emph{non-singleton-cluster unfolding}: 
		if $\lbc C$~is a cluster in~$\sq G$ with $\sizeof{\lb C}=k\ge 2$, 
		then for all $i=1..k$ and $j=1..n$ there are $\lb{z_i}\in \labelsof{\sq G}$ and $\lb{\hz_{i,j}}\in\labelsof{\sq \hG}$ 
    		such that
                we have $\lb{z_i}\unfold\lb{\hz_{i,j}}$ and $\lbc C=\set{\lb{z_1},\ldots,\lb{z_k}}$ and additionally
    		$\lb{\hz_{1,1}}R\cdots R\lb{\hz_{k,1}}R\lb{\hz_{1,2}}R\cdots R\lb{\hz_{k,2}}R\cdots 
		R \lb{\hz_{1,n}} R\cdots R\lb{\hz_{k,n}}$ in $\sq\hG$;
  \item\label{unf:inject} 
  	\emph{injectivity}:
    		if $\lb\hx=\lb\hy$, then $\lb x=\lb y$.
  \end{enumerate}
\end{definition}

Note that every $(n+1)$-unfolding~$\unfold$ is also an $n$-unfolding and that any proper stable sequent  $\sq G$ is an $n$-unfolding of itself for any~$n$, via the diagonal relation $\set{(\lb x, \lb x) \mid \lb x \in \labelsof{G}}$.


To ensure that unfolding only operates with proper sequents, we will now introduce $\labISfd$, a proof system whose rules correspond to applications of several ~$\labISfp$ steps, thus forcing the rules of~$\labISfp$ to be applied in a certain way.

Let $\sq G$ be a sequent $\B, \Left \SEQ \Right$ with $\labels xF$ occurring in~$\Right$. Then we write $\llniftf {(\B, \Left \SEQ \Right)}x{F}$ for $\sq G+\lliftf Gx{F}$. This allows us to define the following inference rules:
$$
\scalebox{.9}{$
  \vlinf{\srig\IMP}{}{\B, \Left \SEQ \Right, \labels x{A \IMP B}}{\llniftf {(\B, \Left \SEQ \Right, \labels x{A \IMP B})}x{A \IMP B}}
  \quad
  \vlinf{\srig\BOX}{}{\B, \Left \SEQ \Right, \labels x{\BOX A}}{\llniftf {(\B, \Left \SEQ \Right, \labels x{\BOX A})}x{\BOX A}}
  $}
$$
For a label~$\lb x$ in a sequent~$\sq G$, 
we define~$\parentsof x$ to be the set of labels~$\lb{x'}$ with $\accsq {x'}x$. For a set~$\lbs X$ of labels, we write $\accs Xy$ for the set $\set{\accs xy\mid \lb x \in\lbs X}$ of relational atoms. With this notation, we can define the rule:
$$
\scalebox{.9}{$
\vlinf{\slef\DIA}{\proviso{$\lb y$  fresh}}{\B, \Left, \labels x{\DIA A} \SEQ \Right}{\B, \accs{\parentsof x}y,\accs yy,\futs yy, \Left,  \labels x{\DIA A}, \labels yA \SEQ \Right}
  $}
$$
Now define \defn{system~$\labISfd$} to be
\begin{itemize}
\item inference rules~$\idr$, $\lef\bot$~and~$\rig\DIA$ as in Fig.~\ref{fig:labIKp};
\item  rule $\slef\BOX$ obtained by setting $\lb x=\lb y$ in $\lef\BOX$ from Fig.~\ref{fig:labIKp};
\item rules  $\klef{\AND}$, $\krig{\AND}$, $\klef{\OR}$, $\krig{\OR}$,~and~$\klef{\IMP}$ that are variants of~$\lef{\AND}$, $\rig{\AND}$, $\lef{\OR}$, $\rig{\OR}$,~and~$\lef{\IMP}$  from Fig.~\ref{fig:labIKp} respectively, where the principal formula is not deleted in the premises, the premises are closed under monotonicity, and, additionally, $\lb x=\lb y$ for $\klef\IMP$; 
\item rules~$\lef{\vax}$~and~$\rig{\vax}$ from Fig.~\ref{fig:AdmRules}; 
\item rules~$\srig\IMP$, $\srig\BOX$,~and~$\slef\DIA$ as shown above.
\end{itemize}
Note that there are no structural rules in~$\labISfd$. 

\begin{proposition}[restate = daggeradm, name = ] 
	\label{prop:daggeradm}
 Every  rule\/~$\rr$ in\/~$\labISfd$ is derivable in\/~$\labISfp$, and if\/ $\sq G$~is the conclusion of an instance of\/~$\rr$, then each premise has the form\/ $\sq G+\sq H$ for some\/~$\sq H$.
\end{proposition}

When writing a premise $\sq G+\sq H$, we assume that~$\sq G$~and~$\sq H$ are disjoint, i.e.,~all labelled formulas and relational atoms of~$\sq H$ 
are 
absent from the conclusion. 


\begin{definition}[Touch]\label{def:touch}
  Let $\rr$ be an instance of  a rule in~$\labISfd$  with a sequent $\sq G$ in its conclusion. 
  We say that $\rr$~\defn{touches} a label $\lb x \in \labelsof{G}$ if{f} for some  premise $\sq G + \sq H$ of~$\rr$,  either we have $\futsqp uxH$ or $\accsqp uxH$ for some label~$\lb u$ or we have $\blackf xBH$ or $\whitef xBH$ for some formula~$\fm B$. We say that $\rr$~\defn{touches a set~$\lbs X$ of labels} if{f} $\rr$~touches some $\lb x\in \lbs X$.    
  \end{definition}

\begin{definition}[Restricted sequent/forbidden label]
\label{def:forbid}
  A {sequent}~$\sq G$ is called \defn{restricted} if{f} it is   equipped with a set $\forbiddenof G\subseteq\labelsof G$ of \defn{forbidden} labels satisfying   two conditions:
 \begin{enumerate}[(a)]
 \item\label{forbid:parent} 
 	if $\lb y \in \forbiddenof G$ and $\accsq xy$, 
	then $\lb x \in \forbiddenof G$, i.e.,~all parents of a forbidden label are forbidden;
 \item\label{forbid:past} 
 	if  $\layerof{x} <_{\sq G} \layerof{y}$ for some $\lb y \in \labelsof{G}$, 
	then $\lb x \in \forbiddenof G$, i.e.,~all labels from the inner layers are  forbidden.
 \end{enumerate}
All labels in $\labelsof G$ that are not forbidden are \defn{allowed}. 
\end{definition}

In the following, we assume that all sequents are restricted, in particular, all sequents occurring in derivations.


  

\begin{definition}[Tidy rule/derivation]
    \label{def:tidy}
\label{const:restr}
Let $\sq G$~be the conclusion of an instance of a rule~$\rr$ in~$\labISfd$. We say that $\rr$~is \defn{tidy} if{f}
 (i)~$\rr$~does not touch any forbidden label in~$\sq G$, and
(ii)~the forbidden labels for each premise~$\sq K$ of~$\rr$ are 
   \begin{equation}
 \label{eq:tidyforbid}
 \forbiddenof{K} =
 \begin{cases}
 \labelsof G  &\text{if } \rr \in\set{\srig\IMP, \srig\BOX,\slef\DIA},
 \\
 \forbiddenof G & \text{otherwise}.
 \end{cases}
   \end{equation}
A derivation~$\deri$  is  \defn{tidy} if{f} it consists of tidy  rule instances.\looseness=-1
 \end{definition}
In other words,  rules in tidy derivations do not touch forbidden labels and either keep all labels and forbidden labels unchanged or introduce new allowed labels turning all old labels into forbidden. This is sufficient, in particular, to ensure that all sequents in a tidy derivation are proper.


\begin{proposition}[restate = properrestrict, name = ] 
\label{prop:properrestric}
Let $\deri$~be a tidy derivation. If the conclusion of~$\deri$ is proper, then every sequent in~$\deri$ is also proper.
  \end{proposition}

\begin{definition}[Unfoldable]
	\label{def:unfoldable}
	A derivation~$\Deri_n$ is called an \defn{$n$-unfolding of a stable set~$\sqset S$} if{f} each premise of~$\Deri_n$ is an $n$-unfolding of some sequent from~$\sqset S$.
	For a formula~$\fm F$, a stable set~$\sqset S$
        is \defn{$\fm F$-unfoldable} if{f} for every $n\in\Nat$ there is
        a tidy derivation~$\Deri_n$ of~$\sq G_0(\fm F)$ (as defined in Step~0 of the algorithm in Fig.~\ref{fig:algorithm}) such that $\der_n$ is an $n$-unfolding of $\sqset S$.
\end{definition}

Clearly, $\sqset S_0'=\set{\sq G_0(\fm F)}$ is $\fm F$-unfoldable. It now has to be shown that all operations performed in the algorithm, in particular, the rewrite relations $\ssatred$, $\sdiared$, and $\Loopred$, preserve this propery. For $\ssatred$, this is straightforward, as we simply apply the inference rules of $\labISfd$ to match the semi-saturation steps.
For $\sdiared$ and $\Loopred$, we need to unfold the newly created clusters. The basic idea is to simply repeat all proof steps that lead to the first occurrence of the cluster $n$-times.\looseness=-1




\begin{lemma}[restate = UnfoldingLemma, name = Unfolding Lemma]
  All sets\/~$\sqset S_i$ of sequents generated by the algorithm from Fig.~\ref{fig:algorithm} are $\fm F$-unfoldable.
\end{lemma}

Hence, when we terminate in Step~2, we can construct a $1$-unfolding of $\sqset S_i$, which constitutes a proof in $\labISfd$ because an unfolding of an axiomatic sequent is always axiomatic.

\begin{theorem}[restate = Provable, name = ] 
	\label{thm:unfolding}
  If the algorithm shown in Fig.~\ref{fig:algorithm} terminates in Step~2, then the formula $\fm F$ is a theorem of\/~$\ISfour$.
\end{theorem}

\section{Termination}\label{sec:termination}

We have already established in Sect.~\ref{sec:algorithm} that every
step in our algorithm (shown in Fig.~\ref{fig:algorithm})
terminates. It remains to show that we cannot run through the main loop
forever, i.e.,   sequence $\sqset S_0,\sqset S_1,\ldots,\sqset S_i,\ldots$ eventually terminates either in Step 2 (we find a proof) or
in Step 4 (we find a countermodel). The basic idea is to restrict the
size of a layer in the sequents of~$\sqset{S}_i$. Then  the number of distinct possible layers is finite,
hence, we will eventually find a simulation (Defs.~\ref{def:simul-layers}, \ref{def:simul-layersII}).

For the remainder of this section, we assume that a formula~$\fm F$ in Step~0 has $n$~subformula occurrences.

We define the \defn{size} $\sizeof{\lb x}$ of a label $\lb x$ in a sequent $\sq G$ to be the number of distinct formula occurrences $\labelsb xA$ or $\labelsw xA$ in $\sq G$. 

\begin{lemma}[restate = sizelabel, name = ] 
	\label{lem:size-label}
  The size of a label occurring in a sequent of some\/ $\sqset{S}_i$ is at most~$n$. And there are\/ $2^{n}$  many equivalence classes of labels with respect to\/~$\lbeq$.
 \end{lemma}


The \defn{size}  $\sizeof{\lb C}$  of a cluster $ \lb C $ is the number of labels in $\lb C$.

\begin{lemma}[restate = sizecluster, name = ] 
	\label{lem:size-cluster}
  The size of a cluster occurring in a sequent of some\/ $\sqset{S}_i$ is at most\/~$2^n$. And there are\/ $2^{2^n}-1$ many equivalence classes of (non-empty) clusters with respect to\/~$\lbeq$.
\end{lemma}


The algorithm visits only stable sequents. A layer $L$ in such a sequent is a tree of clusters. Let $M$ be a branch in $L$, i.e., a sequence $\lb{C_1},\lb{C_2},\ldots,\lb{C_l}$ of clusters from the root to a leaf. 
The \defn{length}~$\sizeof M$ of $M$ is the sum $\sizeof{\lb{C_1}}+\sizeof{\lb{C_2}}+\cdots+\sizeof{\lb{C_l}}$.

\begin{lemma}[restate = lengthbranch, name = ] 
  The length of a branch in a layer in a sequent in a set\/ $\sqset{S}_i$ is bounded, and the bound is determined by~$\fm F$.
\end{lemma}

\begin{theorem}[restate = TerminationAlg, name = Termination] 
  The proof search algorithm given in Fig.~\ref{fig:algorithm} is terminating. 
\end{theorem}
%


\section{Conclusion}
\label{sec:conclusion}

In this paper we have solved a problem open for almost 30 years. Our solution has  two key ingredients.

First, the use of the fully labelled system with relational atoms for both  binary relations enabled us to give a proof system that has only invertible rules and also gives a closer correspondence between sequents and models.

Second, although the identification of labels during during proof search to realize  loops is \emph{a priori} unsound, however, under the right circumstances, we can preserve soundness if we organize the proof search in a certain systematic way.

We conjecture that the same method  can also be applied to $\IKfour$, which is $\ISfour$ without the $\tax$-axiom and which is the other logic in the intuitionistic version of the $\Sfive$-cube for which decidability is an open problem. In fact, the overall argument is the same, but in many definitions and proof arguments, there would be subtle differences due to the absence of reflexivity. For this reason a full treatment of  $\IKfour$ would go beyond the scope of this paper.%

The implementation of the algorithm is object of current work. 
In a future work we would like to investigate the complexity of provability in $\ISfour$.

\medskip\noindent
{\bf Acknowledgements.} Roman Kuznets is supported by the Austrian Science Fund (FWF)  project ByzDEL (P~33600). 
Marianna Girlando is supported by the Horizon 2021 programme, under the
Marie Sk\l odowska-Curie grant CYDER (101064105).



%

\bibliographystyle{abbrvurl}
\bibliography{references}

\ifthenelse{\boolean{arxiv}}{}{\end{document}}

\clearpage

\appendix
\subsection{Examples}
\label{appendix:ex}

This section contains some examples illustrating how the search algorithm (Fig.~\ref{fig:algorithm}) works. Refer to Example~\ref{ex:model_notation} for an explanation of the diagrammatic notation. 

For reasons of space, we represent only one sequent of those generated by the algorithm, that is, either an axiomatic sequent or a sequent from which a countermodel for the formula at the root can be generated. Recall that even if an axiomatic sequent is found the algorithm might continue running: for it to stop (at step 2), all sequents in the set need to be axiomatic. Conversely, the algorithm stops (at step 4) if it finds a sequent containing enough information to build a countermodel. 

\begin{example}[Valid formula]
	\label{ex:provable}
	Let us consider formula $\fm{\Box (\DIA A \AND \DIA b) \IMP \BOT}$, with $ \fm A = \fm{(c \IMP \DIA b ) \IMP \BOT } $, which is valid in $\ISfour$. 
	Figure~\ref{fig:fml_provable} represents one sequent generated by the algorithm. Let $\fmb{\Gamma} $ be the set $ \{\fmb{\Box (\DIA A \AND \DIA b) },\fmb{\DIA A \AND \DIA b }, \fmb{\Diamond A}, \fmb{\DIA b}\}$. To each label in the figure (except $ \lb{1} $) we associate $\fmb{\Gamma} $, plus the formulas explicitly displayed next to the node in the figure.  
	The following $\leq$-relations are not displayed but are present in the sequent: $\futs{3}{10}$, $\futs{4}{11}$, $\futs{6}{12}$, $\futs{6}{15}$, $\futs{7}{13}$ and $\futs{7}{16}$. 
	
	At layer $L_1$ the search on both $R$-branches stops in virtue of option~\ref{def:dia-sat:one} of Def.~\ref{def:dia-sat} (saturation): there,  $\lb 5 $ is replaced with~$ \lb 3$, and $\lb 8$ with $ \lb 6 $. This generates two non-singleton clusters: $\lb{C_1} = \{ \lb{3}, \lb{4} \}$ and $\lb{C_2} = \{ \lb 6, \lb 7\}$. 
	Then, since all labels in $L_1$ are almost happy, lifting saturation is applied: the rewrite ``lifts'' one copy of the layer for each 
	$\fmw{\IMP}$-formula  
	present in the sequent. We only represent one of such layers, $L_2$, generated from formula $\labels{7}{\fmw{c \IMP \DIA b}}$. Since $\lb 7$ belongs to the cluster $\lb{C_2}$, the cluster is duplicated in $L_2$, following Defs.~\ref{def:lifting}--\ref{def:liftingII}.   
	Then, after a semi-saturation step, we have that both $\labels{15}{\fmb{b}}$ and $\labels{15}{\fmw{b}}$. Thus we stop the search along this branch. If all the sequents manipulated by the algorithm eventually result in axiomatic sequents, then the algorithm stops and, according to Theorem~\ref{thm:unfolding}, the formula at the root is a theorem of~$\ISfour$. Indeed, its derivation in~$\labISf$ can be found in Fig.~\ref{fig:ex-deriv} (Left). 
\end{example}

\begin{figure}[t!]
\begin{center}
		\begin{tikzpicture}[thick, every node/.style={scale=1.2}]

		\tikzstyle{node}=[circle,draw]
		\tikzstyle{s-node}=[circle,draw,inner sep=2pt]
		\tikzstyle{nonode}=[inner sep=0pt]
		\tikzstyle{access}=[->,blue]
		\tikzstyle{future}=[->,dashed]
		\tikzstyle{deleted}=[->,gray]
		\tikzstyle{loop}=[->,red]
		
		\node[font = {\Large}] (L0) at (0,0) {$L_0$};
		\node[font = {\Large}] (L1) at (0,2) {$L_1$};
		\node[font = {\Large}] (L2) at (0,6) {$L_2$};
		
		\node[label=right :{$ \rig{\Box (\DIA A \AND \DIA b) \IMP \bot}$}] (1) at (2,0) [node] {$1$};
		
		\node[label=left :{$\rig\bot$}] (2) at (2,2) [node] {$2$};
		\node[label=below :{$\lef A, \rig{c\IMP\DIA b}$}] (3) at (4,2) [node, fill=green!30] {$3$};
		\node[label=below :{$\lef b$}] (4) at (6,2) [node, fill=yellow!30] {$4$};
		\node[] (5) at (8,2) [node,gray,fill=green!10] {$5$};
		\node[label=below :{$\lef b$}] (6) at (4.5,4) [node, fill=yellow!30] {$6$};
		\node[label=below :{$\lef A,\rig{c\IMP\DIA b} $}] (7) at (6.5,4) [node,fill=green!30] {$7$};
		\node[] (8) at (8.5,4) [node, gray,fill=yellow!10] {$8$};

		\node[] (9) at (2,6) [node] {$9$};
		\node[label=below :{$\lef A, \rig{c \IMP \DIA b}$}] (10) at (4,6) [s-node, fill=green!30] {$10$};
		\node[label=below :{$\lef b$}] (11) at (6,6) [s-node, fill=yellow!30] {$11$};
		
		\node[label=above :{$\lef b$}] (12) at (3.5,8) [s-node, fill=yellow!30] {$12$};
		\node[label=below :{$\lef A,\rig{c\IMP\DIA b}$}] (13) at (5,8) [s-node, fill=green!30] {$13$};
		\node[label={[align=center]above :{$\lef A,\rig{c\IMP\DIA b}$\\$\lef c,\rig{\DIA b} ,\rig b$}}] (14) at (6.5,8) [s-node,fill=blue!30] {$14$};
		\node[label=below :{$\lef b,\rig{\DIA b},\rig b$}] (15) at (8,8) [s-node,fill=red!50] {$15$};
		\node[label={[align=center]above :{$\lef A,\rig{c\IMP\DIA b}$\\$\rig{\DIA b},\rig b$}}] (16) at (9.5,8) [s-node, fill=purple!30] {$16$};
		
		\draw[access] (2) -- (3);
		\draw[access] (3) edge[bend right] (4);
		\draw[deleted] (4) -- (5);
		\draw[loop] (4) edge[bend right,below] node{$C_1$} (3);
		\draw[access] (2) -- (6);
		\draw[access] (6) edge[bend right] (7);
		\draw[deleted] (7) -- (8);
		\draw[loop] (7) edge[bend right,below] node{$C_2$} (6);
		
		\draw[access] (9) -- (10);
		\draw[access] (10) edge[bend right] (11);
		\draw[access] (11) edge[bend right] (10);
		\draw[access] (9) -- (12);
		\draw[access] (12) edge[bend right] (13);
		\draw[access] (13) edge[bend right] (12);
		\draw[access] (13) -- (14);
		\draw[access] (14) -- (15);
		\draw[access] (15) edge[bend right] (16);
		\draw[access] (16) edge[bend right] (15);
		
		\draw[future] (1) -- (2);
		\draw[future] (2) -- (9);
		\draw[future] (7) -- (14);
		
	\end{tikzpicture}
\end{center}
	\caption{Formula $\fm{\Box (\DIA A \AND \DIA b) \IMP \BOT}$, with $ \fm A = \fm{(c \IMP \DIA b ) \IMP \BOT} $. 
	}
	\label{fig:fml_provable}
\end{figure}

\begin{figure*}[t!]
	\begin{center}
		\begin{tikzpicture}[thick, every node/.style={scale=1.2}]

		\tikzstyle{node}=[circle,draw]
		\tikzstyle{s-node}=[circle,draw,inner sep=2pt]
		\tikzstyle{nonode}=[inner sep=0pt]
		\tikzstyle{access}=[->,blue]
		\tikzstyle{future}=[dashed,->]
		\tikzstyle{suricata}=[dashed,red]
		\tikzstyle{deleted}=[->,gray]
		\tikzstyle{loop}=[->,red]
		
		\node[font = {\Large}] (L0) at (0,-2) {$L_0$};
		\node[font = {\Large}] (L1) at (0,0) {$L_1$};
		\node[font = {\Large}] (L2) at (0,2) {$L_2$};
		\node[font = {\Large}] (L3) at (0,4) {$L_3$};
		\node[font = {\Large}] (L4) at (0,6) {$L_4$};
		\node[font = {\Large}] (L5) at (0,8) {$L_5$};
		\node[font = {\Large}] (L6) at (0,10) {$L_6$};
		
		\node[label=right :{$\rig{\BOX(A\AND B) \IMP\BOT}$}] (0) at (2,-2) [node] {$0$}; 
		
		\node[label=right :{$\rig{\BOT}$}] (1) at (2,0) [node] {$1$}; 
		
		\node[] (2) at (2,2) [node, fill=blue!30] {$2$};
		\node[label=below right :{$\rig a$}] (3) at (4,2) [node, fill=green!30] {$3$};
		\draw[red] (4,2) circle [radius=1cm];
		\node[red] (C1) at (3,1) {$C_1$};
		
		\node[label=below right :{$\lef a,\rig\bot$}] (4) at (4,4) [node, fill=purple!30] {$4$};
		\node[] (5) at (2,4) [node, fill=blue!30] {$5$};

		\node[label=below right:{$\lef a$}] (6) at (4,6) [node, fill=yellow!30] {$6$};
		\node[label=below right :{$\rig a$}] (7) at (6,6) [node,fill=green!30] {$7$};
		\node[] (8) at (2,6) [node, fill=blue!30] {$8$};

		\node[label=below right :{$\lef a,\rig\bot$}] (9) at (6,8) [node,fill=purple!30] {$9$};
		\node[label=below right :{$\lef a$}] (10) at (4,8) [s-node, fill=yellow!30] {$10$};
		\node[] (11) at (2,8) [s-node, fill=blue!30] {$11$};
		
		\node[label=below right :{$\lef a$}] (12) at (6,10) [s-node, fill=yellow!30] {$12$};
		\node[red] (t) at (6,10.7) {$t$};
		\node[label=below right:{$\rig a$}] (13) at (8,10) [s-node, fill=green!30] {$13$};
		\draw[red] (8,10) circle [radius=1cm];
		\node[red] (C2) at (7,9) {$C_2$};
		\node[label=below right :{$\lef a$}] (14) at (4,10) [s-node,fill=yellow!30] {$14$};
		\node[red] (s) at (4,10.7) {$s$};
		\draw[red] (4,10) circle [radius=1cm];
		\node[red] (Cr) at (3,9) {$C_r$};
		\node[] (15) at (2,10) [s-node, fill=blue!30] {$15$};
		
		\draw[access] (2) -- (3);
		
		\draw[access] (5) -- (4);
		
		\draw[access] (8) -- (6);
		\draw[access] (6) -- (7);
		
		\draw[access] (11) -- (10);
		\draw[access] (10) -- (9);
		
		\draw[access] (15) -- (14);
		\draw[access] (14) -- (12);
		\draw[access] (12) -- (13);
		
		\draw[future] (0) -- (1);
		\draw[future] (1) -- (2);
		\draw[future] (2) -- (5);
		\draw[future] (5) -- (8);
		\draw[future] (8) -- (11);
		\draw[future] (11) -- (15);
		\draw[future] (3) -- (4);
		\draw[future] (4) -- (6);
		\draw[future] (6) -- (10);
		\draw[future] (10) -- (14);
		\draw[future] (7) -- (9);
		\draw[future] (9) -- (12);

	\end{tikzpicture}
	\hspace*{4cm}
	\begin{tikzpicture}[thick, every node/.style={scale=1.2}]
		
		\tikzstyle{node}=[circle,draw]
		\tikzstyle{s-node}=[circle,draw,inner sep=2pt]
		\tikzstyle{nonode}=[inner sep=0pt]
		\tikzstyle{access}=[->,blue]
		\tikzstyle{future}=[dashed,->]
		\tikzstyle{suricata}=[dashed,red]
		\tikzstyle{deleted}=[->,gray]
		\tikzstyle{loop}=[->,red]
		
		\node[font = {\Large}] (L0) at (0,-2) {$L_0$};
		\node[font = {\Large}] (L1) at (0,0) {$L_1$};
		\node[font = {\Large}] (L2) at (0,2) {$L_2$};
		\node[font = {\Large}] (L3) at (0,4) {$L_3$};
		\node[font = {\Large}] (L4) at (0,6) {$L_4$};
		\node[font = {\Large}] (L5) at (0,8) {$L_5$};
		\node[font = {\Large}] (L6) at (0,10) {$L_6'$};
		
		\node[label=right :{$\rig{\BOX(A\AND B) \IMP\BOT}$}] (0) at (2,-2) [node] {$0$}; 
		
		\node[label=right :{$\rig{\BOT}$}] (1) at (2,0) [node] {$1$};
		
		\node[] (2) at (2,2) [node, fill=blue!30] {$2$};
		\node[label=below right :{$\rig a$}] (3) at (4,2) [node, fill=green!30] {$3$};
		
		\node[label=below right :{$\lef a,\rig\bot$}] (4) at (4,4) [node, fill=purple!30] {$4$};
		\node[] (5) at (2,4) [node, fill=blue!30] {$5$};
		
		\node[label=below right:{$\lef a$}] (6) at (4,6) [node, fill=yellow!30] {$6$};
		\node[label=below right :{$\rig a$}] (7) at (6,6) [node,fill=green!30] {$7$};
		\node[] (8) at (2,6) [node, fill=blue!30] {$8$};
		
		\node[label=below right :{$\lef a,\rig\bot$}] (9) at (6,8) [node,fill=purple!30] {$9$};
		\node[label=below right :{$\lef a$}] (10) at (4,8) [s-node, fill=yellow!30] {$10$};
		\node[] (11) at (2,8) [s-node, fill=blue!30] {$11$};
		
		%
		\node[label=below right:{$\rig a$}] (13) at (8,10) [s-node, fill=green!30] {$13$};
		%
		\node[label=below right :{$\lef a$}] (14) at (5,10) [s-node,fill=yellow!30] {$14$};
		\node[] (15) at (2,10) [s-node, fill=blue!30] {$15$};
		
		\draw[access] (2) -- (3);
		
		\draw[access] (5) -- (4);
		
		\draw[access] (8) -- (6);
		\draw[access] (6) -- (7);
		
		\draw[access] (11) -- (10);
		\draw[access] (10) -- (9);
		
		\draw[access] (15) -- (14);
		\draw[access] (14) -- (13);
		
		\draw[future] (0) -- (1);
		\draw[future] (1) -- (2);
		\draw[future] (2) -- (5);
		\draw[future] (5) -- (8);
		\draw[future] (8) -- (11);
		\draw[future] (11) -- (15);
		\draw[future] (3) -- (4);
		\draw[future] (4) -- (6);
		\draw[future] (6) -- (10);
		\draw[future] (10) -- (14);
		\draw[future] (7) -- (9);
		\draw[future] (9) -- (14);
		\draw[future,red,->] (15) edge[bend left] (8);
		\draw[future,red,->] (14)  edge[bend left] (6);
		\draw[future,red,->] (13) edge[bend left] (7);
		
	\end{tikzpicture}
	
	\end{center}
	\caption{Formula $\fm{\BOX ( A \AND B ) \IMP \BOT }$ with $\fm{A} = \fm{\BOX a \IMP \BOT}$ and $\fm{B} = \fm{(a \IMP \BOT) \IMP \BOT}$. A loop saturation step is applied to the leftmost sequent, and the result of the saturation is the rightmost sequent. 
		}
	\label{fig:fml_R_triangle}
\end{figure*}
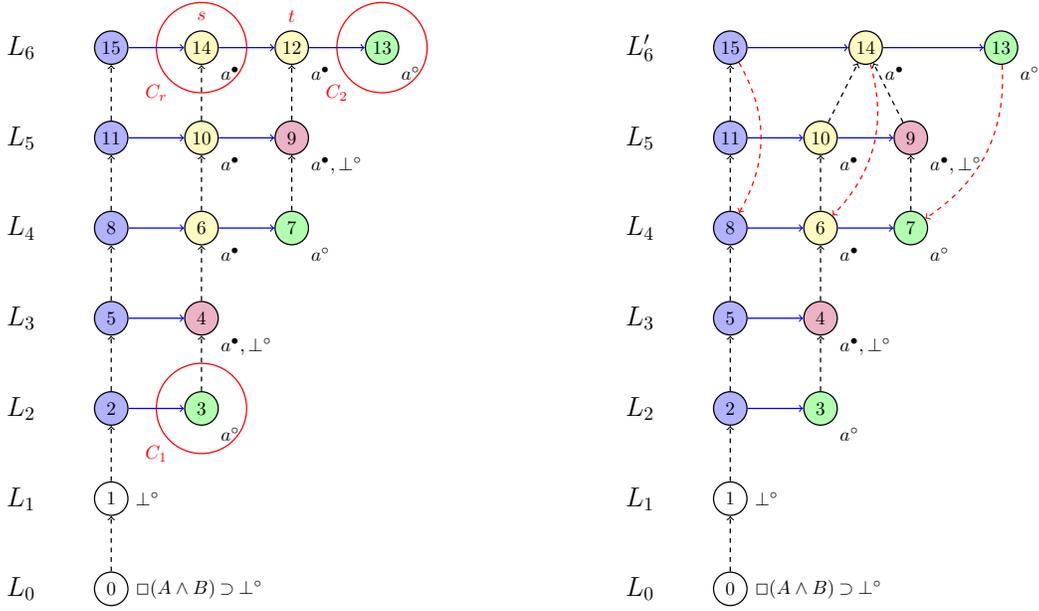

\begin{example}[Non-valid formula, R-triangle loop]	
	\label{ex:R}
	We now consider formula \ref{eq:back-example} from Sect.~\ref{sec:problem}, that is, $\fm{\BOX ( A \AND B ) \IMP \BOT }$ with $\fm{A} = \fm{\BOX a \IMP \BOT}$ and $\fm{B} = \fm{(a \IMP \BOT) \IMP \BOT}$. 
	On the left of Fig.~\ref{fig:fml_R_triangle} we represent one sequent generated by the algorithm.
	Let $\fmb{\Gamma} = \{ \fmb{\BOX (A \AND B)},\fmb{A \AND B}, \fmb{A}, \fmb{B}, \fmw{\Box a}, \fmw{a \IMP \BOT}\}$. To each label in the figure (except $ \lb{0} $), we associate $\fmb{\Gamma} $, plus the formulas explicitly displayed next to the node. 
	To be precise, the figure displays only one $\leq$-branch of the sequent: formulas  $\fmw{\Box a}$ and $\fmw{a \IMP \BOT}$ are in all the labels, and every time they are unhappy they are also lifted, generating a new layer (which is not displayed). 
	
	At layer $L_6$ an unhappy R-triangle loop is detected: employing the terminology from Def.~\ref{def:triangle}, we take  $\lb{C_1} = \{\lb{3}\}$, $\lb p_1=\lb{3}$, $\lb{C_2} = \{\lb{13}\}$, $\lb{s} = \lb{14} $, $\lb{C_r}= \{\lb s\}$ and  $\lb{t} = \lb{12} $. 
	On the right of Fig.~\ref{fig:fml_R_triangle} is represented the sequent resulting from the loop saturation, where $\lb {12}$ is replaced by $ \lb{14} $. 
	Observe that the cluster $\{\lb{14}\}$ remains unchanged. 
	After the loop saturation, proof search along the $\leq$-branch depicted stops: this is because layer $L_{6'}$ can be simulated by layer $L_{4}$. By adding relational atoms $\futs{15}{8}$, $\futs{14}{6}$ and $\futs{13}{7}$ to the sequent (Theorem~\ref{thm:step4}), which are the dashed arrows pointing downwards in the figure, we obtain (a part of) the countermodel for our formula.  
	To complete the countermodel, we need to take into account the layers generated from unhappy $\fmw{\Box}$ and $\fmw{\IMP}$ formulas   present in the sequent. Specifically, $ \labels{8}{\fmw{a \IMP \BOT}} $ and  $ \labels{8}{\fmw{a \IMP \BOT}} $ are unhappy. These formulas give rise to more $\leq$-branches in the sequent (but all branches are finite, and there is only a finite number of them).
\end{example}

\begin{figure*}[t!]

\hspace{2.7cm}	\begin{tikzpicture}[thick, every node/.style={scale=1.2}]

		\tikzstyle{node}=[circle,draw]
		\tikzstyle{s-node}=[circle,draw,inner sep=2pt]
		\tikzstyle{nonode}=[inner sep=0pt]
		\tikzstyle{access}=[->,blue]
		\tikzstyle{future}=[dashed,->]
		\tikzstyle{suricata}=[future,red]
		\tikzstyle{deleted}=[->,gray]
		\tikzstyle{loop}=[->,red]
		
		\node[font = {\Large}] (L0) at (-5.5,0) {$L_0$};
		\node[font = {\Large}] (L1) at (-5.5,2) {$L_1$};
		\node[font = {\Large}] (L2) at (-5.5,4) {$L_2$};
		\node[font = {\Large}] (L3) at (-5.5,6) {$L_3$};
		
		\node[label=left :{}] (1) at (4,0) [node] {$1$};
		
		\node[] (2) at (4,2) [node] {$2$};
		\node[label=below right :{}] (3) at (6,2) [node, fill=yellow!30] {$3$};
		\node[label=below right :{}] (4) at (8,2) [node, fill=green!30] {$4$};
		\node[label=below right :{}] (4d) at (10,2) [node, gray,fill=yellow!10,inner sep=6.3pt] {};
		\node[red] (p1) at (6,1.3) [] {$p_1$};
		\node[red] (C1) at (7,2) {$C_1$};
		
		\node[] (5) at (6,4) [node, fill=blue!30] {$5$};
		\node[label=below right:{}] (6) at (8,4) [node, fill=yellow!30] {$6$};
		\node[label=below right :{}] (7) at (10,4) [node,fill=green!30] {$7$};
		\node[] (8) at (2,4) [node, fill=yellow!30] {$8$};
		\node[label=below right :{}] (9) at (4,4) [node,fill=green!30] {$9$};
		\node[label=below right :{}] (10) at (0,4) [s-node] {$10$};

		\node[] (11) at (8,6) [s-node, fill=blue!30] {$11$};
		\node[label=below right :{}] (12) at (10,6) [s-node, fill=yellow!30] {$12$};
		\node[label=below right:{}] (13) at (12,6) [s-node, fill=green!30] {$13$};
		\node[red] (p2) at (10,6.7) [] {$p_2$};
		\node[red] (C2) at (11,6) {$C_2$};
		\node[label=below right :{}] (14) at (6,6) [s-node,fill=green!30] {$14$};
		\node[red] (t') at (6,6.7) {$t'$};
		\node[] (15) at (4,6) [s-node, fill=yellow!30] {$15$};
		\node[red] (t) at (4,6.7) {$t$};
		\node[] (16) at (2,6) [s-node, fill=purple!30] {$16$};
		\node[] (17) at (0,6) [s-node, fill=green!30] {$17$};
		\node[red] (s') at (0,6.7) {$s'$};
		\node[] (18) at (-2,6) [s-node, fill=yellow!30] {$18$};
		\node[red] (s) at (-2,6.7) {$s$};
		\node[] (19) at (-4,6) [s-node] {$19$};
		

		\draw[access] (2) -- (3);
		\draw[access] (3) edge[bend right] (4);
		\draw[deleted] (4) -- (4d);
		\draw[loop] (4) edge[bend right]  (3);
		
		\draw[access] (10) -- (8);
		\draw[access] (8) edge[bend right] (9);
		\draw[access] (9) edge[bend right]  (8);
		\draw[access] (9) -- (5);
		\draw[access] (5) -- (6);
		\draw[access] (6) edge[bend right] (7);
		\draw[access] (7) edge[bend right]  (6);
		
		\draw[access] (19) -- (18);
		\draw[access] (18) edge[bend right] (17);
		\draw[access] (17) edge[bend right]  (18);
		\draw[access] (17) -- (16);
		\draw[access] (16) -- (15);
		\draw[access] (15) edge[bend right] (14);
		\draw[access] (14) edge[bend right]  (15);
		\draw[access] (14) -- (11);
		\draw[access] (11) -- (12);
		\draw[access] (12) edge[bend right] (13);
		\draw[access] (13) edge[bend right]  (12);
%
%
		
		\draw[future] (1) -- (2);
		\draw[future] (2) -- (10);
		\draw[future] (3) -- (8);
		\draw[future] (3) -- (5);
		\draw[future] (3) -- (6);
		\draw[future] (4) -- (9);
		\draw[future] (4) -- (7);
		\draw[future] (10) -- (19);
		\draw[future] (8) -- (18);
		\draw[future] (9) -- (17);
		\draw[future] (5) -- (16);
		\draw[future] (6) -- (11);
		\draw[future] (6) -- (15);
		\draw[future] (6) -- (12);
		\draw[future] (7) -- (14);
		\draw[future] (7) -- (13);
		
	\end{tikzpicture}
\vspace*{1cm}

\hspace{2.7cm}	\begin{tikzpicture}[thick, every node/.style={scale=1.2}]
		
		\tikzstyle{node}=[circle,draw]
		\tikzstyle{s-node}=[circle,draw,inner sep=2pt]
		\tikzstyle{nonode}=[inner sep=0pt]
		\tikzstyle{access}=[->,blue]
		\tikzstyle{future}=[dashed,->]
		\tikzstyle{suricata}=[future,red]
		\tikzstyle{deleted}=[->,gray]
		\tikzstyle{loop}=[->,red]
		
		\node[font = {\Large}] (L0) at (-5.5,0) {$L_0$};
		\node[font = {\Large}] (L1) at (-5.5,2) {$L_1$};
		\node[font = {\Large}] (L2) at (-5.5,4) {$L_2$};
		\node[font = {\Large}] (L3) at (-5.5,6) {$L_3'$};
		\node[font = {\Large}] (L4) at (-5.5,8) {$L_4$};
		
		\node[label=left :{}] (1) at (4,0) [node] {$1$};
		
		\node[] (2) at (4,2) [node] {$2$};
		\node[label=below right :{}] (3) at (6,2) [node, fill=yellow!30] {$3$};
		\node[label=below right :{}] (4) at (8,2) [node, fill=green!30] {$4$};
		\node[red] (p1) at (6,1.3) [] {$p_1$};
		\node[red] (C1) at (7,2) {$C_1$};
		
		\node[] (5) at (6,4) [node, fill=blue!30] {$5$};
		\node[label=below right:{}] (6) at (8,4) [node, fill=yellow!30] {$6$};
		\node[label=below right :{}] (7) at (10,4) [node,fill=green!30] {$7$};
		\node[] (8) at (2,4) [node, fill=yellow!30] {$8$};
		\node[label=below right :{}] (9) at (4,4) [node,fill=green!30] {$9$};
		\node[label=below right :{}] (10) at (0,4) [s-node] {$10$};
		
		\node[] (11) at (8,6) [s-node, fill=blue!30] {$11$};
		\node[label=below right :{}] (12) at (10,6) [s-node, fill=yellow!30] {$12$};
		\node[label=below right:{}] (13) at (12,6) [s-node, fill=green!30] {$13$};
		\node[] (16) at (6,6) [s-node, fill=purple!30] {$16$};
		\node[] (17) at (4,6) [s-node, fill=green!30] {$17$};
		\node[] (18) at (2,6) [s-node, fill=yellow!30] {$18$};
		\node[] (19) at (0,6) [s-node] {$19$};
		
		\node[] (20) at (10,8) [s-node, fill=blue!30] {$20$};
		\node[] (21) at (12,8) [s-node, fill=yellow!30] {$21$};
		\node[] (22) at (14,8) [s-node, fill=green!30] {$22$};
		\node[red] (p2) at (12,8.7) [] {$p_2$};
		\node[red] (C2) at (13,8) {$C_2$};
		\node[] (23) at (6,8) [s-node, fill=yellow!30] {$23$};
		\node[red] (t) at (6,7.3) {$t$};
		\node[] (24) at (8,8) [s-node, fill=green!30] {$24$};
		\node[red] (t') at (8,7.3) {$t'$};
		\node[] (25) at (4,8) [s-node, fill=purple!30] {$25$};
		\node[red] (t'') at (4,7.3) {$t''$};
		\node[] (26) at (2,8) [s-node, fill=purple!30] {$26$};
		\node[red] (s'') at (2,7.3) {$s''$};
		\node[] (27) at (0,8) [s-node, fill=green!30] {$27$};
		\node[red] (s') at (0,7.3) {$s'$};
		\node[] (28) at (-2,8) [s-node, fill=yellow!30] {$28$};
		\node[red] (s) at (-2,7.3) {$s$};
		\node[] (29) at (-4,8) [s-node] {$29$};
		

		\draw[access] (2) -- (3);
		\draw[access] (3) edge[bend right] (4);
		\draw[access] (4) edge[bend right]  (3);
		
		\draw[access] (10) -- (8);
		\draw[access] (8) edge[bend right] (9);
		\draw[access] (9) edge[bend right]  (8);
		\draw[access] (9) -- (5);
		\draw[access] (5) -- (6);
		\draw[access] (6) edge[bend right] (7);
		\draw[access] (7) edge[bend right]  (6);
		
		\draw[access] (19) -- (18);
		\draw[access] (18) -- (17);
		\draw[access] (17) -- (16);
		\draw[loop] (16) edge[bend right] (18);
		\draw[access] (16) -- (11);
		\draw[access] (11) -- (12);
		\draw[access] (12) edge[bend right] (13);
		\draw[access] (13) edge[bend right]  (12);
		
		\draw[access] (29) -- (28);
		\draw[access] (28) -- (27);
		\draw[access] (27) -- (26);
		\draw[access] (26) edge[bend right] (28);
		\draw[access] (26) -- (25);
		\draw[access] (25) -- (23);
		\draw[access] (23) edge[bend right] (24);
		\draw[access] (24) edge[bend right]  (23);
		\draw[access] (24) -- (20);
		\draw[access] (20) -- (21);
		\draw[access] (21) edge[bend right] (22);
		\draw[access] (22) edge[bend right]  (21);
		

		\draw[future] (1) -- (2);
		\draw[future] (2) -- (10);
		\draw[future] (3) -- (8);
		\draw[future] (3) -- (5);
		\draw[future] (3) -- (6);
		\draw[future] (4) -- (9);
		\draw[future] (4) -- (7);
		\draw[future] (10) -- (19);
		\draw[future] (8) -- (18);
		\draw[future] (9) -- (17);
		\draw[future] (5) -- (16);
		\draw[future] (6) -- (11);
		\draw[future] (6) -- (18);
		\draw[future] (6) -- (12);
		\draw[future] (7) -- (17);
		\draw[future] (7) -- (13);
		\draw[future] (19) -- (29);
		\draw[future] (18) -- (28);
		\draw[future] (17) -- (27);
		\draw[future] (16) -- (26);
		\draw[future] (11) -- (25);
		\draw[future] (12) -- (20);
		\draw[future] (12) -- (23);
		\draw[future] (12) -- (21);
		\draw[future] (13) -- (24);
		\draw[future] (13) -- (22);

	\end{tikzpicture}
\vspace*{1cm}

\hspace{2.7cm}	\begin{tikzpicture}[thick, every node/.style={scale=1.2}]
		
		\tikzstyle{node}=[circle,draw]
		\tikzstyle{s-node}=[circle,draw,inner sep=2pt]
		\tikzstyle{nonode}=[inner sep=0pt]
		\tikzstyle{access}=[->,blue]
		\tikzstyle{future}=[dashed,->]
		\tikzstyle{suricata}=[future,red]
		\tikzstyle{deleted}=[->,gray]
		\tikzstyle{loop}=[->,red]
		
		\node[font = {\Large}] (L0) at (-5.5,0) {$L_0$};
		\node[font = {\Large}] (L1) at (-5.5,2) {$L_1$};
		\node[font = {\Large}] (L2) at (-5.5,4) {$L_2$};
		\node[font = {\Large}] (L3) at (-5.5,6) {$L_3'$};
		\node[font = {\Large}] (L4) at (-5.5,8) {$L_4'$};
		
		\node[label=left :{}] (1) at (4,0) [node] {$1$};
		
		\node[] (2) at (4,2) [node] {$2$};
		\node[label=below right :{}] (3) at (6,2) [node, fill=yellow!30] {$3$};
		\node[label=below right :{}] (4) at (8,2) [node, fill=green!30] {$4$};
		%
		
		\node[] (5) at (6,4) [node, fill=blue!30] {$5$};
		\node[label=below right:{}] (6) at (8,4) [node, fill=yellow!30] {$6$};
		\node[label=below right :{}] (7) at (10,4) [node,fill=green!30] {$7$};
		\node[] (8) at (2,4) [node, fill=yellow!30] {$8$};
		\node[label=below right :{}] (9) at (4,4) [node,fill=green!30] {$9$};
		\node[label=below right :{}] (10) at (0,4) [s-node] {$10$};
		
		\node[] (11) at (8,6) [s-node, fill=blue!30] {$11$};
		\node[label=below right :{}] (12) at (10,6) [s-node, fill=yellow!30] {$12$};
		\node[label=below right:{}] (13) at (12,6) [s-node, fill=green!30] {$13$};
		\node[] (16) at (6,6) [s-node, fill=purple!30] {$16$};
		\node[] (17) at (4,6) [s-node, fill=green!30] {$17$};
		\node[] (18) at (2,6) [s-node, fill=yellow!30] {$18$};
		\node[] (19) at (0,6) [s-node] {$19$};
		
		\node[] (20) at (10,8) [s-node, fill=blue!30] {$20$};
		\node[] (21) at (12,8) [s-node, fill=yellow!30] {$21$};
		\node[] (22) at (14,8) [s-node, fill=green!30] {$22$};
		%
		%
		\node[] (26) at (8,8) [s-node, fill=purple!30] {$26$};
		\node[] (27) at (6,8) [s-node, fill=green!30] {$27$};
		\node[] (28) at (4,8) [s-node, fill=yellow!30] {$28$};
		\node[] (29) at (2,8) [s-node] {$29$};
		

		\draw[access] (2) -- (3);
		\draw[access] (3) edge[bend right] (4);
		\draw[access] (4) edge[bend right]  (3);
		
		\draw[access] (10) -- (8);
		\draw[access] (8) edge[bend right] (9);
		\draw[access] (9) edge[bend right]  (8);
		\draw[access] (9) -- (5);
		\draw[access] (5) -- (6);
		\draw[access] (6) edge[bend right] (7);
		\draw[access] (7) edge[bend right]  (6);
		
		\draw[access] (19) -- (18);
		\draw[access] (18) -- (17);
		\draw[access] (17) -- (16);
		\draw[access] (16) edge[bend right] (18);
		\draw[access] (16) -- (11);
		\draw[access] (11) -- (12);
		\draw[access] (12) edge[bend right] (13);
		\draw[access] (13) edge[bend right]  (12);
		
		\draw[access] (29) -- (28);
		\draw[access] (28) -- (27);
		\draw[access] (27) -- (26);
		\draw[access] (26) edge[bend right] (28);
		\draw[access] (26) -- (20);
		\draw[access] (20) -- (21);
		\draw[access] (21) edge[bend right] (22);
		\draw[access] (22) edge[bend right]  (21);
		

		\draw[future] (1) -- (2);
		\draw[future] (2) -- (10);
		\draw[future] (3) -- (8);
		\draw[future] (3) -- (5);
		\draw[future] (3) -- (6);
		\draw[future] (4) -- (9);
		\draw[future] (4) -- (7);
		\draw[future] (10) -- (19);
		\draw[future] (8) -- (18);
		\draw[future] (9) -- (17);
		\draw[future] (5) -- (16);
		\draw[future] (6) -- (11);
		\draw[future] (6) -- (18);
		\draw[future] (6) -- (12);
		\draw[future] (7) -- (17);
		\draw[future] (7) -- (13);
		\draw[future] (19) -- (29);
		\draw[future] (18) -- (28);
		\draw[future] (17) -- (27);
		\draw[future] (16) -- (26);
		\draw[future] (11) -- (26);
		\draw[future] (12) -- (20);
		\draw[future] (12) -- (28);
		\draw[future] (12) -- (21);
		\draw[future] (13) -- (27);
		\draw[future] (13) -- (22);
		\draw[future,red] (29) edge[bend right] (19);
		\draw[future,red] (28) edge[bend right] (18);
		\draw[future,red] (27) edge[bend right] (17);
		\draw[future,red] (26) edge[bend right] (16);
		\draw[future,red] (20) edge[bend right] (11);
		\draw[future,red] (21) edge[bend right] (12);
		\draw[future,red] (22) edge[bend right] ( 13);

	\end{tikzpicture}
	\caption{Formula $\fm{ \BOX( \DIA D \AND \DIA c )\IMP \BOT }$, with $\fm{D} = \fm{(a \IMP b) \IMP \BOT }$. A loop saturation step is applied to the topmost sequent, and the result of the saturation is represented in the middle sequent. Then, a loop saturation is also applied to the middle sequent, and the result is the lowermost sequent. 
	}
	\label{fig:fml_U_triangle}
\end{figure*}
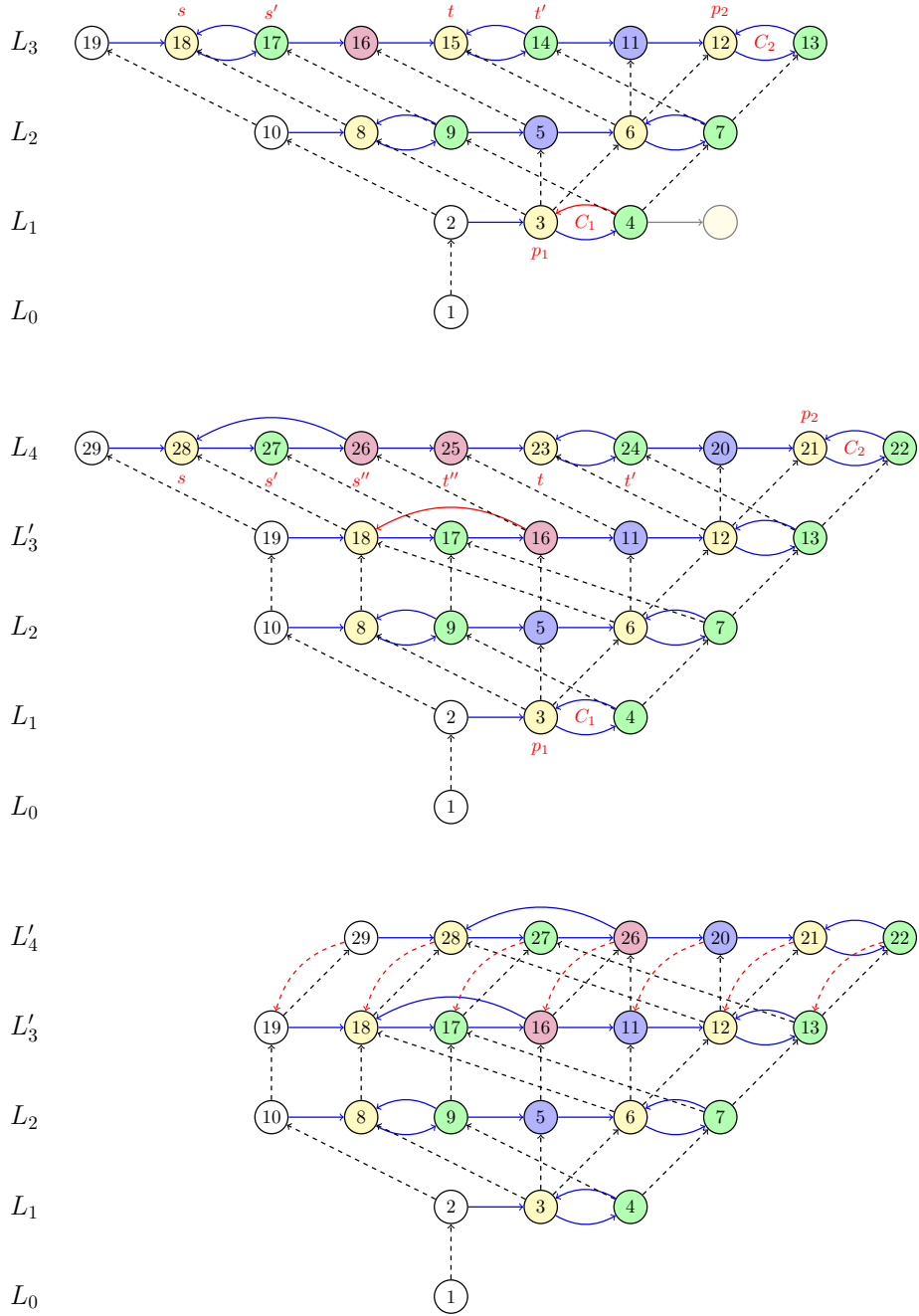

\begin{example}[Non-valid formula, U-triangle loop]	
	\label{ex:U}
	Let us consider formula  $\fm{ \BOX( \DIA D \AND \DIA c )\IMP \BOT }$, with $\fm{D} = \fm{(a \IMP b) \IMP \BOT }$. Figure~\ref{fig:fml_U_triangle} illustrates one sequent generated by the search algorithm. Let $\fmb{\Gamma} = \{\fmb{\BOX(\DIA D \AND \DIA c)}, \fmb{\DIA D \AND \DIA c}, \fmb{\DIA D}, \fmb{\DIA c}\}$. 
	To each label we associate the following sets, color-coded in the figure:
	
	\begin{tabular}{l | l}
		$ \lb 1 $ & $\fmb{\Gamma} \cup \{\fmw{ \BOX( \DIA D \AND \DIA c )\IMP \BOT }\} $\\
		\hline 
		$ \lb 2 $ & $\fmb{\Gamma} \cup \{\fmw{\BOT}\}$\\
		\hline 
		$ \lb {10} \lbeq \lb{19} \lbeq \lb{29}   $ & $\fmb{\Gamma} $\\
		\hline 
 		$ \lb 3 \lbeq \lb 8 \lbeq \lb 6 \lbeq  $ & \multirow{3}{*}{ $\fmb{\Gamma} \cup \{ \fmb{D}, \fmw{a \IMP b}\}$ } \\
 		$ \lb{18} \lbeq\lb{15} \lbeq  \lb{12} \lbeq $ & \\
 		$\lb{28} \lbeq   \lb{23} \lbeq \lb{21} $ & \\ 
 		\hline 
 		$\lb{4} \lbeq  \lb{9} \lbeq \lb{7} \lbeq $ & \multirow{3}{*}{ $\fmb{\Gamma} \cup \{ \fmb{c}\}$ }  \\
 		$\lb{17} \lbeq  \lb{14} \lbeq \lb{13} \lbeq $ & \\ 	
 		$\lb{27} \lbeq  \lb{24} \lbeq \lb{22} \lbeq $ & \\	
 		\hline 
		$\lb{16} \lbeq  \lb{26} \lbeq \lb{25}  $ &  $\fmb{\Gamma} \cup \{ \fmb{D}, \fmb{a}, \fmw{a \IMP b}\}$\\ 
		\hline 
		$\lb{5} \lbeq  \lb{11} \lbeq \lb{20}  $ &  $\fmb{\Gamma} \cup \{ \fmb{D}, \fmb{a}, \fmw{b}, \fmw{a \IMP b}\}$\\ 
	\end{tabular}
	
	\noindent As in the previous example, only one $\leq$-branch is represented. 
	Moreover, for reasons of space, we have not pictured a second $R$-branch originating from $\lb{2}$ by saturation, to make formula $\labels{2}{\fmb{\DIA c}}$ happy. The labels originated by lifting this branch are present in all layers.

	As in Example~\ref{ex:provable}, in layer $L_1$, after saturation, a non-singleton loop is created, $ \{\lb{3}, \lb{4}\}$. Then, layer $L_2$ is generated by lifting saturation, which duplicates the cluster. The process is repeated to generate $L_3$.  
	At $L_3$, an unhappy U-triangle loop is found. Using the terminology from Def.~\ref{def:u-triangle}, take $\lb{C_1} = \{\lb{3}, \lb{4}\}$, $\lb{C_2} = \{\lb{12}, \lb{13}\}$, $\lb{s} = \lb{18}$ and $\lb{t} =\lb{15}$. Thus, we substitute $ \lb{15} $ with $\lb{18}$. Moreover, a second unhappy U-triangle loop is found, by taking  $\lb{s'} = \lb{17}$ and $\lb{t'} =\lb{14}$ (refer to the topmost sequent of Fig.~\ref{fig:fml_U_triangle}). Layer $L_{3'}$ is the result of the loop saturation: it contains a cluster $\{\lb{18}, \lb{17}, \lb{16}\}$. The algorithm produces a sequent consisting of layers $L_0$ - $L_2$, $L_{3'}$, and continues (middle sequent in  Fig.~\ref{fig:fml_U_triangle}).
	
	Then, after a lifting saturation, layer $L_4$ is generated. Here, three unhappy U-triangle loops are present: take $\lb{C_1} =\{\lb 3, \lb 4\}$, $\lb{C_2} =\{\lb {21}, \lb {22}\}$, and then  $\lb{s} = \lb{28}$ and $\lb{t} =\lb{23}$, $\lb{s'} = \lb{27}$ and $\lb{t'} =\lb{24}$ and $\lb{s''} = \lb{26}$ and $\lb{t''} =\lb{25}$. After the loop saturation, we obtain layer $L_{4'}$, which can be simulated by layer $L_{3'}$ (lowermost sequent in Fig.~\ref{fig:fml_U_triangle}). 
	By adding relational atoms $\futs{29}{19}$, $\futs{28}{18}$, $\futs{27}{17}$, $\futs{26}{16}$, $\futs{20}{11}$, $\futs{21}{12}$ and $\futs{22}{13}$ we have a (partial) countermodel for the formula at the root. To complete the countermodel, we need to include the layers generated by lifting the layers of the sequent in  correspondence to each remaining unhappy formula $\fmw{(a \IMP b)}$. 	
\end{example}



\subsection{Proofs from Section \ref{sec:sequents_models}}

\completeness*

\begin{proof}
	The worlds~$W$ of~$\modelof{G}$ are the labels of the sequent. Conditions~$\fone$~and~$\ftwo$, the transitivity and reflexivity of~$\sle G$ and $\srel G$, and the monotonicity of~$V$ all follow by construction due to structural saturation. Thus, $\modelof{G}$ is a transitive and reflexive birelational model. It only remains to show the two properties about forcing, that we prove  by mutual induction on the size of~$\fm A$, proceeding by case analysis on the main connective of~$\fm A$:
	
	\begin{itemize}
		\item $\blackf x{\BOT}G$: it is not possible for a happy sequent.
		
		\item $\whitef x{\BOT}G$: we have $\nforce{\modelof{G}}x\BOT$ by definition.
		
		\item $\blackf xaG$: by Definition~\ref{dfn:modelofsequent}, $\force{\modelof{G}}xa$.
		
		\item $\whitef xaG$: it is not the case that $\blackf xaG$ by happiness of~$\lb x$, hence,   $\nforce{\modelof{G}}{x}{a}$ by Definition~\ref{dfn:modelofsequent}.
		
		\item $\blackf x{B \AND C}G$: by happiness of~$\lb x$,  both $\blackf xBG$ and $\blackf xCG$. Then $\force{\modelof{G}}xB$ and  $\force{\modelof{G}}xC$ by~IH. Therefore,  $\force{\modelof{G}}x{B \AND C}$.
		
		\item $\whitef x{B \AND C}G$: by happiness of~$\lb x$, either $\whitef xBG$ or $\whitef xCG$. Then either $\nforce{\modelof{G}}xB$ or $\nforce{\modelof{G}}xC$ by~IH. Therefore,  $\nforce{\modelof{G}}x{B \AND C}$.
		
		\item Cases for $\blackf x{B \OR C} G$ and $\whitef x{B \OR C} G$ are analogous. 
%
		
		\item $\blackf x{B \IMP C}G$: consider any $\lb y$ with $\futsq xy$.
		By ($\monl$)-structural saturation,  $\blackf y{B \IMP C}G$.
		By happiness of~$\lb y$, either $\whitef yBG$ or $\blackf yCG$. By~IH, either $\nforce{\modelof{G}}yB$ or $\force{\modelof{G}}yC$. Thus,   $\force{\modelof{G}}yB$ implies $\force{\modelof{G}}yC$ for all $\lb y$ with $\futsq xy$. Therefore,  $\force{\modelof{G}}x{B \IMP C}$.
		
		
		\item $\whitef x{B \IMP C}G$: by happiness of~$\lb x$, there is a world~$\lb{y}$ such that $\futsq xy$,  $\blackf yB G$,  and $\whitef yC G$. By IH,  $\force{\modelof{G}}yB$ and $\nforce{\modelof{G}}{y}{C}$. Therefore, $\nforce{\modelof{G}}x{B \IMP C}$.
		
		\item $\blackf x{\DIA B} G$: by happiness of $\lb x$, there is a world~$\lb{y}$ such that $\accsq xy$ and $\blackf yBG$. By IH, we have $\force{\modelof{G}}yB$ and, therefore,  $\force{\modelof{G}}x{\DIA B}$.
		
		\item $\whitef x{\DIA B} G$:
		by happiness of~$\lb x$, we have $\whitef yBG$ for all worlds~$\lb y$ such that $\accsq xy$. Thus, by~IH,  $\nforce{\modelof{G}}yB$ whenever $\accsq xy$. Therefore, $\nforce{\modelof{G}}x{\DIA B}$.
		
		
		\item $\blackf x{\BOX B} G$: 
		 consider arbitrary $\lb y$ and $\lb z$ with $\futsq xy$ and $\accsq yz$.
		By ($\monl$)-structural saturation, $\blackf y{\BOX B}G$. By happiness of $\lb y$, we have $\blackf zBG$. Thus, by IH, $\force{\modelof{G}}zB$ whenever $\futsq xy$ and $\accsq yz$. Therefore, $\force{\modelof{G}}{x}{\BOX B}$.

		\item $\whitef x{\BOX B} G$: by happiness of $\lb x$, there exist~$\lb{y}$~and~$\lb{z}$ such that $\futsq xy$, $\accsq yz$, and $\whitef zBG$. By IH, $\nforce{\modelof{G}}zB$. Therefore, $\nforce{\modelof{G}}x{\BOX B}$.
		\qedhere
	\end{itemize}
	
\end{proof}

\subsection{Proofs from Section \ref{sec:clusters}}

We start with a proposition  that is not in the main text but simplifies many arguments.

\begin{proposition}
	\label{prop:hat_satur}
	Structurally saturated sequents\/~$\sq G$  are also saturated w.r.t.~$\lrel G$:
	\begin{enumerate}[leftmargin=3.5em]
		\item[$(\fhat)$] if $\accsqh xy $ and  $ \futsq xz $, then there is~$\lb u$ such that $ \futsq yu $ and $\accsqh zu$.
	\end{enumerate}
\end{proposition}

\begin{proof}
	There must be a sequence $\lb{x_0}, \dots, \lb{x_n}$ of labels with $n\ge0$ such that $\lb{x_0}=\lb x$, $\lb{x_n}=\lb y$, and  for each $0 \le i\le n-1$, either $\accsq{x_i}{x_{i+1}}$ or $\accsq{x_{i+1}}{x_{i}}$. We use induction on $n$. If $n=0$, i.e., $\lb x = \lb y$, then $\lb u \colonequals \lb z$ suffices. Otherwise, by IH, there is~$\lb v$ such that $\futsq{x_{n-1}}v$ and $\accsqh zv$. If $\accsq{x_{n}}{x_{n-1}}$, there is $\lb u$ by $(\fone)$-structural saturation  such that  $\futsq{x_n}u$ and $\accsq uv$. If  $\accsq{x_{n-1}}{x_n}$, there is $\lb u$ by $(\ftwo)$-structural saturation such that $\futsq{x_n}u$ and $\accsq vu$. Either way, $\futsq yu$ and $\accsqh zu$.
\end{proof}

\OrderLayers* 

\begin{proof}
	For reflexivity, consider any layer~$L$. For any label $\lb x \in L$, by ($ \Lref $)-structural saturation, $\futsq xx$. Hence, $L \le L$ by Definition~\ref{def:lay-sq}.
	
	For transitivity, let $L_1 \le L_2$  and $L_2 \le L_3$ for layers~$L_1$, $L_2$,~and~$L_3$.  By Definition~\ref{def:lay-sq}, $\futsq x{y'}$ and $\futsq yz$ for some labels~$\lb x \in L_1$, $\lb {y'}, \lb y \in L_2$, and $\lb z \in L_3$. Since $\accsqh {y'}y$, by ($\fhat$)-structural saturation (Proposition~\ref{prop:hat_satur}), there is a label $\lb {z'}$ such that $\futsq {y'}{z'}$ and $\accsqh {z'}z$. The latter means that $\lb {z'} \in L_3$.  By ($ \Ltr $)-structural saturation, $\futsq x{z'}$. Hence, $L_1 \le L_3$ by Definition~\ref{def:lay-sq}.
	
	
	For antisymmetry, let $L_1\le L_2$ and $L_1\neq L_2$.
	By Definition~\ref{def:lay-sq}, there are labels $\lb{x} \in L_1$ and $\lb{x'} \in L_2$ such that  $\futsq{x}{x'}$. For arbitrary labels $\lb{y} \in L_1$ and $\lb{y'} \in L_2$, we have $\accsqh xy$ and $\accsqh{x'}{y'}$. Since $L_1 \neq L_2$, we have $\lb x \neq \lb{x'}$. Hence, $\nfutsq{y'}y$ by~\ref{def:lay-sq-it2}) of Definition~\ref{def:lay-sq}. Since $\nfutsq{y'}y$ for any $\lb{y'} \in L_2$ and $\lb{y} \in L_1$, it is not the case that $L_2 \leq L_1$. 
\end{proof}

\OrderClusters*

\begin{proof}
	For reflexivity of~$\grel G$, consider any cluster~$\lbc C$. For any label $\lb x \in \lbc C$, by ($ \Rref $)-structural saturation, $\accsq xx$. Hence, $\accsq CC$ by Definition~\ref{def:cluster}. 
	For transitivity of~$\grel G$, let $\accsq{C_1}{C_2}$  and $\accsq{C_2}{C_3}$ for clusters~$\lbc C_1$, $\lbc C_2$,~and~$\lbc C_3$.  By Definition~\ref{def:cluster}, $\accsq xy$ and $\accsq uz$ for some labels~$\lb x \in \lbc{C_1}$, $\lb y, \lb u \in \lbc{C_2}$, and $\lb z \in \lbc{C_3}$. Since $\accsq yu$, by $(\Rtr)$-structural saturation $\accsq xz$. Thus, $\accsq{C_1}{C_3}$ by Definition~\ref{def:cluster}. 
	For antisymmetry of~$\grel G$, let $\accsq{C_1}{C_2}$ and $\accsq{C_2}{C_1}$.  
	By Definition~\ref{def:cluster}, there are labels $\lb{x}, \lb{y} \in \lbc{C_1}$ and $\lb{x'},\lb{y'} \in \lbc{C_2}$ such that  $\accsq{x}{x'}$ and $\accsq{y'}{y}$. For arbitrary labels $\lb{u} \in \lbc{C_1}$ and $\lb{v} \in \lbc{C_2}$, we have $\accsq ux$, and $\accsq{x'}v$, hence, by $(\Rtr)$-structural saturation, $\accsq uv$. Similarly, $\accsq vu$ because $\accsq v{y'}$ and $\accsq yu$. Since both $\accsq uv$ and $\accsq vu$  for all $\lb{u} \in \lbc{C_1}$ and $\lb{v} \in \lbc{C_2}$, all labels in these two clusters form one equivalence class w.r.t.~$\grel G \cap \grel{G}^{-1}$, i.e.,~$\lbc{C_1} = \lbc{C_2}$. 
	
	For reflexivity of~$\sle G$, consider any cluster~$\lbc C$. For every label $\lb y \in \lbc C$, by ($ \Lref $)-structural saturation, $\futsq yy$. Hence, $\futsq CC$ by Definition~\ref{def:cluster}. 
	For transitivity of~$\sle G$, let $\futsq{C_1}{C_2}$  and $\futsq{C_2}{C_3}$ for clusters~$\lbc C_1$, $\lbc C_2$,~and~$\lbc C_3$.  By Definition~\ref{def:cluster}, for every  $\lb z \in \lbc{C_3}$, there is $\lb y \in \lbc{C_2}$ such that $\futsq yz$. In its turn, for this~$\lb y$, there is $\lb x \in \lbc{C_1}$ such that $\futsq xy$. By $(\Ltr)$-structural saturation $\futsq xz$ for this~$\lb x$. Thus, $\futsq{C_1}{C_3}$ by Definition~\ref{def:cluster}. Hence, $\sle G$ is a preorder.
	
	Assume now additionally that $\sq G$ is layered.
	For antisymmetry of~$\sle G$, let $\futsq{C_1}{C_2}$ and $\lbc{C_1}\neq\lbc{C_2}$, i.e., $\lbc{C_1} \cap \lbc{C_2} = \varnothing$. To show that $\nfutsq{C_2}{C_1}$, we consider any label $\lb x \in \lbc{C_1}$ and show that $\nfutsq yx$ for all~$\lb y \in \lbc{C_2}$.
	By Definition~\ref{def:cluster}, for each $\lb y \in \lbc{C_2}$, there is some  label $\lb{x'} \in \lbc{C_1}$ such that  $\futsq{x'}y$. We have $\accsqh{x'}x$ because they belong to the same cluster, $\accsq yy$, and hence $\accsqh yy$, by $(\Rref)$-structural saturation, and $\lb{x'} \ne \lb y$ because they are from disjoint clusters. Hence,  $\nfutsq yx$ by Definition~\ref{def:lay-sq}. 
	Since $\lb y$ was chosen arbitrarily, it follows that $\nfutsq{C_2}{C_1}$ and $\sle G$~is an order.	
\end{proof}

\subsection{Proofs from Section \ref{sec:algorithm}}

 For simplicity, when $w\le v$, then we say that $w$ is \emph{a past of $v$} and that $v$ is \emph{a future of $w$}. 

\SaturationStable*

\begin{proof}
	To prove~\ref{lemma:ssat:stable} assume that set~$\sqset{S}$ is finite and all sequents $\sq G \in \sqset{S}$ are  stable, i.e., tree-layered and tree-clustered (in particular, layered and structurally saturated) with happy inner layers. The size of~$\sqset{S'}$ is larger than that of $\sqset S$ by at most one sequent, hence, $\sqset{S'}$ is also finite. Since all labels in the inner layers of sequents from $\sqset S$ are happy, labels of all newly added labelled formulas, be it label~$\lb x$ itself or labels $\grel G$-accessible from~$\lb x$, must be in a topmost layer of some sequent from~$\sqset S$. Given that this sequent is layered, the labels of all new labelled formulas have no non-reflexive futures, making ($ \monl $)~trivial for them. 
	Since  rewrites neither introduce new labels nor add relational atoms, all other conditions of structural saturation, as well as being being tree-layered and tree-clustered,  remain true.  
	Since no labelled formulas have been removed and all inner layers of all sequents remain unchanged,   the inner layers remain happy (see Remark~\ref{rem:tobereferred}). Therefore, all sequents in~$\sqset{S'}$ are stable, and so is~$\sqset{S'}$ itself.

	
	To prove \ref{lemma:ssat:term}, observe that each rewrite 
	adds a labelled formula to a sequent~$\sq{G} \in \sqset{S}$ if and only if the formula is not already in~$\sq G$. 
	Moreover, the added formulas are subformulas
	of formulas in the sequent, and no new labels are introduced.  
	Since each~$\sq{G}$~contains finitely many formulas and finitely many labels, only finitely many formulas can be added to the sequent. Thus, if there are finitely many sequents in~$\sqset S$, relation~$\ssatred$ is terminating.

	Finally, to prove \ref{lemma:ssat:ssat}, suppose first that $\sqset S$~is semi-saturated. Then, by definition, $\sqset S$~is finite, all sequents in~$\sqset S$ are stable, in particular,  have happy inner layers, and have naively happy topmost layers. Since all these finitely many sequents are stable, $\sqset S$~is stable. Since all their layers are (at least) naively happy,  no rewrite step is applicable to~$\sqset{S}$, so $\sqset{S}$~is in normal form w.r.t.~$\ssatred$. Conversely, suppose that  $\sqset S$~is stable and in normal form w.r.t.~$\ssatred$. Then $\sqset S$~is a finite set of stable sequents. Since no rewrite is applicable to any sequent from~$\sqset S$, all (topmost) layers of all these sequents are naively happy. Hence, all finitely many sequents from~$\sqset{S}$ are semi-saturated, and so is~$\sqset S$ itself.%
\end{proof}

The following two lemmas are not in the main text and are needed to prove Lemma~\ref{lemma:sat}.

\begin{lemma}
	\label{lemma:satA}
	The following statements hold:
	\begin{enumerate}[i)]
		\item
		\label{lemma:sat:onestep}
		If a sequent\/~$\sq G$ is semi-saturated and\/ $\sq G \diared \sq{G'}$, then the sequent\/~$\sq{G'}$ is  stable.
		\item
		\label{lemma:sat:optiononedone}
		If a sequent\/~$\sq G$ is semi-saturated and\/ $\sq G \diared \sq{G'}$ according to option~\ref{def:dia-sat:one}, then  the sequent\/~$\sq{G'}$ is semi-saturated.
	\end{enumerate}
\end{lemma}

\begin{proof}
	To prove \ref{lemma:sat:onestep} assume that a sequent~$\sq G$ is semi-saturated, in particular, stable (i.e., structurally saturated, tree-layered, and tree-clustered  with happy inner layers). Let $\sq G\diared\sq G'$. For either option from Def.~\ref{def:dia-sat}, the unhappy label $\lb y$ can only be in a topmost layer of~$\sq G$. Both labels $\lb x$  from option~\ref{def:dia-sat:one} or $\lb z$ from option~\ref{def:dia-sat:two} belong to the same topmost layer. Hence, in either case neither labels nor most labelled formulas from the inner layers are affected and, therefore, they remain happy by Remark~\ref{rem:tobereferred}. For option~\ref{def:dia-sat:two}, no formulas or labels are removed so all white $\fmw{A \IMP B}$ and $\fmw{\BOX A}$ formulas from the inner layers remain happy. For option~\ref{def:dia-sat:one},  formulas $\fmw{A \IMP B}$ from inner layers remain happy because $y$ has no past. Finally, it is possible that  $\whitef k{\BOX A}{G}$ from some inner layer is happy because of $\whitef y{A}{G}$, $\futsq k w$, and $\accsq w y$. But since $\lb x \lbeq \lb y$, also $\whitef x{A}{G'}$. Note that $\lb w \ne \lb y$ since $\lb y$ has no past. Thus, $\futsqpr kw$ and $\accsqpr wx$ means that $\whitef k{\BOX A}{G'}$ remains happy. For option~\ref{def:dia-sat:one}, there are no new labels and $\lb y$ has no past in~$\sq G$, hence, the only change in $\le$-relational atoms is that $\futs yy$ turns into $\futs xx$; 
	for option~\ref{def:dia-sat:two}, the only new label is $\lb z$ with $\futs zz$ being the only new $\le$-relational atom. Either way, the $(\Lref)$- and $(\Ltr)$-structural saturation is ensured and the set of irreflexive $\le$-relational atoms remains unchanged. The latter implies that the  $(\monl)$-, \mbox{$ (\fone) $-,} and $(\ftwo)$-structural  saturation of~$\sq{G'}$ follows from the same properties of~$\sq G$. The $(\Rtr)$-structural  saturation is explicitly enforced. So is  $(\Rref)$-structural  saturation for option~\ref{def:dia-sat:two}, where it is necessary for the added label~$\lb z$. 	
	This completes the proof that $\sq{G'}$~is {structurally saturated}. 
	As the rewrite does not affect the  layer structure of the sequent, $\sq{G'}$~being {tree-layered} immediately follows from the same property of~$ \sq G $.  
	To show that  $\sq{G'}$ is {tree-clustered}  we consider the two options separately. In option~\ref{def:dia-sat:one} of Def.~\ref{def:dia-sat}, the rewrite performs a substitution of label~$\lb x$ for label~$ \lb y $ and applies the transitive closure. By considering the shortest chains justifying new $\grel{G'}$-links, and using the structural saturation of both $\sq G$ and $\sq{G'}$ along with $\accsq xy$, it follows that:
	\begin{align}
	\label{newlinktox}
	\accsqpr ux \quad&\Longrightarrow\quad \accsq uy;
	\\
	\label{newlinkfromx}
	\accsqpr xv  \quad&\Longrightarrow\quad \accsq xv;
	\\
	\label{newlinknox}
	\accsqpr uv  \text{ but not } \accsq uv \quad&\Longrightarrow\quad \accsq uy \& \accsq xv.
	\end{align}
	Let $\lbc{C'_x}$ be the cluster in $\sq{G'}$ that contains $\lb x$. For any cluster $\lbc C$ in $\sq G$, either $\lb y \in \lbc C$ and  $\lbc C \setminus \{\lb y\}\subseteq \lbc{C'_x}$ or $\lb y \notin \lbc C$ and either $\lbc C \subseteq \lbc{C'_x}$ or   $\lbc C$ remains a cluster in~$\sq{G'}$. The case of $\lb y \in \lbc C$ is easy.  
	Suppose $\lb y \notin \lbc C$, which is not a cluster  in~$\sq{G'}$. Since all $\grel G$-links  unrelated to~$\lb y$ are preserved in $\sq{G'}$, cluster~$\lbc C$ can only grow. So there must be some labels $\lb u \in \lbc C$ and $\lb v\notin \lbc C$ such that   $\accsqpr uv$, $\accsqpr vu$, but either not $\accsq uv$ or not $\accsq vu$. By~\eqref{newlinknox}, in the former case, $\accsq uy$ and $\accsq xv$, hence, $\accsqpr ux$ and $\accsqpr xv $, while in the latter case, $\accsq vy$ and $\accsq xu$, hence, 	$\accsqpr vx$ and $\accsqpr xu $. Either way,  by structural saturation of~$\sq{G'}$, $\lb u \in \lbc{C'_x}$, and $\lbc C \subseteq \lbc{C'_x}$. 
	To show that $\sq{G'}$ is tree-clustered (in option~\ref{def:dia-sat:one} of Def.~\ref{def:dia-sat}), we consider 
	three clusters $\lbc{C_1}$, $\lbc{C_2}$,~and~$\lbc{C'}$  from $\sq{G'}$ such that $\accsqpr{C_1}{C'}$ and $\accsqpr{C_2}{C'}$ and show that $\lbc{C_1}$~and~$\lbc{C_2}$~are $\grel{G'}$-comparable. If any two of them are equal, this is trivial, so  assume they are pairwise distinct. If none of them is~$\lbc{C'_x}$, then they are also clusters in $\sq G$ and it is sufficient to show that there is some cluster~$\lbc{C^*}$ in~$\sq G$ such that $\accsq{C_1}{C^*}$ and $\accsq{C_2}{C^*}$. Then $\lbc{C_1}$~and~$\lbc{C_2}$~are $\grel G$-comparable because $\sq G$ is tree-clustered and the connection remains in $\sq{G'}$, meaning that they are also $\grel{G'}$-comparable. 
	If both $\accsq{C_1}{C'}$ and $\accsq{C_2}{C'}$, then $\lbc{C^*} = \lbc{C'}$. If neither $\accsq{C_1}{C'}$ nor $\accsq{C_2}{C'}$, then
	it follows from~\eqref{newlinknox} that $\accsq{C_1}{C_y}$ and $\accsq{C_2}{C_y}$ where $\lbc{C_y}$ is the cluster in $\sq G$ that contains $\lb y$. So $\lbc{C^*} = \lbc{C_y}$. Finally, if, w.l.o.g.\ $\accsq{C_1}{C'}$ but not $\accsq{C_2}{C'}$, then  $\accsq{C_2}{C_y}$ and $\accsq{C_x}{C'}$ by~\eqref{newlinknox} where $\lbc{C_x}$~is the cluster in~$\sq G$ that contains~$\lb x$. Since $\sq G$ is tree-clustered, either $\accsq{C_x}{C_1}$ or $\accsq{C_1}{C_x}$. In the latter case, $\accsq{C_1}{C_y}$, so $\lbc{C^*}=\lbc{C_y}$. In the former case, we derive the $\grel{G'}$-connection $\accsqpr{C_2}{C_1}$ directly from $\accsq{C_2}{C_y}$ and $\accsq{C_x}{C_1}$. 
	It remains to consider the case when exactly one of the three clusters is~$\lbc{C'_x}$  and the other two are clusters of $\sq G$. 
	If $\lbc{C'} = \lbc{C'_x}$, then $\accsq{C_1}{C_y}$ and $\accsq{C_2}{C_y}$ by~\eqref{newlinktox}, so the above argument with $\lbc{C^*}=\lbc{C_y}$ suffices.
	Finally, if w.l.o.g.\ $ \lbc{C_1} =\lbc{C'_x} $, then $\accsq{C_x}{C'}$ by~\eqref{newlinkfromx}. If $\accsq{C_2}{C'}$, then  $\lbc{C_x}$ is $\grel G$-comparable to $\lbc{C_2}$ because $\sq G$ is tree-clustered, and $\lbc{C'_x}$ is $\grel{G'}$-comparable to $\lbc{C_2}$. If $\accsqpr{C_2}{C'}$ but not $\accsq{C_2}{C'}$, then $\accsq{C_2}{C_y}$ by~\eqref{newlinknox}, and $\accsqpr{C_2}{C'_x}$. The situation in option~\ref{def:dia-sat:two} of Def.~\ref{def:dia-sat} is much simpler: there  a single new cluster $\lbc{\{z\}}$ is added to the clusters of~$\sq G$ and $\accsqpr{C}{\{z\}}$ if{f} $\lbc{C} = \lbc{\{z\}}$ or $\accsq{C}{C_y}$. The only non-trivial new case to consider is when $\lbc{C'}=\lbc{\{z\}}$, $\lbc{C_1} \ne \lbc{\{z\}}$, and $\lbc{C_2}\ne \lbc{\{z\}}$, in which case $\accsq{C_1}{C_y}$ and $\accsq{C_2}{C_y}$, and $\lbc{C_1}$ and $\lbc{C_2}$ are $\grel{G'}$-comparable because they are $\grel G$-comparable.
	This completes the proof that 	
	$\sq{G'}$ is tree-clustered and overall stable.

	To prove~\ref{lemma:sat:optiononedone}, since most of the conditions were just shown in \ref{lemma:sat:onestep}, it is sufficient to additionally show that all topmost layers of $\sq{G'}$ are naively happy. Since option~\ref{def:dia-sat:one} adds neither new formulas nor new labels, for most types of formulas their (naive) happiness is inherited from $\sq G$. The only exceptions are $\fmb{\BOX}$- and $\fmw{\DIA}$-formulas. The argument is the same for both. We show it for  $\fmb{\BOX}$-formulas. Suppose $\accsqpr uv$ and $\blackf u{\BOX C}{G'}$. Since no formulas are added in option~\ref{def:dia-sat:one}, also $\blackf u{\BOX C}{G}$. There are two possibilities: either $\accsq uv$ already in $\sq G$ or it was added to ensure the structural saturation of $\sq{G'}$. In the former case, $\blackf v{C}{G'}$ and $\blackf v{\BOX C}{G'}$ by the naive happiness of this layer in~$\sq G$. In the latter case, we have $\accsqpr uv$ but not $\accsq uv$, hence,  by~\eqref{newlinknox}, both $\accsq uy$ and $\accsq xv$. Therefore, $\blackf y{\Box C}{G}$  naive happiness of the layer in~$\sq G$, so $\blackf x{\Box C}{G}$ because  $\lb x\lbeq\lb y$ in $\sq{G}$, and $\blackf v{C}{G}$ and $\blackf v{\BOX C}{G}$, again due to naive happiness. Both~$\fmb{C}$~and~$\fmb{\BOX C}$ remain at $\lb v$ in~$\sq{G'}$. 	%
\end{proof}

\begin{lemma}\label{lemma:satB}
	Let\/ $\sq G$ be a semi-saturated sequent, let\/ $\sq G \diared \sq{G'}$, and let\/ $\weirdS$ be a semi-saturation of $\set{\sq G'}$. Then all happy\/ $\fmb{\DIA}$-formulas are preserved by the transition from\/~$\sq{G}$ to\/ $\weirdS$, i.e.,~for all sequents\/ $\sq{G^*} \in \weirdS$ and for all labels~$\lb u$ occurring in\/~$\sq{G^*}$,   if\/ $\blackf u{\DIA C} G$ is happy, then so is\/ 
	$\blackf u{\DIA C}{G^*}$.
\end{lemma}

\begin{proof}
	Note  that semi-saturation does not remove formulas or labels, nor does it add $R$-relational atoms. Hence, semi-saturation in option~\ref{def:dia-sat:two} cannot spoil happiness of any $\fmb{\DIA}$-formula, and it is sufficient to show that no happy $\fmb{\DIA}$-formula of $\sq{G}$ becomes unhappy in $\sq{G'}$. For option~\ref{def:dia-sat:two}, this is trivial as, again, nothing is removed. For option~\ref{def:dia-sat:one}, the only problems relate to the removal of $\lb y$ and could potentially occur if  $\blackf u{\DIA C}G$ was happy because of $\accsq uy$ and $\blackf yCG$. However, in this case, $\accsqpr ux$ and $\blackf xC{G'}$ because $\lb x\lbeq\lb y$ in $\sq{G}$, thus, happiness persists.
\end{proof}

\DiaSaturationStable*

\begin{proof}
	To prove~\ref{lemma:sat:stable}
	assume that set~$\sqset{S}$ is finite and all sequents in~$ \sqset{S}$ are  semi-saturated (in particular, stable). Observe that  $\sqset S'= \bigl(\sqset S\setminus\set{\sq G}\bigr)\cup \weirdS$ for some $\sq G\diared\sq G'$  where $\weirdS$ is as in Lemma~\ref{lemma:satB}.
        In Option~\ref{def:dia-sat:one}, we replace one semi-saturated sequent $\sq G$ with another one sequent~$\sq{G'}$, which is semi-saturated by~Lemma~\ref{lemma:satA}.\ref{lemma:sat:optiononedone}. Hence, the resulting set is semi-saturated. In option~\ref{def:dia-sat:two}, sequent~$\sq{G'}$ is stable by~Lemma~\ref{lemma:satA}.\ref{lemma:sat:onestep}. By Lemma~\ref{lemma:ssat},  set $\weirdS$ is semi-saturated. Hence, so is $\sqset S'$. 
	
	To prove \ref{lemma:sat:term}, consider a sequence of sets $\sqset{S} \Diared \sqset{S}_1 \Diared \sqset{S}_2 \Diared \dots$. Let us divide all labels occurring in this sequence into \emph{old}, i.e., occurring in $\sqset{S}$, and \emph{new}, i.e., introduced later. We also call unhappy $\fmb{\DIA}$-formulas old or new based on their label. The number of old unhappy $\fmb{\DIA}$-formulas  can never increase, and each  $\Diared$ transition makes at least one $\fmb{\DIA}$-formula (old or new) happy. Hence, it is sufficient to show that it is impossible to continue creating new labels indefinitely. 
	Using tree-clusteredness, we define the \emph{new-cluster-depth} of a new label~$\lb x$ (from a topmost layer of a sequent from one of the sets) to be the length of the shortest chain $\lb{o} \grel{G} \lb{x_1} \grel{G} \dots \grel{G} \lb{x_n} \grel{G} \lb x$ such that $\lb o$~is an old label while labels $\lb{x_0}, \dots, \lb{x_n}$ are new. In particular, the new-cluster-depth of any old label~$\lb o$  is~$0$. To prove termination we consider a Derschowitz--Manna-style ordering on multisets of new-cluster-depths of all new unhappy $\fmb{\DIA}$-formulas from~$\sqset{S}_i$. Let us denote such a multiset of integers by $\unhs{S}{i}$ and pair it with  the number of old unhappy $\fmb{\DIA}$-formulas from~$\sqset{S}_i$. Let the $\sqset{S}_i\Diared\sqset{S}_{i+1}$ be performed  based on an unhappy $\fmb{\DIA A}$ at label~$\lb y$ with new-cluster-depth~$k$. If $k=0$, i.e., $\lb y$~is an old label, then the number of old unhappy $\fmb{\DIA}$-formulas decreases. It is easy to show by induction on $i$ that for new $\lb{y}$, i.e., for $k>0$, label  $\lb y$ forms a singleton cluster~$\set{\lb y}$ any non-trivial $R$-children of $\lb y$ must be $R$-leaves in the respective topmost layer. Thus, since by part~\ref{lemma:satB} all $\fmb{\DIA}$-formulas outside of $\lb y$ preserve happiness, all formulas in~$\lb y$ disappear, and other clusters with unhappy $\fmb{\DIA}$-formulas preserve their new-cluster-depth, multiset $\unhs{S}{i+1} \subsetneq \unhs{S}{i}$, i.e.,  at least one unhappy $\fmb{\DIA}$-formula, $\fmb{\DIA A}$ at label~$\lb y$,  disappeared, no additional unhappy $\fmb{\DIA}$-formulas appeared, and no unhappy $\fmb{\DIA}$-formulas decreased their new-cluster-depth. Alternatively, if  $\sqset{S}'$ is obtained via option~\ref{def:dia-sat:two}, then, again by~\ref{lemma:satB}, no additional unhappy  $\fmb{\DIA}$-formulas appear in labels present in $\sqset{S}_i$ and some unhappy ones may become happy. However, the semi-saturation  can create new unhappy  $\fmb{\DIA}$-formulas in the newly created label $\lb z$, which has cluster-depth $k+1$ (it is easy to see that other labels are unaffected by the semi-saturation). Thus, 
	$\unhs{S}{{i+1}} \subseteq \unhs{S}{i} \setminus\{k\}\cup\{k+1,\dots,k+1\}$ where the multiplicity of added $k+1$ depends on how many unhappy $\fmb{\DIA}$-formulas are in~$z$ after  semi-saturation. To ensure termination, it remains to note that there is a global upper bound on new-cluster-depths of unhappy $\fmb{\DIA}$-formulas. Indeed,  let $M$ be the number of formulas that are subformulas of~$\sqset{S}$. We show that no unhappy  $\fmb{\DIA}$-formula can have new-cluster-depth exceeding $2^M+1$. This follows from the fact that, due to the subformula property, there are at most~$2^M$ distinct sets of formulas assigned to one label. All new labels~$\lb z$ created in option~\ref{def:dia-sat:two} have no past, and $\fmb{\DIA}$-formulas in a child of such~$\lb z$ are not used for saturation  until $\lb z$~is almost happy. Consider two such new labels~$\lb{z_1}$~and~$\lb{z_2}$   such that $\accsq{z_1}{z_2}$. Suppose an unhappy $\fmb{\DIA A}$ in $\lb{z_2}$ was chosen for the next saturation step. Then $\set{\lb{z_2}}$ is a singleton cluster and all its non-trivial parents, including $\lb{z_1}$ are almost happy. If $\lb{z_2} \lbeq\lb{z_1} $, then option~\ref{def:dia-sat:one} would dictate to identify $\lb{z_2}$ with $\lb{z_1}$.  Thus, for $\lb{z_2}$ to remain a separate cluster, it has to be different from all the labels in its strictly preceding clusters of new labels. Thus, for  any sequence of distinct clusters $\lbc{C_1}\grel{G}\dots\grel{G}\lbc{C_k}$ of new labels, there must be labels $\lb{z_1}\grel{G}\dots\grel{G}\lb{z_k}$ such that $\lb{z_j} \not\lbeq\lb{z_l}$ for any $j \ne l$. 
	It means that after at most $k\geq 2^M$ clusters of new labels in a row, the (semi-saturated) new label~$\lb z$ would be equivalent to some of the previously created labels and option~\ref{def:dia-sat:one} would prevent extending the sequence further. Since all new unhappy $\fmb{\DIA}$-formulas are pushed in the direction of  increasing their new-cluster-depth, which cannot exceed the global bound of $2^M+1$ and the number of old unhappy $\fmb{\DIA}$-formulas only decreases, 
	we conclude that $\Diared$ terminates.

	To prove \ref{lemma:sat:sat}, observe that $\sqset{S}$ being saturated by definition means that  it is semi-saturated and all topmost layers of its sequents are almost happy. A layer is almost happy if and only if it is naively happy and  all $\fmb{\DIA A}$-formulas occurring in it are happy.  Saturation~$\Diared$ can be applied to a semi-saturated set if{f} it has an unhappy $\fmb{\DIA A}$-formula (in its topmost layer). Hence, $\sqset{S}$ is saturated if{f} it is semi-saturated and  in normal form w.r.t.~$\Diared$.
\end{proof}

\LiftingSaturationStable*

\begin{proof}
	Assume that $ \sq G $ is saturated. Then it is stable, i.e., structurally saturated, tree-layered, and tree-clustered with happy inner layers,  and, in addition, has  almost happy topmost layers. 
	By definition, $\liftsat{\sq{G}}$ consists of $\sq {G}$ to which we add one topmost layer $ \hL $ for each unhappy formula $ \labels{x}{\fmw F} $ with $ \fm{F} = \fm{A \IMP B} $ or $ \fm F = \fm{\Box A}$ occurring in some not simulated topmost layer $ L $ of $ \sq G $ (these formerly topmost layers become inner in $\liftsat{\sq{G}}$). 
	If  $\lbc{\set{  x}} $ is a singleton cluster,  such a layer~$ \hL $ is a copy of $ L $ (with an additional label~$\lb z$ if $ \fm F = \fm{\Box A}$); as can be seen in  Fig.~\ref{fig:mm-lifting}, if $ \lb x $ belongs to a non-singleton cluster $ \lb{C_x} $, layer $ \hL $ additionally contains a copy of $ \lb{C_x}$ and a label $ \lb \hx $ (and an additional label~$\lb z$ if $ \fm F = \fm{\Box A}$).
	Since each of these added layers is topmost, all conditions can be verified for each of them independently. 
	As mentioned earlier, Construction~\ref{def:lifting} (and Construction~\ref{def:liftingII} for the additional label~$\lb z$) explicitly ensures that $\liftsat{\sq{G}}$ is layered,  that the new layers $ \hL $ of $\liftsat{\sq{G}}$ enjoy the  $(\Rref)$- and $(\Rtr)$-structural saturation, and that $\liftsat{\sq{G}}$ satisfies the $(\Lref)$-, \mbox{$(\Ltr)$-,} $ (\monl)$-, \mbox{$(\fone)$-,} and $(\ftwo)$-structural saturation. 
	The fact that $\liftsat{\sq{G}}$ is tree-layered follows from the fact that $ \sq G $ is tree-layered and, for each added topmost layer~$ \hL $, we have $L' < \hL$ if{f} $L' \le L$. 
	Showing that $ \hL $ is tree-clustered in case $ \lb x $ belongs to a singleton cluster is immediate, as the $R$-structure of $ \hL $ is either isomorphic to that of the tree-clustered $ L $ or is obtained from it by adding one singleton cluster~$\set{\lb z}$ that is a leaf w.r.t.~$R$.  
	In case $ \lb x $ belongs to a non-singleton cluster~$ \lb{C_x}$, it is easy to see from Construction~\ref{def:lifting} and Construction~\ref{def:liftingII} that $\hL$ is partitioned into the following clusters: $\lbc{\hCone} = \{\lb{\hx'_j} \mid \lb{x_j} \in \lbc{C_x}\}$, $\lbc{\hCtwo} = \{\lb{\hx''_j} \mid \lb{x_j} \in \lbc{C_x}\}$, $\lbc{\set{\hx}}$, which we call $\lb x$-clusters, as well as clusters $\lbc\hC \colonequals \{\lb{\hy_j} \mid \lb{y_j} \in \lbc{C}\}$ for each cluster $\lbc C \ne \lbc{C_x}$ of $L$, which we call $\lb y$-clusters and, in case $ \fm F = \fm{\Box A}$, additionally cluster $\lbc{\set{ z}}$ (see Fig.~\ref{fig:mm-lifting}). It is equally easy to see for arbitrary $\lb y$-clusters $\lbc{\hC}$ and $\lbc{\hC^*}$ that: 
	\begin{enumerate}[a)]
		\item $\accsqar{\hC}{\hC^*}$ if{f} $\accsq{C}{C^*}$, 
		\item $\accsqar{\hC}{\hCone}$ if{f} $\accsqar{\hC}{\hCtwo}$ if{f} $\accsqar{\hC}{\set{\hx}}$ if{f} $\accsq{C}{C_x}$ (also if{f} $\accsqar{\hC}{\set{z}}$ in presence of $\lb z$),
		\item $\accsqar{\hCone}{\hC}$ if{f} $\accsqar{\hCtwo}{\hC}$ if{f} $\accsqar{\set{\hx}}{\hC}$ if{f} $\accsq{C_x}{\hC}$ (but never $\accsqar{\set{z}}{\hC}$),
		\item $\accsqar{\hCone}{\hCtwo}$,  $\accsqar{\hCone}{\set{\hx}}$,  $\accsqar{\set{\hx}}{\hCtwo}$ are the only non-reflexive relations among $\lb x$-clusters, 
		\item in presence of $\lb z$: $\accsqar{\hCone}{\set{z}}$, $\accsqar{\set{\hx}}{\set{z}}$, and $\accsqar{\set{z}}{\set{z}}$ are the only remaining relations involving $\lbc{\set{z}}$.
	\end{enumerate}
	Consider any three of these clusters $ \lb{C_1} $, $ \lb{C_2} $,~and~$ \lb{C_3} $ from $ \hL $ such that $ \accsqar{C_1}{C_3} $ and $ \accsqar{C_2}{C_3} $. We need  to  show that $ \lb{C_1} $ and $ \lb{C_2} $ are comparable in $\liftsat{\sq{G}}$. Omitting some trivial cases, assume that the three clusters are pairwise distinct. 
	We distinguish cases depending on whether  these clusters are $\lb x$-clusters, $\lb y$-clusters, etc.
	The case of $\lbc{C_3} = \lbc{\set{z}}$ reduces to $\lbc{C_3} = \lbc{\set{\hx}}$ because $\accsqar{C_i}{\set{z}}$ if{f} $\accsqar{C_i}{\set{\hx}}$ or $\lbc{C_i} = \lbc{\set{z}}$, the latter case being trivial. All other cases involving $\lbc{\set{z}}$ are also trivial. The case when both $\lbc{C_1}$ and $\lbc{C_2}$ are $\lb x$-clusters is trivial because they are all pairwise comparable in $\liftsat{\sq{G}}$. Thus, it is sufficient to consider the cases when none of the clusters is~$\lbc{\set{z}}$ and, w.l.o.g., $\lbc{C_1}$ is a $\lb y$-cluster.
	All  non-$\lbc{\set{z}}$  clusters in $\hL$ have origin clusters in $L$: $\lbc{C}$ for a $\lb y$-cluster~$\lbc{\hC}$ or $\lbc{C_x}$~for each of the $\lb x$ clusters. Let us call the origins of $\lbc{C_1}$, $\lbc{C_2}$, and $\lbc{C_3}$ clusters $\lbc{C^o_1}$, $\lbc{C^o_2}$, and $\lbc{C^o_3}$ respectively. Considering all four possibilities on whether $\lbc{C_2}$ and $\lbc{C_3}$ are $\lb x$-clusters or $\lb y$-clusters,  one can see that $ \accsq{C^o_1}{C^o_3} $ and $ \accsq{C^o_2}{C^o_3} $. Since $\sq G$ is tree-clustered, $\lbc{C^o_1}$ and $\lbc{C^o_2}$ are comparable in $\sq G$. And this implies that $\lbc{C_1}$ and $\lbc{C_2}$ are comparable in $\liftsat{\sq{G}}$ in each of the four cases.
	This concludes the proof that  $\liftsat{\sq{G}}$~is tree-clustered. 
	
	Finally, since $\sq G$ is saturated, all its topmost layers are almost happy, that is, all formulas that are not of the shape $\fmw{A \IMP B}$ or $\fmw{\BOX A}$ are happy. By adding new topmost layers without removing any formulas or any labels we make all of them happy, without making any other formula of $\sq G$ unhappy (here we rely on the local formulation of happiness for $\blackf x{A \IMP B} G$ in Def.~\ref{def:happyformula}). Thus,  all inner layers of~$\liftsat{\sq{G}}$ are happy, and   $\liftsat{\sq{G}}$ is stable. 
\end{proof}

\LoopSaturationStable*

\begin{proof}
	To prove \ref{lemma:loop-sat:sat} assume that $ \sqset{S} $ is saturated. Let us consider a sequent $ \sq G \in \sqset{S}$ such that $ \sq G \loopred \sq G' $.  Then, $ \sq G $ contains either an unhappy R-triangle loop or an unhappy U-triangle loop. Sequent $ \sq G' $ is obtained from~$ \sq G $ by substituting  label $ \lb s $ for all occurrences of label $ \lb t $ and then closing $R$ under transitivity (Definition~\ref{def:loop-sat}).  
	By assumption, $ \sq G $ is stable, i.e., it is structurally saturated, tree-layered, and tree clustered with happy inner layers.
	Irrespective of which kind of triangle loop occurs in $ \sq G $, labels $ \lb s $ and  $ \lb t $ belong to the same topmost layer of the sequent. Hence, neither labels nor labelled formulas from inner layers are affected by the rewrite and,  by Remark~\ref{rem:tobereferred}, their happiness is preserved for most formulas. There are two types of formulas that might be exceptions. We consider them separately. For the first type, if $\whitef k{A \IMP B}{G}$ from some inner layer is happy because $\futsq k t$, $\blackf tAG$, and $\whitef tBG$, then both $\blackf sA{G'}$ and $\whitef sB{G'}$ because $\lb s \lbeq \lb t$. Thus, $\whitef k{A \IMP B}{G'}$ remains happy because $\futsqpr ks$. For the other type, consider the situation where $\whitef k{\BOX A}{G}$ from some inner layer is happy because $\whitef v{A}{G}$, $\futsq k w$, and $\accsq wv$ where one or both of labels $\lb w$ and $\lb v$ are $\lb t$. If only $\lb v = \lb t$, then $\whitef sA{G'}$ because $\lb s \lbeq \lb t$, thus, $\futsqpr kw$ and $\accsqpr ws$ 
	make $\whitef k{\BOX A}{G'}$ happy. If only $\lb w = \lb t$, then $\futsqpr ks$ and $\accsqpr sv$ make $\whitef k{\BOX A}{G'}$ happy. Finally, if both $\lb v = \lb w =\lb t$, then $\whitef sA{G'}$ because $\lb s \lbeq \lb t$ and $\futsqpr ks$ an $\accsqpr ss$ make $\whitef k{\BOX A}{G'}$ happy. In all cases, all formulas in inner layers remain happy.
	The rewrite does not introduce new labels, and the only changes in $ \le $-relational atoms are that $\futs{t}{t}$ turns into $ \futs{s}{s} $ and, for any $ \lb k \ne \lb t $, if $\futsq{k}{t}$, then $ \futsqpr{k}{s} $. The $(\Rtr)$-structural  saturation of $\sq{G'}$ is explicitly enforced, while the $(\Lref)$- and  $(\Rref)$-structural saturation are not affected by the removal of label $\lb t$. The only non-trivial case for $(\Ltr)$-structural saturation is when $\futsqpr wk$ and $\futsqpr ks$ because $\futsq kt$ where $\lb k \ne \lb s$. Then $\lb k$ and $\lb w$ belong to unaffected layers, hence, $\futsq wk$. Thus, $\futsq wt$ by $(\Ltr)$-structural saturation of $\sq G$, and $\futsqpr ws$ by construction. The  $(\monl)$-structural  saturation follows from $\lb s \lbeq \lb t$.
	Finally,  for \mbox{$ (\fone) $-,} and $(\ftwo)$-structural  saturation of~$\sq{G'}$, the most non-trivial case is when, for one of the two, $\accsqpr kw$ and $\futsqpr ks$ because $\futsq kt$  where $\lb k \ne \lb s$. Again $\lb k$ and $\lb w$ belong to unaffected layers, hence, $\accsq kw$. By the same structural saturation property of $\sq G$, there exists a label $\lb z$ such that $\futsq wz$ and $\accsq tz$. If $\lb t \ne \lb z$, then $\accsqpr sz$ and $\futsqpr wz$, so the same label $\lb z$ fulfills the conditions. If $\lb t = \lb z$, then $\futsqpr ws$, which together with $\accsqpr ss$ means that $\lb s$ fulfills the conditions. The remaining case is symmetric.
	This concludes the proof that $ \sq G' $ is structurally saturated. 
	Since the rewrite does not affect the layer structure of the sequent, we immediately conclude that $\sq G'$ is tree-layered. 
	The proof that $ \sq G' $ is tree-clustered proceeds in the same way as the proof of case~\ref{lemma:sat:stable} of Lemma~\ref{lemma:sat}, in the case of option \ref{def:dia-sat:one} of Definition~\ref{def:dia-sat}. 
	This concludes the proof that $ \sq G' $ is stable. 
	By assumption, $ \sq G $ is saturated. This means that all  topmost layers of $ \sq G $ are almost happy, that is, all formulas that are not of the shape $\fmw{A \IMP B}$ or $\fmw{\BOX A}$ are happy. 
	It remains to show that all topmost layers of $ \sq G' $ are almost happy. Almost happiness of labelled propositional formulas of $ \sq G' $ is  local and, hence, is not affected by the new $R$- and $\le$-links. Happiness of $ \fmb{\Box} $- and $ \fmw{\DIA} $-formulas is shown as in the proof of case~\ref{lemma:sat:stable} of Lemma~\ref{lemma:sat}. 
	
	To prove \ref{lemma:loop-sat:term}, observe that, unlike in the case of $ \Diared $ (Definition~\ref{def:dia-sat}), $ \Loopred $ never  introduces new labels in the sequents on which it operates. Whenever $ \loopred $ is applied to a sequent $ \sq G $, the number of labels occurring in one of the topmost layers of $ \sq G $ is reduced by one because $ \lb s \ne \lb t $. 
	Therefore,  for any topmost layer $ L $ in $ \sq G $  containing $ n \geq 1 $ labels,  transformation $ \Loopred $ can be applied   at most $ n-1 $ times to the labels occurring in $ L $. Since saturated set $\sqset S$ is finite, the rewrite $ \Loopred $ is terminating.
\end{proof}

\subsection{Proofs from Section \ref{sec:countermodel}}

\countermodel*

\begin{proof}
%
  
If the algorithm terminates in Step~4, then $\liftsat {\sq G_i}=\sq G_i$ for some non-axiomatic sequent $\sq G_i \in \sqset{S}_i$. It is easy to see that label $\lb r$, the only label in~$\sqset{S}_0$, is never removed by the algorithm because it always remains in the root cluster of the root layer of every sequent in every set. Indeed: semi-saturation does not remove labels; saturation does not remove labels from root clusters; lifting saturation does not remove labels; finally, looping saturation does not remove labels from the root layer. Hence, $\whitef rF{G_i}$.  For each unhappy topmost layer $L$ of $\sq{G_i}$, there is some inner layer $L'$ simulating~$L$  (see Definition~\ref{def:simul-layers}) via a simulation $\simul_L$. Let $\sq G$ be obtained by first adding to $\sq{G_i}$ all relational atoms $\futs x{x'}$  such that $\lb {x'}\simul_L \lb x$ for some unhappy topmost layer $L$ and then closing the result under transitivity of $\le$. It is sufficient to show that $\sq G$ is happy. Then by Theorem~\ref{thm:completeness},  $\nforce{\modelof{G}}rF$. Therefore, by Theorem~\ref{thm:plotkin}, $\fm F$~is not a theorem in $\ISfour$.%

	We need to show that $\sq G$ is structurally saturated and all its formulas are happy. Since no new labels were added, $(\Lref)$- and $(\Rref)$-structural saturation is preserved. $(\Rtr)$-structural saturation is preserved because no new $R$-links were added. $(\Ltr)$-structural saturation is explicitly enforced. To show that $(\monl)$-, $(\fone)$-, and $(\ftwo)$-structural saturation are preserved, it is sufficient to demonstrate them for each of the  $\le$-links added before the transitive closure. For the \mbox{$(\fone)$-,} and $(\ftwo)$-structural saturation these are exactly the simulation conditions $\rn{S1}$ and $\rn{S2}$ from Definition~\ref{def:simul-layers}. For the $(\monl)$-simulation this follows from the fact that $\simul_L \subseteq \lbeq$ for all  $L$, which means that whenever $\blackf{x}AG$ and $\futsq x{x'}$ because $\lb{x'} \simul_L \lb x$ for some $L$, we have $\lb x \lbeq \lb{x'}$ and, hence, $\blackf{x'}AG$. This completes the proof that $\sq G$ is structurally saturated. Since no formulas or links were removed, all formulas happy in $\sq{G_i}$ remain happy in $\sq G$ (here we again rely on the local formulation of happiness for $\blackf x{A \IMP B} G$ in Def.~\ref{def:happyformula}). All formulas from the inner layers of saturated sequent $\sq{G_i}$ were happy, as were all formulas apart from possibly some $\fmw{F}$ of one of the two shapes $\fmw{A \IMP B}$ or $\fmw{\BOX A}$ in topmost layers.   For any such  unhappy $\whitef x{F} {G_i}$ from some topmost layer~$L$ of $\sq{G_i}$, there is an inner layer $L'$ and, by Prop.~\ref{prop:sim_bij}, there is a label $\lb{x'} \in L'$ such that $\lb{x'}\simul_L\lb x$, hence, $\futsq x{x'}$.  At the same time $\lb x \lbeq \lb{x'}$ so $\whitef{x'}{F} {G_i}$  is happy in its inner layer. If $\fmw{F}=\fmw{A\IMP B}$, we must have $\blackf{z}{A}{G_i}$, $\whitef{z}{B}{G_i}$, and $\futsqp{x'}{z}{G_i}$. Then $\futsq x z$ by $(\Ltr)$-structural saturation of $\sq G$, which makes $\whitef{x}{F} {G}$ happy. Finally, if $\fmw{F}=\fmw{\Box A}$, we must have $\whitef{z}{A}{G_i}$, $\futsqp{x'}{u}{G_i}$, and $\accsqp{u}{z}{G_i}$. Then $\futsq x u$ by $(\Ltr)$-structural saturation of $\sq G$ and $\accsq{u}{z}$ by construction, which makes $\whitef{x}{F} {G}$ happy. 
This completes the proof that $\sq G$ is a happy sequent.
\end{proof}

\subsection{Proofs from Section \ref{sec:unfolding}}

\daggeradm*

\begin{proof}
  Whenever the rule of $\labISf$ that corresponds to $\rr$ removes the principal formula, we use first the contraction rule to create a second copy. Then, rules of $\labISfd$ adding formulas to the antecedent are emulated by the respective rule of $\labISf$ followed by $\monl$ to propagate the antecedent formula upward. 

The reflexive relational atoms $\futs xx$ in $\lef{\IMPk}$ and $\lef{\BOXk}$ are present in the conclusion  by the assumption that the conclusion is a proper sequent.  

The composite rule $\slef{\DIAk}$  of $\labISfd$ is emulated by using $\lef{\DIA}$ to create label $\lb \hx$, followed by a sequence of $\ftwo$ rules to create other labels from $\hU$ (it is important to perform these rules in the order compatible with the order on layers, i.e., without skipping intermediate layers), followed by $\Rref$ and $\Lref$ for all these created labels, followed by instances of $\Ltr$ to add all requisite $\leq$-relational atoms  (relying on the $\Ltr$-structural saturation of the conclusion sequent), followed by  instances of $\Rtr$ to add all requisite $R$-relational atoms, followed by instances of admissible $\monl$ to propagate $\fm A$ upward.

The composite rule $\srig{\IMP}$  of $\labISfd$ is emulated by using $\rig{\IMP}$ to create label $\lb \hx$, followed by a sequence of $\fone$ and $\ftwo$ rules to create other labels from $\hL$ (it is important to perform these rules in the order of increasing distance from $\lb x$ in layer~$L$, i.e., without skipping intermediate labels), followed by $\Rref$ and $\Lref$ for all these created labels, followed by instances of $\Rtr$ to add all requisite $R$-relational atoms  (relying on the $\Rtr$-structural saturation in layer $L$), followed by  instances of $\Ltr$ to add all requisite $\leq$-relational atoms, followed by instances of admissible $\monl$ to lift all succedent formulas of $L$ into $\hL$.

The composite rule $\srig{\BOXk}$  of $\labISfd$ is emulated in the same way, only using $\rig{\BOX}$ to create $\lb \hx$ and $\lb \hz$ in place of $\rig{\IMP}$. Note that here instances of $\Rtr$ would additionally add $\accs \hu \hz$ whenever $\accs u x$ was present in the conclusion sequent because $\accs \hu \hx$ is created as in the case of $\rig{\IMPk}$, while $\accs \hx \hz$ is created by rule~$\rig{\BOX}$.
\end{proof}

\properrestrict*
\begin{proof}
  We also prove that for every instance of a unary or binary rule\/ $\rr\notin\set{\srig\IMP, \srig\BOX}$  in~$\deri$, the label of the principal formula   is in a topmost layer (of both the conclusion and the premise(s)).

  The proposition and this statement are
  proved by mutual induction  on the given tidy derivation. It is sufficient to consider each rule  $\rr$ of $\labISfd$ separately. Let $\sq G$ be the conclusion of an instance of~$\rr$ in $\deri$ and $\sq H$ be (one of) its premise(s).
\begin{description}
\item[$\rr \in \set{\lef{\ANDk}, \rig{\ANDk}, \lef{\ORk}, \rig{\ORk}, \lef{\IMPk}, \lef{\BOX}, \rig{\DIA}, \lef{\vax}, \rig{\vax}}$] 
These  rules create neither new labels nor new relational atoms, and have $\forbiddenof H = \forbiddenof G$ by~\eqref{eq:tidyforbid}. Hence, most aspects of properness for $\sq H$ are trivially inherited from $\sq G$, including verticality, lack of non-singleton clusters, tree-clusteredness and tree-layeredness, as well as most aspects of structural completeness, with the exception of $\monl$, which potentially could have been violated by adding formulas to inner layers. However, the tidiness of $\deri$ means that all added formulas have been added to labels that are  allowed, which in restricted sequents can only be  in topmost layers. In other words, whenever rule $\rr$ adds a labelled formula $\labelsb xA$, label $\lb x$ has no non-trivial futures, i.e., there are no labels $\lb y \ne \lb x$ such that $\futsqp xy{H}$. Thus, sequent $\sq H$ is $\monl$-saturated as well, and, hence, proper. Finally, since the label of the principal formula for all these rules is in the same layer as the label(s) to which formulas are added (the same label for  $\lef{\ANDk}$, $\rig{\ANDk}$, $\lef{\ORk}$, $\rig{\ORk}$, and $\lef{\IMPk}$  or a parent label for $\lef{\BOX}$, $\rig{\DIA}$, $\lef{\vax}$,~and~$\rig{\vax}$), the principal formula is also in a topmost layer.
\item[$\rr=\slef{\DIA}$] 
Let $\labelsb x{\DIA A}$ be the principal formula of the rule. Although this rule creates a new label $\lb y$ forming a singleton cluster, it is added to the existing layer $\layerof x$ and has no past in~$\sq H$ (nor non-trivial futures). Thus, tree-layeredness, verticality, and singleton-cluster conditions are all preserved.  Note that by~\eqref{eq:tidyforbid} the new label $\lb y$  must be allowed in $\sq H$, which by definition of restrictedness implies that $\layerof x$ is topmost (in both $\sq G$ and $\sq H$). The relational atoms added by the rule clearly ensure tree-clusteredness, $\Rtr$-, $\Rref$-, and $\Lref$-structural completeness for the newly created label, whereas $\Ltr$-structural completeness is trivial. $\monl$-structural completeness is satisfied since no formulas have been added to labels existing in $\sq G$, whereas the only new $\leq$-relational atom is $\futsqp yyH$ for which $\monl$-structural completeness is trivial. $\fone$- and $\ftwo$-structural completeness for this new atom $\futsqp yyH$ are similarly trivial. As for the new $R$-atoms, they all are of the form $\accsqp zyH$ where both $\lb z$ and $\lb y$ are in the topmost layer $\layerof{x}$ of $\sq H$. Since neither has non-trivial futures, $\fone$- and $\ftwo$-structural completeness is fulfilled.
\item[$\rr\in \set{\srig{\IMP}, \srig{\BOX}}$] 
Let $\labelsw x{A}$ be the principal formula of the rule and $\lb{x'}$ be one of the new labels created by the rule such that $\futsqp x{x'}H$. All the labels created by this rule are in the same layer $\layerof{x'}$ as $\lb{x'}$ so that $\layerof{x} \sle{\sq{H}} \layerof{x'}$ and this layer becomes a new topmost layer. All newly created labels (save, in the case of~$\srig{\BOX}$, for one label) have past. Clearly, tree-layeredness, verticality, $\fone$-, $\ftwo$-, $\monl$-, $\Lref$- and $\Ltr$-structural completeness for the newly created labels are all ensured.  The tree-clusteredness for the new layer, as well as, $\Rtr$- and $\Rref$-structural completeness for the new layer $\layerof{x'}$  is inherited from those for $\layerof x$ (modulo an additional label $\lb y$ created by $\srig{\BOX}$ in a way that ensures all these properties). Note that, since the rule only touches newly created labels, which are created in a new, topmost layer, it is not necessary that $\lb x$ is in a topmost layer. \qedhere
\end{description}
\end{proof}

In order to proof Lemma~\ref{lem:semi-sat-sound}, we need another definition and an auxiliary lemma.

\begin{definition}
Let a derivation~$\der_n$ be an $n$-unfolding of a stable set~$\sqset S$, with premises $\sq\hG_1,\ldots, \sq\hG_k$ being $n$-unfoldings of a specific sequent~$\sq G \in \sqset S$. Let $\unfold_i \subseteq \labelsof{G} \times \labelsof{\hG_i}$ be $n$-unfolding relations witnessing that $\sq \hG_i$~is an $n$-unfolding of~$\sq G$ for each $i=1..k$.
We say that $\deri_n$~is \defn{allowed} for a label~$\lb x \in \labelsof{G}$ if{f} for every $i=1..k$, every label~$\lb\hx\in\labelsof{\hG_i}$ such that $\lb x\unfold_i\lb\hx$ is allowed in $\sq\hG_i$.
\end{definition}

\begin{lemma}
  \label{lem:semi-sat-sound-aux}
  Assume that\/ $\sqset S\ssatred\sqset S'$, that\/ $\sq G\in\sqset S$ is the sequent  replaced in the rewriting step, and that $\labels xC$~is the unhappy formula that is expanded according to Def.~\ref{def:sat-sqset}. If\/  $\sqset S$ is \hbox{$\fm F$-unfoldable} such that for every $N\in \Nat$ there is an $N$-unfolding~$\Deri_N$ of\/~$\sqset S$ that is allowed for~$\lb x$, then\/  $\sqset S'$~is also \hbox{$\fm F$-unfoldable}.
\end{lemma}

\begin{proof}
  We need to show that  $\sqset S'$ is $\fm F$-unfoldable, i.e., for every $n$ there is a tidy derivation~$\der'_n$ of $\sq G_0(\fm F)$ such that each premise of $\der'_n$ is an $n$-unfolding of some sequent from~$\sqset S'$.

  It is sufficient to consider each way that $\sqset S'$ can be obtained from~$\sqset{S}$ by  steps  listed in Def.~\ref{def:sat-sqset}.
 For each case of Def.~\ref{def:sat-sqset}, we pick some $N\geq n$ and extend $\der_N$ to $\der'_n$.
 The only sequent from $\sqset S$ that is not present in~$\sqset S'$ is $\sq G$, which is replaced in~$\sqset S'$ either with sequent~$\sq G'$ or with two sequents~$\sq G'$~and~$\sq G''$, depending on the step.  Hence, the only premises of $\der_N$ that have to be extended are the $N$-unfoldings of~$\sq G$. All the other premises are $N$-unfoldings and, hence, $n$-unfoldings of  some (stable) sequent from $\sqset{S}$ that is still present in $\sqset S'$.

  Since $\der_N$~is allowed for~$\lb x$, for each premise~$\sq \hG$ of~$\der_N$ that is an $N$-unfolding of $\sq G$, there is an $N$-unfolding relation~$\unfold\subseteq\labelsof G\times\labelsof\hG$ such that $\lb\hx \notin \forbiddenof{\hG}$ whenever $\lb x \unfold \lb \hx$.
  We will break up the steps of Def.~\ref{def:sat-sqset} into substeps that only add one or two formulas to one particular label $\lb y$ with $\accsq xy$.

  The only unfolding condition from Def.~\ref{def:unfolding}  that prevents~$\unfold$ from making $\sq\hG$ an $N$-unfolding   of $\sq G'$ (and of~$\sq G''$)  is ~\ref{unf:formula}, and only with regard to labels $\lb \hy$ such that $\lb y\unfold\lb \hy$ because $\lb y$~is the only  label of~$\sq G$ with formulas added. Additionally, when $\unfold$ is modified, \ref{unf:singleton}--\ref{unf:nonsingleton} need to checked. Every time we extend the derivation $\der_N$, the set of forbidden labels for the premise of each added rule must be the same as for the rule's conclusion to preserve tidiness. \looseness=-1
 	\begin{enumerate}
	\item 
	  Suppose $\blackf x{A\AND B} G$ and the missing $\labelsb xA$ and $\labelsb xB$ were added to obtain $\sq G'$. We use $N=n$, set $\unfold' \colonequals \unfold$, which trivially ensures \ref{unf:singleton}--\ref{unf:nonsingleton}.
          Using the fact that $\lb x \unfold \lb\hx$ implies that $\lb \hx$~is allowed and that $\blackf \hx{A \AND B}{\sq \hG}$  by~\ref{unf:formula} for~$\sq G$, to obtain $\sq \hG'$ we  extend premise $\sq\hG$ of~$\der_n$   by rules \looseness=-1
		\[
			\vlderivation{\vlin{\lef{\ANDk}}{}{\B, \Left, \labels \hx{A \AND B} \SEQ \Right}{\vlhy{\B, \Left, \labels \hx{A \AND B}, \labels \hx A, \labels \hx B \SEQ \Right }}}
			\]
for each label $\lb{\hx}$ such that  $\lb x \unfold \lb\hx$.  
It is easy to see that $\unfold$~makes $\sq \hG'$ and $n$-unfolding of~$\sq G'$ because  $\lbeql{\hx}{\sq\hG'}{x}{\sq G'}$ whenever $\lb x \unfold \lb\hx$. 
	\item 
		The case when  $\whitef x{A\OR B} G$ and $\labels x{\rig{A}}$ and $\labels x{\rig{B}}$ were added  is analogous, with $\rig{\ORk}$ used instead of $\lef{\ANDk}$.
	\item 
		Suppose $\blackf x{\BOX A} G$. We further break this step into substeps consisting of adding the missing~$\labelsb yA$~and~$\labelsb y{\BOX A}$ for only one label $\lb y$ such that $\accsq{x}{y}$. 
		Note that 
		$\accsqp uu{\hG}$ 
		 for each label $\lb u \in \labelsof{\hG}$ due to $\sq \hG$ being proper. 
		The sequence of rules 
		depends on the  position of~$\lb x$ relative to $\lb y$:
		\begin{enumerate}
 		\item 
			If $\lb x = \lb y$, use $N=n$ and set $\unfold'\colonequals \unfold$. Again, the only problem to be fixed is~\ref{unf:formula}. Each~$\lb \hx$ such that $\lb x \unfold \lb \hx$ is allowed and, by \ref{unf:formula} for~$\sq G$, we have $\blackf \hx{\BOX A} \hG$. To obtain $\sq \hG'$, we extend premise $\sq\hG$ of $\der_n$   by rules
					\[
			\vlderivation{
			\vlin{\lef{\BOXk}}{}{\B,  
			\accs \hx\hx, \Left, \labels \hx{\BOX A} \SEQ\Right}
			{\vlhy{\B,  
			\accs \hx \hx, \Left, \labels \hx{\BOX A}, \labels \hx A \SEQ \Right }}}
			\]
			for each $\lb \hx$ with $\lb x \unfold \lb \hx$.	Thus, $\lbeql{\hx}{\sq\hG'}{x}{\sq G'}$ whenever $\lb x \unfold \lb\hx$.	\looseness=-1
		\item 
			If $\lb x$  and $\lb y$ are not in the same cluster, again use $N=n$ and set $\unfold'\colonequals \unfold$. By~\ref{unf:singleton}--\ref{unf:nonsingleton}, there exists~$\lb{\hx}$ such that $\lb x \unfold \lb{\hx}$.  Since $\der_n$ is allowed for $\lb x$, label~$\lb\hx$ is allowed in $\sq \hG$. By \ref{unf:R}, $\accshat{\hx}{\hy}$  for each~$\lb{\hy}$ such that $\lb y \unfold\lb\hy$. By Def.~\ref{def:forbid}\ref{forbid:parent}, all these $\lb \hy$ are also allowed in~$\sq \hG$. By \ref{unf:formula} for $\sq G$, we have  $\blackf \hx{\BOX A} \hG$.  Again \ref{unf:formula} is the only problem to be fixed.  To obtain~$\sq \hG'$, we extend premise $\sq\hG$ of $\der_n$   by rules		
			\[
			\vlderivation{\vlin{\lef{\BOXk},\lef{\vax}}{}{\B, 
			\accs{\hx}{\hy},  \Left, \labels \hx{\BOX A} \SEQ \Right}
			{\vlhy{\B, 
			\accs{\hx}{\hy}, \Left, \labels \hx{\BOX A}, \labels \hy {\BOX A},\labels \hy A \SEQ \Right }}}
			\]
			for each label $\lb{\hy}$ such that $\lb y \unfold \lb\hy$.   This ensures that $\lbeql{\hy}{\sq\hG'}{y}{\sq G'}$ whenever $\lb y \unfold \lb\hy$.
		\item
			If $\lb x\ne \lb y$  are in the same cluster $\lb{C}$, it is not a singleton cluster. Let $\sizeof{\lbc C}=k\geq 2$. Use $N=n\mathord+1$. 
			By \ref{unf:nonsingleton} for $\sq G$, sequent $\sq \hG$ contains $n\mathord+1$ unfolded copies ${\hC_1},\dots,{\hC_{n\mathord+1}}$ of all labels from cluster $\lbc C$ (the unfoldings are not themselves clusters as $\sq\hG$ is a proper sequent). 
			In particular, there is $\lb{\hx_1} \in {\hC_1} \subseteq \labelsof{\sq\hG}$ such that $\lb x \unfold\lb{\hx_1}$ and there are $\lb{\hy_j} \in \hC_j \subseteq \labelsof{\sq\hG}$ for $j=2..n\mathord+1$ such that  $\lb{y} \unfold\lb{\hy_j}$ for  $j=2..n\mathord+1$. 
Since $\der_{n+1}$ is allowed for $\lb x$, label $\lb{\hx_1}$ is allowed in $\sq \hG$. 
			By~\ref{unf:formula} for~$\sq G$, we have  $\blackf {\hx_1}{\BOX A} \hG$. Since the proper sequent~$\sq\hG$ must be $\Rtr$-saturated, it follows from \ref{unf:nonsingleton} for $\sq G$ that $\accshat{\hx_1}{\hy_j}$ for  $j=2..n\mathord+1$. Hence, by Def.~\ref{def:forbid}\ref{forbid:parent}, all $\lb{\hy_j}$ for $j=2..n+1$ are also allowed. Here, we restore~\ref{unf:formula} only for $n$~labels $\lb{\hy_2},\dots,\lb{\hy_{n\mathord+1}}$ unfolding~$\lb y$. To obtain~$\sq \hG'$, we extend premise $\sq\hG$ of $\der_{n+1}$   by rules	 	
					\[
			\vlderivation{\vlin{\lef{\BOXk},\lef{\vax}}{}{\B, 
			\accs{\hx_1}{\hy_j},  \Left, \labels{\hx_1}{\BOX A} \SEQ \Right}
			{\vlhy{\B, 
			\accs{\hx_1}{\hy_j}, \Left, \labels{\hx_1}{\BOX A}, \labels{\hy_j}{\BOX A},\labels{\hy_j}A \SEQ \Right }}}
			\]
for each $j=2..n\mathord+1$. Since $\accshat{\hx_1}{\hy_1}$ need not generally be the case, for label~$\lb{\hy_1}$, as well as for all other unfoldings of $\lb y$, \ref{unf:formula} cannot be guaranteed, necessitating their removal from the unfolding relation, i.e., 
	\[
	\unfold' \colonequals \unfold \setminus \Bigl\{(\lb y,\lb \hu) \mid \lb \hu \notin\{\lb{\hy_2},\dots, \lb{\hy_{n\mathord+1}}\}\Bigr\}.
	\]
	For each singleton cluster of~$\sq G'$, \ref{unf:singleton} for this $\unfold'$ follows from \ref{unf:singleton} for the same cluster  of $\sq G$ for $\unfold$. All non-singleton clusters of $\sq G'$  other than $\lbc C$ have $n\mathord+1$ unfolded copies by \ref{unf:nonsingleton} for the $(n\mathord+1)$-unfolding of~$\sq G$, and any $n$ of them satisfy \ref{unf:nonsingleton} requisite for an $n$-unfolding of $\sq G'$. Finally,  for cluster $\lbc C$ of $\sq G'$, label sets $\hC_2,\dots,\hC_{n\mathord+1}$ serve as the requisite $n$ unfolded copies, in particular, 		$\lbeql{\hy_j}{\sq\hG'}{y}{\sq G'}$ for $j=2..n\mathord+1$.
	\end{enumerate}

		\item The case when $\whitef x{\DIA A} G$ and  $\labelsw y{{A}}$ and $\labelsw y{{\DIA A}}$ were added for some label $\lb y$ such that $\accsq{x}{y}$ is analogous, with $\rig{\DIA}$ and $\rig{\vax}$ used instead of $\lef{\BOXk}$ and $\lef{\vax}$ respectively. 
		\item Suppose $\blackf x{A \OR B} G$ and $\labelsb xA$ was added to obtain~$\sq G'$ and $\labelsb xB$ was added to obtain $\sq G''$. The sequence of rules depends on the size of the cluster that $\lb x$ belongs to in $\sq G$:
		\begin{enumerate}
			\item If $\lbc{\{ x\}}$ is a singleton cluster, use $N=n$ and set \mbox{$\unfold'\colonequals \unfold$}.  By \ref{unf:singleton} for $\sq G$, there is a unique  $\lb \hx$ such that $\lb x \unfold \lb \hx$ and $\lb \hx$ is allowed in $\sq \hG$. By \ref{unf:formula} for $\sq G$, we have $\blackf \hx{A \OR B} \hG$ for this~$\lb \hx$. Again \ref{unf:formula} is the only problem to be fixed, which for this rule can be done in two ways: by  either adding $\lef A$ towards an $n$-unfolding of $\sq G'$ or adding~$\lef B$ towards an $n$-unfolding of $\sq G''$. We extend premise~$\sq\hG$ of $\der_n$   by one rule 
			\[
			\scalebox{0.9}{
				$
				\vlderivation{\vliin{\lef{\ORk}}{}{\B, \Left, \labels \hx{A \OR B} \SEQ \Right}{\vlhy{\B, \Left, \labels \hx{A \OR B}, \labels \hx{A} \SEQ \Right}}{\vlhy{\B, \Left, \labels \hx{A \OR B}, \labels \hx{B} \SEQ \Right}}}
				$
			}
			\]
for the unique label $\lb \hx$. The resulting left premise~$\sq \hG'$ and right premise~$\sq \hG''$ can easily be seen to be $n$-unfoldings of $\sq G'$ and  $\sq G''$ respectively, in particular, 
\[
\lbeql{\hx}{\sq\hG'}{x}{\sq G'} \qquad \text{and} \qquad \lbeql{\hx}{\sq\hG''}{x}{\sq G''}.
\]
			
			\item If $\lb x \in \lbc C$ for some cluster $\lbc C$ with $\sizeof{\lbc C} = k\geq 2$, use $N=2n\mathord-1$.  By \ref{unf:nonsingleton} for $\sq G$, sequent $\sq \hG$ contains $2n\mathord-1$ unfolded copies ${\hC_1},\dots,{\hC_{2n\mathord-1}}$ of all labels from cluster~$\lbc C$. In particular, there are $\lb{\hx_j} \in \hC_j \subseteq \labelsof{\hG}$ for $j=1..2n\mathord-1$ such that  $\lb{x}  \unfold\lb{\hx_j}$ for  $j=1..2n\mathord-1$. 
			Since $\der_{2n\mathord-1}$ is allowed for $\lb x$, all labels $\lb{\hx_j}$ for $j=1..2n\mathord-1$ are allowed in $\sq \hG$. 
			By~\ref{unf:formula} for $\sq G$, we have $\blackf {\hx_j}{A \OR B} \hG$ for  $j=1..2n\mathord-1$. We will not be able to fix~\ref{unf:formula} for all $2n\mathord-1$ labels~$\lb{\hx_j}$, but only for (at least) $n$~of them.
			We set $\sqset{T}_0 \colonequals \{\sq\hG\}$ to be the set of $2^0=1$ premises of the current derivation. Note that $\sq \hG$~is the special case (with $l=0$) of sequents of the form
			\begin{equation}
			\label{eq:disj_nonsingle_form}
			\B, \Left, \labels{\hx_1}{D_1},\dots,\labels{\hx_l}{D_l} \SEQ \Right
			\end{equation}
			where $\fm{D_i} \in \{\fm A, \fm B\}$ for each $i=1..l$  and $\labels{\hx_j}{A \OR B} \in \Left$ for all $j=1..2n\mathord-1$. Starting with $k=0$ and repeating the process until  $k=2n\mathord-2$, we extend each  of the $2^k$ derivation premises of form~\eqref{eq:disj_nonsingle_form} for $l=k$ in $\sqset{T}_k$ as follows
						\[
			\scalebox{0.62}{
				$
				\vlderivation{\vliin{\lef{\ORk}}{}{\B, \Left, \labels{\hx_1}{D_1},\dots,\labels{\hx_k}{D_k} \SEQ \Right}{\vlhy{\B, \Left, , \labels{\hx_1}{D_1},\dots,\labels{\hx_k}{D_k},\labels{\hx_{k+1}}{A} \SEQ \Right}}{\vlhy{\B, \Left, , \labels{\hx_1}{D_1},\dots,\labels{\hx_k}{D_k}, \labels{\hx_{k+1}}{B} \SEQ \Right}}}
				$
			}
			\]
thus, obtaining $2^{k\mathord+1}$  derivation premises of form~\eqref{eq:disj_nonsingle_form} for $l=k\mathord+1$ that form the set $\sqset{T}_{k\mathord+1}$. In the end, one premise $\sq \hG$ spawns the  set $\sqset{T}_{2n\mathord-1}$ of $2^{2n\mathord-1}$ premises of form~\eqref{eq:disj_nonsingle_form} for $l=2n\mathord-1$. It remains to show how, for each of the premises from $\sqset{T}_{2n\mathord-1}$, to restrict relation $\unfold$ in a way that produces an $n$-unfolding of either~$\sq G'$~or~$\sq G''$. Consider any $\sq H \in \sqset{T}_{2n\mathord-1}$. It has the form
			\[
			\B, \Left, \labels{\hx_1}{D_1},\dots,\labels{\hx_{2n\mathord-1}}{D_{2n\mathord-1}} \SEQ \Right
			\]
Since all $2n\mathord-1$ formulas $\fm{D_j}$ are repetitions of only two distinct formulas $\fm A$ and $\fm B$, by the pigeonhole principle, one of these two formulas must be repeated at least $n$~times. In other words
			 there is $\fm{E}\in\{\fm A, \fm B\}$ such that there are $n$ indices $1 \leq i_1 < \dots < i_n \leq 2n\mathord-1$ with $\fm{D_{i_j}}=\fm E$ for all $j=1..n$. In this case, for this premise $\sq H$, we set
				\[
	\unfold_{\sq H} \colonequals \unfold \setminus \Bigl\{(\lb x,\lb \hu) \mid \lb \hu \notin\{\lb{\hx_{i_1}},\dots, \lb{\hx_{i_n}}\}\Bigr\}.
	\]
	Let $\sq G^{\sq H}$ denote $\sq G'$ if $\fm E = \fm A$ or $\sq G''$ if $\fm E = \fm B$.
		For each singleton cluster of~$\sq G^{\sq H}$, \ref{unf:singleton} for~$\unfold_{\sq H}$ follows from \ref{unf:singleton} for the same cluster of $\sq G$ for $\unfold$. All non-singleton clusters of $\sq G^{\sq H}$  other than $\lbc C$ have $2n\mathord-1$ unfolded copies by \ref{unf:nonsingleton} for the $(2n\mathord-1)$-unfolding of~$\sq G$, and any $n$ of them satisfy \ref{unf:nonsingleton} requisite for an $n$-unfolding of $\sq G^{\sq H}$. Finally, for cluster $\lbc C$ of $\sq G^{\sq H}$, label sets $\hC_{i_1},\dots,\hC_{i_n}$ serve as the requisite $n$ unfolded copies of $\lbc C$ in $\sq G^{\sq H}$, in particular, $\lbeql{\hx_{i_j}}{\sq H}{x}{\sq G^{\sq H}}$ for $j = 1..n$.  Thus, all $2^{2n\mathord-1}$ premises spawned by $\sq \hG$ are $n$-unfoldings of either $\sq G'$ or $\sq G''$.

		\end{enumerate}
		\item The case when $\whitef x{A \AND B} G$ and $\labelsw x{A}$ was added to obtain $\sq{G'}$ and $\labelsw x{B}$ was added to obtain $\sq{G''}$
		 is analogous, with $\rig{\ANDk}$ used instead of~$\lef{\ORk}$.
		\item The case when $\blackf x{A \IMP B} G$ and $\labelsw xA $ was added to obtain $\sq{G'}$ and $\labelsb xB $ was added to obtain $\sq{G''}$ is also analogous, with $\lef{\IMPk}$ used instead of $\lef{\ORk}$. 
	\end{enumerate}
	For each of the steps, we have extended the given tidy derivation $\der_N$ to a derivation $\der'_n$ in such a way that no labels, forbidden labels, or relational atoms were modified. Since all added rules touch only allowed labels, the resulting derivation is tidy. Thus, derivation $\der'_n$ of $\sq G_0(\fm F)$ is an $n$-unfolding of $\sqset S'$. Moreover, whenever $\der_N$ was allowed for some label $\lb z$, so is $\der'_n$ because no label has changed its forbidden/allowed status in the premises. 
\end{proof}

\begin{lemma}[restate = SoundSemiSat, name = ] 
	\label{lem:semi-sat-sound}
  \begin{enumerate}[(i)]
  \item  If\/ $\sqset S'_0\tssatred\sqset T$, then\/ $\sqset T$~is $\fm F$-unfoldable.
  \item
  For each~$i$, if\/ $\sqset S_i$~is \hbox{$\fm F$-unfoldable} and\/ $\sqset S'_{i+1}\tssatred\sqset T$ for some  semi-saturated\/~$\sqset T$, then\/ $\sqset T$~is $\fm F$-unfoldable.
  \end{enumerate}
\end{lemma}

\begin{proof}
\leavevmode
\begin{enumerate}[(i)]
\item\label{semicaseone}
 If $\sqset S'_0\tssatred\sqset T$, then  $\sqset S'_0=\sqset T_0\ssatred\sqset T_1\ssatred \cdots\ssatred\sqset  T_k=\sqset T$ for stable sets $\sqset T_0$, \ldots, $\sqset  T_k$ for some $k\ge 0$. Note all sequents $\sq G$  appearing in these sets are proper with  $\labelsof{G} = \set{\lb r}$   because semi-saturation steps neither add labels to the only sequent $\sq G_0(\fm F)$ of $\sqset T_0$ with the single label $\lb r$ nor create clusters. 
  
  To show the base of induction, i.e., that $\sqset T_0 = \set{\sq G_0(\fm F)}$ is $\fm F$-unfoldable, it is sufficient to consider a tidy derivation~$\der^0$ of the restricted version of~$\sq G_0(\fm F)$ where label~$\lb r$ is allowed and no rules are applied (note that the only label $\lb r$ in each of the sequents forms the only layer, which is, therefore, topmost and not inner). As discussed earlier, this one derivation is an $n$-unfolding of $\sqset T_0$ for each $n \in \Nat$ via the binary relation $\unfold \colonequals \set{(\lb r, \lb r)}$ that is an $n$-unfolding relation for each $n \in \Nat$. Note that $\der^0$ is clearly allowed for $\lb r$.
  
  Step of induction. Let derivations $\der^i_n$ of~$\sq G_0(\fm F)$ for $n \in \Nat$  be $n$-unfoldings of $\sqset T_i$ that are allowed for $\lb r$. For the semi-saturation step $\sqset T_i\ssatred\sqset  T_{i\mathord+1}$  we can apply  Lemma~\ref{lem:semi-sat-sound-aux} for $\lb x = \lb r$ to obtain derivations $\der^{i\mathord+1}_n$ of~$\sq G_0(\fm F)$ for $n \in \Nat$ that are $n$-unfoldings of $\sqset T_{i\mathord+1}$ and are allowed for $\lb r$.\footnote{Since all sequents involved in sets $\sqset T_i$ are proper, a closer observation of Lemma~\ref{lem:semi-sat-sound-aux} shows  that one derivation can serve as $n$-unfoldings of $\sqset T_i$ for all $n \in \Nat$.}
\item\label{semicasetwo}
 Recall that according to step~5) of the algorithm, Def.~\ref{def:lsat}, and Constr.~\ref{def:liftingIII}, $\sqset S_{i+1}' = \bigl(\sqset S_i\setminus\set{\sq G}\bigr)\cup\set{\liftsat{\sq G}}$ for some fully saturated non-axiomatic $\sq G\in\sqset S_i$, where 
 \[
 \liftsat{\sq G}=\sq G+\lliftf G{x_1}{F_1}+\cdots+\lliftf G{x_h}{F_h}
 \] 
 and $\labelsw{x_1}{F_1},\ldots,\labelsw{x_h}{F_h}$ are all the  unhappy formulas in some of the topmost layers of~$\sq G$ and each $\fmw{F_j}$ is either a $\IMP$- or a $\BOX$-formula.
  Also note that since $\sqset S_i$ is semi-saturated, we only need to semi-saturate subset $\set{\liftsat{\sq G}} \subseteq \sqset S_{i+1}'$ to obtain $\sqset T$.

It might seem natural to split~\ref{semicasetwo} into several finer-grained statements stating that unfolding can be preserved by smaller steps, including many of the steps already covered by Lemma~\ref{lem:semi-sat-sound-aux} and adding to them new steps corresponding to  Construction~\ref{def:liftingII} that create one new layer based on one  of $\labelsw{x_j}{F_j}$. Unfortunately, these intermediate stages cannot be formulated as unfoldings due to the sequent ceasing to be stable because as soon as the first of $\labelsw{x_j}{F_j}$ in a layer is made happy, this layer turns into an inner one, so the sequent does not become stable again until all remaining $\labelsw{x_j}{F_j}$ in this layer are  happy.

At the same time,  applying all requisite instances of $\srig\IMP$ and $\srig\BOX$ to obtain an $n$-unfolding of $\sqset S_{i+1}'$ and then proceeding as in Case~\ref{semicaseone}, repeatedly applying Lemma~\ref{lem:semi-sat-sound-aux} is met with another technical problem. It would  violate the tidiness conditions (see Def.~\ref{def:tidy}): each $\srig\IMP$- and $\srig\BOX$-instance creates a new layer and renders the layers created by the previous $\srig\IMP$- or $\srig\BOX$-instances forbidden, so that the rules needed to unfold the semi-saturation steps for those layers cannot be applied anymore. For this reason, we have to apply the necessary rules for unfolding the semi-saturation after each $\srig\IMP$- or $\srig\BOX$-instance.

Thus, we will still follow individual smaller steps in a specific order maintaining conditions \ref{unf:formula}--\ref{unf:inject} for all $n \in \Nat$ after each, but are forced to lump them together into~\ref{semicasetwo}, at the end of which the result can once again be called an unfolding. 

  Let $L_1,\ldots,L_h$ be the layers of $\liftsat{\sq G}$ that are introduced by $\lliftf G{x_1}{F_1},\ldots,\lliftf G{x_h}{F_h}$ respectively.
  Since $\sqset T$ is semi-saturated and each rewriting step of $\ssatred$ is local to a layer, we can therefore without loss of generality assume that the rewrite sequence   $\sqset S'_{i+1}\tssatred\sqset T$ is adding formulas layerwise, i.e., all formulas that are added to layer $L_j$ are added before adding formulas to layer $L_{j'}$, whenever $j<j'$.

For each step semi-saturating the next layer $L_j$ we preserve \ref{unf:formula}--\ref{unf:inject} for all $n \in \Nat$ the same way as in Case~\ref{semicaseone} using the fact that $L_j$ is created of  allowed labels that are not made forbidden while unfolding  semi-saturation steps. However, formally we cannot rely on Lemma~\ref{lem:semi-sat-sound-aux} because of the lack of stability mentioned before. Instead, we simply repeat the arguments from the proof of Lemma~\ref{lem:semi-sat-sound-aux} noting that the only part of stability used in its proof was $\Rtr$-structural completeness, which does hold upon creating layer $L_j$.

The only remaining steps to cover are the creation of layers $L_j$. This is done by simply applying the $\srig\IMP$- or $\srig\BOX$-rule to $\labelsw{x_j}{F_j}$. However, in order to preserve the  \ref{unf:singleton} and~\ref{unf:nonsingleton} conditions, we have to start with an $(2n+1)$-unfolding of $\sqset S_i$ in order to obtain an $n$-unfolding of $\sqset T$. For every lifted cluster that does not contain $\lb{x_j}$ the additional $n+1$ copies in the unfolding are discarded. But when $\lb{x_j}$ is in a cluster, then $L_j$ contains 2 copies of that cluster (see Figure~\ref{fig:mm-lifting}), whose $n$ repetitions are given to us by the copies $1..n$ and $n+1..2n+1$ of our lifting. In the $n$th copy, only the lifting of $\lb{x_j}$ is part of the unfolding relation.

It is easy to check that at the end of this process all the conditions of $\fm F$-unfoldability are restored.
%
\qedhere
  \end{enumerate}
\end{proof}

The next step is to show that $\sdiared$ preserves the property of being unfoldable. For this, we need to repeat parts of the proof. 
To simplify the process of repeating, we use the notion of embedding.

\begin{definition}[Embedding]
  Let $\sq G$~and~$\sq H$ be restricted proper sequents. An \defn{embedding} $\emb\colon\sq G\to\sq H$ is an injective function $\emb\colon\labelsof G\to\labelsof H$ obeying the following conditions for all $\lb x,\lb y\in\labelsof G$:
  \begin{enumerate}[($\emb$1),leftmargin=2.2em]
  \item\label{emb:formula} 
  	$\lb{x}\lbeq{\emb(\lb x)}$;
  \item\label{emb:R}  
  	$\accsq xy$ if{f} $\embof x\grel H\embof y$;
  \item\label{emb:layer}  
  	$\layerof x\sle G\layerof y$ if{f} $\layerof{\embof x}\sle H\layerof{\embof y}$.
  \item\label{emb:nopast}  
  	$\lb x$~has no past in~$\sq G$  if{f} $\emb(\lb x)$~has no past in~$\sq H$;
  \item\label{emb:forbid}  
  	$\lb x\in\forbiddenof G$ if{f} $\emb(\lb x)\in \forbiddenof H$;
  \end{enumerate}
\end{definition}

\begin{proposition}
	\label{prop:embed}
        Embeddings are closed under composition and preserve unfoldings.
\end{proposition}

\begin{proof}
First, we prove that a composition of two embeddings is an embedding. Let $\sq G$, $\sq H$, and $\sq J$ be restricted proper sequents. Let  $\emb \colon \sq G \to \sq H$ and $\emb' \colon \sq H \to \sq J$ be  embeddings from $\sq G$ to $\sq H$ and from $\sq H$ to $\sq J$ respectively. We need to show that $\emb' \emb$ is an embedding from $\sq G$ to $\sq J$, where $(e'e)(\lb x) \colonequals \emb'\bigl(\emb(\lb x)\bigr)$ for any $\lb x \in \labelsofx{\sq G}$. Since $\emb\colon \labelsofx{\sq G} \to \labelsofx{\sq H}$ and $\emb' \colon \labelsofx{\sq H} \to \labelsofx{\sq J}$ are  injective functions, it is immediate that $\emb'\emb \colon \labelsofx{\sq G} \to \labelsofx{\sq J}$  is an injective function. We have for all  $\lb x, \lb y \in \labelsofx{\sq G}$:
\begin{itemize}
\item[\ref{emb:formula}]
	${}_{\sq G}\lb x \lbeq {}_{\sq H}\emb(\lb x) \lbeq {}_{\sq J}\emb'\bigl(\emb(\lb x)\bigr)$.
\item[\ref{emb:R}]
	$\accsq xy$ 	
	\quad if{f}\quad 
	$\accsqp {\emb(x)}{\emb(y)}H$ 	
	\quad if{f}\quad 
	$\accsqp {\emb'\bigl(\emb(x)\bigr)}{\emb'\bigl(\emb(y)\bigr)}J$.
\item[\ref{emb:layer}]
	$\layerof x \sle G \layerof y$ 	
	\,\,if{f}\,\,
	$\layerof {\emb(x)} \sle H \layerof {\emb(y)}$ 	
	\,\,if{f}\,\,
	$\layerof {\emb'(\emb(x))} \sle J \layerof {\emb'(\emb(y))}$.	
\item[\ref{emb:nopast}]
	$\lb x$ has no past in $\sq G$ 
	\quad if{f}\quad
	$\emb(\lb x)$ has no past in $\sq H$ 
	\quad if{f}
	$\emb'\bigl(\emb(\lb x)\bigr)$ has no past in $\sq J$.	
\item[\ref{emb:forbid}]
	$\lb x \in \forbiddenofx{\sq G}$ 
	\quad if{f}\quad 
	$\emb(\lb x) \in \forbiddenofx{\sq H}$
	\quad if{f}\quad
	$\emb'\bigl(\emb(\lb x)\bigr) \in \forbiddenofx{\sq J}$.
\end{itemize}

Secondly, let $\unfold \subseteq \labelsof{\sq G} \times \labelsof{\sq \sq \hG}$ be an $n$-unfolding relation witnessing that $\sq \hG$ is an $n$-unfolding of $\sq G$. We show that binary relation $\emb\unfold\subseteq \labelsof{\sq G} \times \labelsof{\sq \sq \hH}$ witnesses that $\sq \hH$ is an $n$-unfolding of $\sq G$, where
\[
\lb x (\emb\unfold) \lb \hy 
\qquad \Longleftrightarrow \qquad
(\exists \lb \hx) \Bigl(\lb x \unfold \lb \hx \text{ and } \lb \hy = \emb(\lb \hx)\Bigr).
\]
Let $\lb x\, (\emb\unfold)\, \emb(\lb\hx)$  because  $\lb x \unfold \lb \hx$ and  $\lb{x'}\, (\emb\unfold)\, \emb(\lb{\hx'}) $ because  $\lb{x'} \unfold \lb{\hx'}$.
\begin{itemize}
\item[\ref{unf:formula}] 
	$ \lbeql{x}{\sq G}{\hx}{\sq \hG}$ by \ref{unf:formula} for $\unfold$ and  $ \lbeql{\hx}{\sq \hG}{\emb(\lb{\hx})}{\sq \hH}$ by \ref{emb:formula} for $\emb$.
\item[\ref{unf:R}] 
	Let $\lb x$ and~$\lb{x'}$ be from different clusters. Then $\accsq x{x'}$ if{f} $\accshat \hx{\hx'}$ by \ref{unf:R} for $\unfold$ and the latter if{f} $\accsqp {\emb(\hx)}{\emb(\hx')}{\hH}$ by \ref{emb:R} for $\emb$.
\item[\ref{unf:layer}] 
	$\layerof x\sle{\sq G}\layerof{x'}$ if{f} $\layerof \hx\sle{\sq \hG}\layerof{\hx'}$ by \ref{unf:layer} for $\unfold$, and the latter if{f} $\layerof{\emb(\hx)}\sle{\sq \hH}\layerof{\emb(\hx')}$ by \ref{emb:layer} for $\emb$.
\item[\ref{unf:nopast}] 
	$\lb{x}$ has no past in $\sq G$ if{f} $\lb{\hx}$ has no past in $\sq \hG$ by \ref{unf:nopast} for~$\unfold$, and the latter if{f} $\lb{\emb(\hx)}$ has no past  in $\sq \hH$ by \ref{emb:nopast} for $\emb$.		
\item[\ref{unf:singleton}]
	If $\lb x$ forms a singleton cluster in $\sq G$, then $\lb \hx$ is the only label in $\sq \hG$ such that $\lb x \unfold \lb \hx$ by \ref{unf:singleton} for $\unfold$. Hence, $\emb(\lb \hx)$ is the only label in $\sq \hH$ such that $\lb x\, (\emb\unfold)\, \emb(\lb \hx)$.
\item[\ref{unf:nonsingleton}] 
	If $\lbc C$  is a non-singleton cluster in $\sq G$ with $\sizeof{\lb C}=k\ge 2$, then by \ref{unf:nonsingleton} for $\unfold$ there are $\lb{x_i}\in \labelsof{\sq G}$ and $\lb{\hx_{i,j}}\in\labelsof{\sq \hG}$ for
    $i=1..k$ and $j=1..n$
    such that $\lbc C=\set{\lb{x_1},\ldots,\lb{x_k}}$, and $\lb{x_i}\unfold\lb{\hx_{i,j}}$ for $i=1..k$, $j=1..n$, and  in~$\sq\hG$
    \[
    \lb{\hx_{1,1}}R\cdots R\lb{\hx_{k,1}}R\lb{\hx_{1,2}}R\cdots R\lb{\hx_{k,2}}R\cdots R \lb{\hx_{1,n}} R\cdots R\lb{\hx_{k,n}}.
    \]
   In this case, $\lb{x_i}\,(\emb\unfold)\,\emb(\lb{\hx_{i,j}})$  for $i=1..k$, $j=1..n$ and in~$\sq\hH$ by \ref{emb:R} for $\emb$
    \begin{multline*}
    \emb(\lb{\hx_{1,1}})\,R\cdots R\,\emb(\lb{\hx_{k,1}})\,R\,\emb(\lb{\hx_{1,2}})\,R\cdots R\,\emb(\lb{\hx_{k,2}})\,R\cdots
    \\ 
    R\, \emb(\lb{\hx_{1,n}})\, R\cdots R\,\emb(\lb{\hx_{k,n}}).
   \end{multline*}
   Thus, $\emb(\lb{\hx_{i,j}}) \in \labelsof{\sq \hH}$ for $i=1..k$, $j=1..n$ provide the requisite $n$ unfoldings for the labels of $\lbc C$ in $\sq \hH$.
\item[\ref{unf:inject}]
	If $\emb(\lb{\hx})=\emb(\lb{\hx'})$, then $\lb{\hx} = \lb{\hx'}$ by the injectivity of $\emb$, and $\lb x = \lb{x'}$ by \ref{unf:inject} for $\unfold$.\qedhere
\end{itemize}
\end{proof}

\begin{lemma}[Embedding Lemma]
	\label{lem:embed}
  Assume we have a tidy derivation~$\der$ with a restricted proper endsequent\/~$\sq G$ and premises\/ $\sq G_1,\ldots,\sq G_n$. If there is an embedding\/ $\emb\colon\sq G\to\sq H$ into a restricted proper sequent\/~$\sq H$, then there is a tidy derivation~$\der'$ with conclusion\/~$\sq H$ and premises\/ $\sq H_1,\ldots,\sq H_n$ such that there are embeddings\/ $\emb_i\colon\sq G_i\to \sq H_i$ for $i=1..n$.\looseness=-1
\end{lemma}

\begin{proof}
  Let us first consider the case where $\deri$ consists of only one tidy
  inference rule instance $\rr$. Let $\sq G_i$ be a premise of
  $\rr$. Then, by Proposition~\ref{prop:daggeradm}, we have $\sq G_i=\sq
  G+\sq G_i'$ for some $\sq G_i'$. If $\rr$ does not add new labels,
  then $\labelsofx{\sq G_i}=\labelsof G$, and we have $\labelsofx{\sq
    H_i}=\labelsof H$ and $\emb_i=\emb$. This gives us a correct rule
  application satisfying the lemma. Otherwise, if $\rr$ adds new
  labels, then $\rr$ is one of $\slef\DIA$, $\srig\IMP$, or $\srig\BOX$, there is only
  one premise, and  the new labels only occur in $\sq G_1'$. 
  \begin{itemize}
  \item $\rr$ is $\slef\DIA$: The rule applies to $\labelsb x{\DIA A}$
    in $\sq G$, creating a new label $\lb y$, and we can apply the
    same rule to $\labelsb {\embof x}{\DIA A}$ in~$\sq H$, creating a
    new label $\lb {y'}$. We can define $\emb_1(\lb y)=\lb{y'}$, and
    for all other $\lb u\in\labelsofx{\sq G_1}$ we define $\emb_1(\lb
    u)=\embof u$.
  \item $\rr$ is $\srig\IMP$ or $\srig\BOX$: The rule applies to
    $\labelsw x{F}$ in $\sq G$, creating a new layer $L'$ whose labels
    all occur only in $\sq G_1'=\lliftf GxF$.  We can apply the same
    rule to $\labelsw {\embof x}{F}$ in $\sq H$, creating a new layer
    $L''$ and adding $\sq H_1'=\lliftf H{\embof x}F$ to $\sq H$. In
    general, $L''$ can contain more labels than $L'$. But $\emb$
    defines an embedding of $\layerof x$ into $\layerof{\embof x}$,
    which canonically defines a mapping $\emb'\colon L'\to L''$ as follows: for
    $\lb y'\in L'$, we define $\emb'(\lb{ y'})=\lb{y''}\in L''$ if{f} either there is
    a $\lb v\in\layerof x$ with $\futsqp v{y'}{G_{\rm1}}$ and $\futsqp{\embof
      v}{y''}{H_{\rm1}}$, or $\lb y'$ has no past in $\sq G_1$ and $\lb y''$~has
    no past in $\sq H_1$.  We can then define our embedding
    $\emb_1\colon\sq G_1\to\sq H_1$ as $\emb_1(\lb u)=\emb'(\lb u)$ if
    $\lb u\in L'$ and $\emb_1(\lb u)=\embof u$ otherwise.
  \end{itemize}
  We can now prove the lemma for arbitrary $\der$ with a straightforward induction on the height of $\der$.
\end{proof}

\begin{lemma}
\label{lem:optiononesinglecluster}
Whenever for a semi-saturated sequent $\sq G$, the algorithm in step 1) applies Option~\ref{def:dia-sat:one}) of Def.~\ref{def:dia-sat} by substituting label~$\lb x$ for label~$\lb y$ in $\sq G$  and applying the transitive closure of~$R$ to obtain~$\sq G'$ such that $\sq G \diared \sq G'$, label $\lb y$ forms a singleton cluster in $\sq G$.
\end{lemma}
\begin{proof}
Since the application of Option~\ref{def:dia-sat:one}) means that  $\lb y$ has no past in $\sq G$, it follows that the algorithm created $\lb y$ as a singleton cluster (unless $\lb y = \lb r$, in which case it was present from the very beginning, but was a singleton cluster at first). It is not hard to check that every time the algorithm creates a new nonsingleton  cluster,either $\lb z$ have a past   for all  label $\lb z$  in the cluster or all labelled formulas $\labelsb z {\DIA B}$  become happy  for all  label $\lb z$  in the cluster. Since $\labelsb y{\DIA A}$ has to be unhappy and have no past to apply Option~\ref{def:dia-sat:one}), label $\lb y$ is not part of a nontrivial cluster.
\end{proof}

\begin{definition}
For a label $\lb x$ of a sequent $\sq G$ we define 
\[
\coffsp G x \colonequals \{\lb z \mid \accsq x z\}.
\]
\end{definition}

\begin{lemma}[restate = SoundSat, name = ] 
  \label{lem:sdiared-sound}
  If $\sqset S\sdiared\sqset\hS$ and $\sqset S$ is semi-saturated and \hbox{$\fm F$-unfoldable}, then $\sqset\hS$ is $\fm F$-unfoldable.
\end{lemma}

\begin{proof}
 Recall that  according to Def.~\ref{def:dia-sat}, $\sqset\hS$ here is a semi-saturation of a set $ \bigl(\sqset S\setminus\set{\sq G}\bigr)\cup\set{\sq G'}$  for some $\sq G\in\sqset S$ such that $\sq G \diared \sq G'$.  Note that $\sq G'$ is a stable sequent by Lemma~\ref{lemma:satA}\eqref{lemma:sat:onestep}. The semi-saturation is handled as in Lemma~\ref{lem:semi-sat-sound} using the fact that $\sq G$ was semi-saturated and, hence, the only label that semi-saturation is going to touch is a new label, which is  created as allowed. Thus, what remains is to explain how to turn $N$-unfoldings of $\sq G$ into $n$-unfoldings of $\sq G'$ for appropriate $N\geq n$.
	According to Definition~\ref{def:dia-sat} there are two options:
	\begin{enumerate}
		\item  In Option 1, we have $\blackf y{\DIA A}G$ where $\lb y$ forms a singleton cluster by Lemma~\ref{lem:optiononesinglecluster}, and we have  some $\lb x$ from a topmost layer with
		$\lb x\lbeq \lb y$, $\accsq xy$, and $\lb x\neq \lb y$ such that all $\lb u \in \coffsp G x$ have no past.  Then $\sq G'$
		is obtained by substituting $\lb x$ for~$\lb y $  and closing under
		$R$-transitivity. Note that the resulting~$\sq G'$  is  semi-saturated, hence, $\sqset\hS= \bigl(\sqset S\setminus\set{\sq G}\bigr)\cup\set{\sq G'}$.
		Consider any premise $\sq\hG$ of $\der_N$ that is an $N$-unfolding of  $\sq G$ via $\unfold_1$. By \ref{unf:singleton}, there is a unique $\lb \hy$ such that $\lb y \unfold_1 \lb \hy$. 
		The choice of $N$ and the extension of~$\der_N$ depend  on the size of the cluster that $\lb x$~belongs to in $\sq G$:
\begin{enumerate}
\item  If $\lbc{\{ x\}}$ is a singleton cluster,  set $N \colonequals n$. By \ref{unf:singleton}, there is a unique~$\lb\hx$ such that $\lb x \unfold_1 \lb \hx$. By \ref{unf:R}, $\accshat \hx\hy$. Since $\lb x$~has no past by assumption, by \ref{unf:nopast} neither has $\lb\hx$. It follows from tree-layeredness of $\sq\hG$ that all labels in $\coffsp \hG \hx$, including $\lb \hy$, have no past. Hence, all labels in $\coffsp \hG \hx$, with the possible exception of $\lb \hx$ itself, are created by rule $\slef{\DIA}$. Let $\sq K_1$ be the premise of the first (i.e., lowermost) instance of rule $\slef{\DIA}$ on the branch of~$\der_n$ leading to $\sq \hG$ that creates a non-trivial child  of $\lb \hx$, let us call this child $\lb u_1$. Since $\der_n$ is tidy, label $\lb u_1$ is the only allowed label in $\sq K_1$. Let $\labelsb \hx {\DIA C}$ be the principal formula of this instance of  $\slef{\DIA}$. Then $\blackf {u_1}C{K_{\mathrm1}}$ is the only labelled formula in $\lb {u_1}$. Since $\blackf \hx {\DIA C}{K_{\mathrm1}}$, also  $\blackf \hx {\DIA C}\hG$ higher up the branch, hence, $\blackf x {\DIA C}G$ by \ref{unf:formula}. Therefore, $\blackf y {\DIA C}G$ because $\lb x \lbeq \lb y$ and $\blackf \hy {\DIA C}\hG$ by \ref{unf:formula}. Since $\lb \hy$ is in a topmost layer of $\sq\hG$, a tidy rule $\slef{\DIA}$ can be applied to $\sq \hG$ with $\labelsb \hy {\DIA C}$ as its principal formula. Let the premise of this rule be $\sq K_2$ and $\lb{u_2}$ be the new label created by this rule, the only allowed label in $\sq K_2$. Consider the following function $\emb_1 \colon \labelsof{K_{\mathrm1}} \to \labelsof{K_{\mathrm2}}$:
\[
\emb_1 \colonequals \set{(\lb\hx,\lb\hy), (\lb{u_1},\lb{u_2})} \cup \set{(\lb w, \lb w) \mid \lb w \in \labelsof{K_{\mathrm1}} \setminus\set{\lb{\hx},\lb{u_1}}}.
\]
It is easy to see that $\emb_1$ is an embedding.  The tidy derivation subtree rooted at $\sq K_1$ can be applied to $\sq K_2$ via this embedding resulting in a tidy derivation  by Lemma~\ref{lem:embed} with all the premises of the extension being $n$-unfoldings of the same sequents from $\sqset S$ as the respective premises of~$\der_n$ by Prop.~\ref{prop:embed}. In particular, for  premise~$\sq \hG$ of the subtree, embedding $\emb_1$ will be extended to en embedding into some premise $\sq \hG_2$ of the extended derivation with $\emb_1(\hy) = \hy_2$ such that $\lbeql{\hy_2}{\sq\hG_2}{\hy}{\sq\hG}$ (note that   $\lbeql{\hy}{\sq K_{\mathrm2}}{\hy}{\sq\hG}$). By  Prop.~\ref{prop:embed}, we can construct an $n$-unfolding $\unfold_2$ of $\sq G$ into $\sq \hG_2$ such that $\lb w \unfold_1 \lb \hw$ if{f} $\lb w \unfold_2 \lb \hw$ for all labels $\lb \hw \in \labelsof{K_{\mathrm1}} \setminus\set{\lb{\hx},\lb{u_1}} = \labelsof{\hG} \setminus\set{\lb{\hx}}$, including for all non-trivial parents of $\lb \hx$. Further, $\lb x \unfold_2 \lb \hy$ and, whenever $\accsq x v$ and $\lb v\unfold_2 \lb \hv$ we have $\accsqp {\hy}{\hv}{\hG_{\mathrm2}}$. In particular, $\lb y \unfold_2 \lb {\hy_2}$ for some $\lb{\hy_2}$ such that $\accsqp {\hy}{\hy_2}{\hG_{\mathrm2}}$ and  $\blackf {\hy_2} {\DIA C}{\hG_{\mathrm{2}}}$. 

We can apply $\slef{\DIA}$ to obtain $\sq K_3$ with a new label $\lb{u_3}$, take an embedding
\[
\emb_2 \colonequals \set{(\lb\hy,\lb{\hy_2}), (\lb{u_2},\lb{u_3})} \cup \set{(\lb w, \lb w) \mid \lb w \in \labelsof{K_{\mathrm2}} \setminus\set{\lb{\hy},\lb{u_2}}}.
\]
 from $\sq K_2$ into $\sq K_3$ and repeat the process.
 
After  $n-1$  such repetitions, we will get a premise~$\sq \hG_{n\mathord+1}$ and an $n$-unfolding $\unfold_{n\mathord+1}$ of $\sq G$ into $\sq \hG_{n\mathord+1}$ such that $\lb w \unfold_1 \lb \hw$ if{f} $\lb w \unfold_{n\mathord+1} \lb \hw$ for all labels $\lb \hw \in  \labelsof{\hG} \setminus\set{\lb{\hx}}$, including for all non-trivial parents of $\lb \hx$. Further, $\lb x \unfold_{n\mathord+1} \lb{\hy_n}$ and, whenever $\accsq x v$ and $\lb v\unfold_{n\mathord+1} \lb \hv$ we have $\accsqp {\hy_n}\hv{\hG_{\mathrm{n\mathord+1}}}$. In particular, $\lb y \unfold_{n\mathord+1} \lb {\hy_{n\mathord+1}}$ for some $\lb{y_{n\mathord+1}}$ such that $\accsqp {\hy_n}{\hy_{n\mathord+1}}{\hG_{\mathrm{n\mathord+1}}}$.

Once again, all   other premises of this derivation expanded $n$ times are $n$-unfoldings of sequents from $\sqset S$. It remains to turn $\sq \hG_{n\mathord+1}$ into an $n$-unfolding of $\sq G'$. We do this by modifying $\unfold_{n+1}$ for the newly created cluster $\lbc{C_x} = \{\lb u \ne \lb y \mid \lb x R_{\sq G} \lb u R_{\sq G} \lb y\}$ in $\sq G'$ and keeping unfoldings of all other labels of $\sq G$ as in $\unfold_{n\mathord+1}$. For each label $\lb u \in\lbc{C_x}$ and each $j=1..n$ there is  at least one label~$\lb{\hu_j}$ such that $\lb u \unfold_j\lb{\hu_j}$ and $\lb {\hy_{j\mathord-1}} R_{\sq \hG_{\mathrm{n\mathord+1}}} \lb{\hu_j} R_{\sq \hG_{\mathrm{n\mathord+1}}} \lb {\hy_{j}}$. These labels $\lb{\hu_1},\dots,\lb{\hu_n}$ for each $\lb u \in \lbc{C_x}$ can, thus, be taken as the requisite $n$ unfolding copies of $\lbc{C_x}$. Indeed, whenever $\accsq w x$ and $\lb w \unfold_{n+1} \lb \hw$ for some $\lb w \ne \lb x$, as discussed, we have $\lb {\hw} R_{\sq \hG_{\mathrm{n\mathord+1}}} \lb{\hx} R_{\sq \hG_{\mathrm{n\mathord+1}}}  \lb{\hu_j}$. On the other hand, whenever $\accsq x v$ and $\lb v \unfold_{n+1} \lb \hv$ for some $\lb v \notin \lbc{C_x} \cup \set{\lb y}$, as discussed, we have $\lb{\hu_j} R_{\sq \hG_{\mathrm{n\mathord+1}}} \lb {\hy_n} R_{\sq \hG_{\mathrm{n\mathord+1}}} \lb{\hv}$. Note that children of $\lb x$ that are not in $\lbc{C_x}$  in $\sq G'$ must be unfolded into labels that are children of all $\lb{\hu_j}$, which is why we needed to create $n+1$ successful unfoldings  rather than $n$: the first $n$ unfoldings are used to create $n$ unfolded copies of $\lbc{C_x}$ whereas the last $(n\mathord+1)$th provides unfoldings for children of $\lb x$ that are not in its cluster.
\item	 
If $\lb x$ belongs to a non-singleton cluster $\lbc{C_x}$, we use a similar expansion to the case of singleton clusters just discussed. Hence, we will only outline the differences. Firstly, instead of $\der_n$ we start with $\der_N$ for $N=\max(n,2)$. The fact that $N \geq n$  guarantees $n$-unfoldings for all premises other than $\sq G$ and for all clusters of $\sq G$ other than  the newly created $\lbc{C'_x}$ that contains $\lbc{C_x}$ and all labels in between $\lbc{C_x}$ and $\lb y$ excluding $\lb y$.  

Among the $N \geq 2$ unfolding copies of $\lb x$ in $\sq \hG$ existing by \ref{unf:nonsingleton}, use the penultimate, $(N\mathord-1)$th unfolding copy as $\lb \hx$ and the unique unfolding of $\lb y$ as $\lb \hy$. For these $\lb \hx$~and~$\lb \hy$ perform the expansion as in the case for a singleton cluster. The only difference to that case is how the final unfolding $\unfold_{n\mathord+1}$ is to be modified to account for the newly created cluster $\lbc{C'_x}$. For all labels outside of $\lbc{C_x}$, the construction remains the same. For each label $\lb u \in \lbc{C_x}$, its last, $N$th unfolding copy from~$\unfold_{1}$ is among labels repeated in the segment (note that the same need not be true about the $(N\mathord-1)$th unfolding copy, which is why we had to start one unfolding copy earlier). Hence, this $N$th copy is now repeated $n\mathord+1$ times, and the first $n$ of these are taken as new unfoldings of $\lb u$.

		\end{enumerate}
		\item In Option 2, we have $\blackf y{\DIA A}G$ and create a child~$\lb z$ of~$\lb y$ such that $\blackf z{A}{G'}$. This case is similar to the steps taken for the semi-saturation in Lemma~\ref{lem:semi-sat-sound-aux}, which is also used to perform the semi-saturation after $\sq G$ is replaced by $\sq G'$, as discussed earlier. 
		We provide an abbreviated description how to expand $\sq \hG$ that is an $N$-unfolding of $\sq G$ into an $n$-unfolding of $\sq G'$. Once again, the exact procedure depends on the size of the cluster that $\lb y$ belongs to in $\sq G$:
\begin{enumerate}
			\item If $\lbc{\{ y\}}$ is a singleton cluster, use $N=n$.  By \ref{unf:singleton}, there is a unique $\hy$ such that $y \unfold \hy$. By \ref{unf:formula}, $\blackf \hy{\DIA A } \hG$ for this  $\lb \hy$. It is sufficient to apply a tidy instance of~$\slef{\DIA}$ to this $\labelsb \hy{\DIA A}$ creating a fresh label $\lb\hz$, the only allowed label in the premise
and set $\unfold' \colonequals \unfold \cup \{(\lb z,\lb \hz)\}$. It is easy to check that all unfolding conditions are satisfied for the resulting premise to be an $n$-unfolding of $\sq G'$.
	\item	
		If $\lb y \in \lbc{C}$ for some cluster $\lbc C$ with $\sizeof{\lbc C} = k\geq 2$, use $N=n\mathord+1$.  By \ref{unf:nonsingleton}, sequent $\sq \hG$ contains $n\mathord+1$ unfolded copies $\lb{\hy_j}$ for $j=1..n\mathord+1$ in~$\sq \hG$ such that  $\lb{y} = \unfold(\lb{\hy_j})$ for  $j=1..n\mathord+1$.  By \ref{unf:formula}, $\blackf {\hy_{n\mathord+1}}{\DIA A } \hG$. Here we apply a tidy instance of~$\slef{\DIA}$ to this $\labelsb {\hy_{n\mathord+1}}{\DIA A}$ creating a fresh label $\lb\hz$, the only allowed label in the premise 
and modify $\unfold$ as follows:
\begin{itemize}
\item $\lb \hz$ is made the unique unfolding of $\lb z$;
\item unfoldings of all labels from $\labelsof{G}\setminus \lbc{C}$ remain intact;
\item for the unfoldings of $\lbc{C}$ we only keep the first $n$ unfoldings guaranteed by \ref{unf:nonsingleton} for $\unfold$ removing all others.
\end{itemize}
		\end{enumerate}
		The removal of the extra unfoldings in the last item is necessary to make sure that $\lb \hz$ is a child of all unfoldings of cluster $\lbc{C}$. Once again,  it is easy to check that all unfolding conditions are satisfied for the resulting premise to be an $n$-unfolding of $\sq G'$.\qedhere
	\end{enumerate}
\end{proof}

\begin{lemma}[restate = SoundLoop, name = ] 
	\label{lem:loopred-sound}
	 If $\sqset S\Loopred\sqset\hS$ and $\sqset S$ is saturated and $\fm F$-unfoldable, then $\sqset\hS$ is $\fm F$-unfoldable.
\end{lemma}

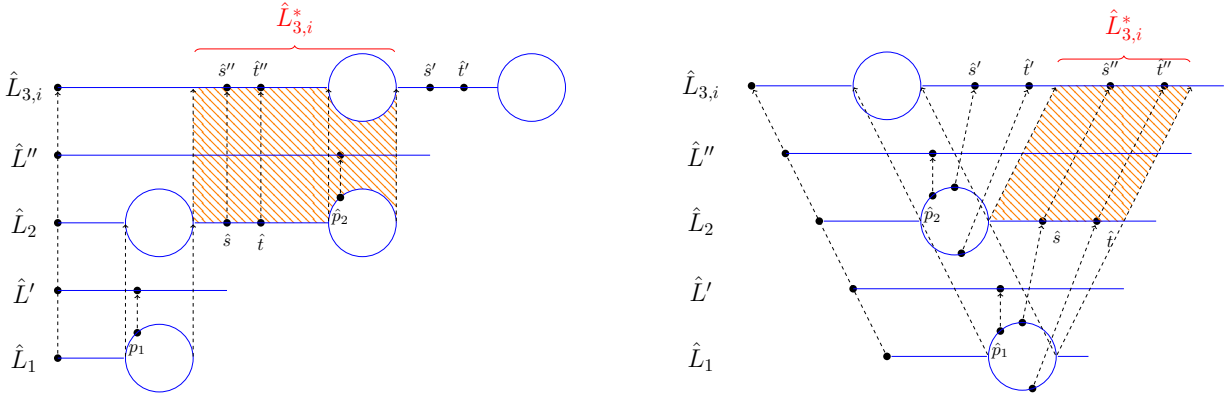
\begin{figure*}
		\begin{tikzpicture}[thick, every node/.style={scale=1.5}]

		\tikzstyle{node}=[circle,fill=black,inner sep=1.5pt]
		\tikzstyle{nonode}=[inner sep=0pt]
		\tikzstyle{access}=[blue]
		\tikzstyle{future}=[dashed,->]
		
		\node[font = {\Large}] (L1) at (2,2) {$\hat L_1$};
		\node[font = {\Large}] (L') at (2,4) {$\hat L'$};
		\node[font = {\Large}] (L2) at (2,6) {$\hat L_2$};
		\node[font = {\Large}] (L'') at (2,8) {$\hat L''$};
		\node[font = {\Large}] (L3) at (2,10) {$\hat L_{3,i}$};
		
		\node[label=below:{$p_1$}] (p1) at  (5.35,2.75)  [node] {};
		%
		\node (l1) at (3,2) [node] {};
		\node (x1) at (5,2) [nonode] {};
		\node (y1) at (7,2) [nonode] {};
		
		\node[label=above:{}] (sur') at (5.35,4) [node] {};
		\node (l') at (3,4) [node] {};
		
		%
		\node (l2) at (3,6) [node] {};
		\node (x2) at (5,6) [nonode] {};
		\node (y2) at (7,6) [nonode] {};
		\node (x2') at (11,6) [nonode] {};
		\node (y2') at (13,6) [nonode] {};
		
		\node[label=above:{}] (sur'') at (11.35,8) [node] {};
		\node (l') at (3,8) [node] {};
		
		\node (l3) at (3,10) [node] {};
		\node (y2'') at (7,10) [nonode] {};
		\node (x3) at (11,10) [nonode] {};
		\node (y3) at (13,10) [nonode] {};
		\draw[draw=none,pattern=north west lines, pattern color=orange] (y2) rectangle (y3);
		\draw[fill=white,draw=blue] (12,6) circle [radius=1cm];

		\node[label=below:{$\hat s$}] (s) at (8,6) [node] {};
		\node[label=below:{$\hat t$}] (t) at (9,6) [node] {};
		\node[label=below:{$\hat p_2$}] (p2) at (11.35,6.75) [node] {};
		\node[label=above:{$\hat s''$}] (s'') at (8,10) [node] {};
		\node[label=above:{$\hat t''$}] (t'') at (9,10) [node] {};
		\node[label=above:{$\hat s'$}] (s') at (14,10) [node] {};
		\node[label=above:{$\hat t'$}] (t') at (15,10) [node] {};
		
		\draw[access] (3,2) -- (x1);
		\draw[access] (6,2) circle [radius=1cm];
		
		\draw[access] (3,4) -- (8,4);

		\draw[access] (3,6) -- (x2);
		\draw[access] (6,6) circle [radius=1cm];
		\draw[access] (y2) -- (11,6);
		
		\draw[access] (3,8) -- (14,8);
		
		\draw[access] (3,10) -- (x3);
		\draw[fill=white,draw=blue] (12,10) circle [radius=1cm];
		\draw[access] (y3) -- (16,10);
		\draw[access] (17,10) circle [radius=1cm];
		\draw [red,decorate, decoration = {brace,raise=30pt,amplitude=5pt}, thick] (y2'') --  (y3) node[pos=.5,above=25pt]{\Large $\hat L_{3,i}^*$};

		\draw[future] (l1) -- (l3);
		\draw[future] (p1) -- (sur');
		\draw[future] (p2) -- (sur'');
		\draw[future] (y2) -- (y2'');
		\draw[future] (x1) -- (x2);
		\draw[future] (y1) -- (y2);
		\draw[future] (x2') -- (x3);
		\draw[future] (y2') -- (y3);
		\draw[future] (s) -- (s'');
		\draw[future] (t) -- (t'');
	\end{tikzpicture}

\hspace*{3cm}

\begin{tikzpicture}[thick, every node/.style={scale=1.5}]
	
	\tikzstyle{node}=[circle,fill=black,inner sep=1.5pt]
	\tikzstyle{nonode}=[inner sep=0pt]
	\tikzstyle{access}=[blue]
	\tikzstyle{future}=[dashed,->]
	
	\node[font = {\Large}] (L1) at (-3.5,2) {$\hat L_1$};
	\node[font = {\Large}] (L') at (-3.5,4) {$\hat L'$};
	\node[font = {\Large}] (L2) at (-3.5,6) {$\hat L_2$};
	\node[font = {\Large}] (L''') at (-3.5,8) {$\hat L''$};
	\node[font = {\Large}] (L3) at (-3.5,10) {$\hat L_{3,i}$};
	
	\node[label=below:{$\hat p_1$}] (p1) at (5.35,2.75) [node] {};
	\node[] (x1) at (6,3) [node] {};
	\node[] (x3) at (6.3,1.05) [node] {};
	\node[] (y1) at (2,2) [node] {};
	\node[] (y2) at (5,2) [nonode] {};
	\node[] (y3) at (7,2) [nonode] {};
	\node[] (y4) at (8,2) [nonode] {};
	%
	
	\node[] (y1') at (1,4) [node] {};
	\node[] (sur') at (5.35,4) [node] {};
	
	\node[label=below:{$\hat p_2$}] (p2) at (3.35,6.75) [node] {};
	\node[label=below right:{$\hat s$}] (s) at (6.6,6) [node] {};
	\node[label=below right:{$\hat t$}] (t) at (8.2,6) [node] {};
	\node[] (y1h) at (0,6) [node] {};
	\node[] (y2h) at (3,6) [nonode] {};
	\node[] (y3h) at (5,6) [nonode] {};
	\node[] (y4h) at (10,6) [nonode] {};
	\node[] (x1h) at (4,7) [node] {};
	\node[] (x3h) at (4.2,5.05) [node] {};
	
	\node[] (y1'') at (-1,8) [node] {};
	\node[] (sur'') at (3.35,8) [node] {};
	
	\node[] (y1h'') at (-2,10) [node] {};
	\node[] (y2h'') at (1,10) [nonode] {};
	\node[] (y3h'') at (3,10) [nonode] {};
	\node[] (y4h'') at (7,10) [nonode] {};
	\node[] (y5h'') at (11,10) [nonode] {};
	\node[] (y6h'') at (12,10) [nonode] {};
	\node[label=above:{$\hat s'$}] (s') at (4.6,10) [node] {};
	\node[label=above:{$\hat t'$}] (t') at (6.2,10) [node] {};
	\node[label=above:{$\hat s''$}] (s'') at (8.6,10) [node] {};
	\node[label=above:{$\hat t''$}] (t'') at (10.2,10) [node] {};
	\draw [red,decorate, decoration = {brace,raise=20pt,amplitude=5pt}, thick] (y4h'') --  (y5h'') node[pos=.5,above=20pt]{\Large $\hat L_{3,i}^*$};
	
	\draw[draw=none,pattern=north west lines, pattern color=orange] (y3h) -- (7,10) -- (7,6) -- cycle;
	\draw[draw=none,pattern=north west lines, pattern color=orange] (7,6) rectangle (9,10);
	\draw[draw=none,pattern=north west lines, pattern color=orange] (9,6) -- (9,10) -- (11,10) -- cycle;
	
	\draw[access] (y1) -- (y2);
	\draw[access] (6,2) circle [radius=1cm];
	\draw[access] (y3) -- (y4);
	
	\draw[access] (1,4) -- (9,4);
	
	\draw[access] (y1h) -- (y2h);
	\draw[access] (4,6) circle [radius=1cm];
	\draw[access] (y3h) -- (y4h);
	
	\draw[access] (-1,8) -- (11,8);
	
	\draw[access] (y1h'') -- (y2h'');
	\draw[access] (2,10) circle [radius=1cm];
	\draw[access] (y3h'') -- (y6h'');
	
	\draw[future] (x1) -- (s);
	\draw[future] (s) -- (s'');
	\draw[future] (x1h) -- (s');
	\draw[future] (p1) -- (sur');
	\draw[future] (p2) -- (sur'');
	\draw[future] (x3) -- (t);
	\draw[future] (t) -- (t'');
	\draw[future] (x3h) -- (t');
	\draw[future] (y1) -- (y1h'');
	\draw[future] (y2) -- (y2h'');
	\draw[future] (y3) -- (y3h'');
	\draw[future] (y3h) -- (y4h'');
	\draw[future] (y3) -- (y5h'');
	
\end{tikzpicture}
	\caption{Left: Unfolding of an unhappy R-triangle loop
		\qquad Right: Unfolding of an unhappy U-triangle loop
	}
	\label{fig:rloop-unfold}
	\label{fig:uloop-unfold}
\end{figure*}

\begin{proof}
	Assume that $\sq G\loopred \sq G'$ and $\sq G\in \sqset S$
	saturated and $\sqset S$ is $\fm F$-unfoldable, and $\sqset
	S'=\sqset S\setminus\set{\sq G}\cup\set{\sq G'}$. We want to
	construct an $n$-unfolding for $\sqset S'$. $\sq G'$~is obtained
	from $\sq G$ by closing an unhappy triangle loop. Let
	$L_1,L_2,L',\lb{C_1},\lb{C_2},\lb{p_1},\lb s,\lb t$
        as in Definitions~\ref{def:triangle} and~\ref{def:u-triangle}.
        We also have
	a $\lb{p_2}$ with $\lb{p_2}\in\lbc{C_2}$ and
	$\lb{p_2}\lbeq\lb{p_1}$, that we will use in the proof.
	
	If $\lb s$ and $\lb t$ are in the same cluster, then every
	$n$-unfolding of $\sqset S$ is also an $n$-unfolding of $\sqset
	S'$. Now, assume that $\lb s$ and $\lb t$ are not in the same
	cluster.

        As in case 1b) of the previous lemma, we start from an $N$-unfolding $\Deri_N$ of $\sqset S$, where $N=\max(n,2)$ and
	let $\sq\hG$ be a premise of $\Deri_N$ that is an unfolding of
	$\sq G$. Let $\unfold$ be the corresponding unfolding relation and let $\hL_1,\hL_2,\hL'$ be the unfoldings of
	$L_1,L_2,L'$, respectively.
	
        We now let $\cT$ be the subtree of $\deri_N$ rooted at the sequent where $\hL'$ first occurs. Let $\sq \hG_0$ be that sequent, let $\rr$ be the instance of the inference rule (an $\srig\BOX$ or $\srig\IMP$) that introduced $\hL'$, let $\sq H$ be the conclusion of that rule instance, and let $\lb{\hp_1}$ be the label in $\sq H$ containing the formula to which that rule was applied. Then $\lb{p_1}\unfold\lb{\hp_1}$. Let $\lb{\hp_2}$ be a label in $\sq\hG$ with $\lb{p_2}\unfold\lb{\hp_2}$. If $\lb{p_2}$ is a singleton, then $\lb{\hp_2}$ is uniquely defined. Otherwise, we pick for $\lb{\hp_2}$ the $k$th repetition of the cluster $\lb{C_2}$, where $k$ is the repetion of $\lb{C_1}$ in which $\lb{\hp_1}$ occurred. We can apply the same rule $\rr$ to $\lb{\hp_2}$ to obtain the layer $\hL''$. Let the new sequent be~$\sq \hG_1$. 
        
        We also let $\lb{\hp'_1}$ be the future of $\lb{\hp_1}$ in $\hL'$, and let $\lb{\hp'_2}$ be the future of $\lb{\hp_2}$ in $\hL''$. This allows us to define an embedding $\emb\colon\sq\hG_0\to\sq\hG_1$ acting like the identity on all layers, except $\hL'$ which is embedded into $\hL''$ mapping each label to its unique future, except for $\childrenof{\hp_1}$ which is embedded into $\childrenof{\hp_2}$. We are now going to apply the Embedding Lemma~\ref{lem:embed} to repeat the proof tree $\cT$. More precisely, let  $\sq H_1,\ldots,\sq H_m$ be the leaves of $\cT$ and let $\sq H_1,\ldots,\sq H_l$ for some $l\le m$ be the leaves of $\cT$ that are $N$-unfoldings of $\sq G$ (and $\sq\hG$ is among them).
        Each of $\sq H_i$ has a topmost layer $\hL_{2,i}$ that is an $N$-unfolding of $L_2$, and that has the labels $\lb{\hs_i}$, $\lb{\htt_i}$, and $\lb{\hp_{2,i}}$, corresponding to $\lb s$, $\lb t$, and $\lb{p_2}$ in the unfolding (if $\lb s$ is in a cluster, we pick for $\lb{\hs_i}$ the first repetition; if $\lb t$ is in a cluster, we pick for $\lb{\htt_i}$ the last repetition).

	We proceed now similarly to the proof of Lemma~\ref{lem:sdiared-sound} by plugging in a copy of the subtree
	$\cT$ at each leaf $\sq H_1,\ldots,\sq H_l$. Note that whenever a new layer is created and the previous one is lifted using the $\srig\IMP$ and $\srig\BOX$-rules, also the gap between the old an the new triangle is lifted (see Fig.~\ref{fig:rloop-unfold}). 
		We call $\hL_{3,i}$ the new topmost layer (which is a copy of $\hL_{2,i}$ with and additional part $\hL^\ast_{3,i}$ that is a copy of labels in $\hL_{2,i}$ but that have no counterpart in $\hL_1$. 
                If a rule instance in $\cT$ is adding a formula to one of the labels $\lb u$ with $\accslt sut{h_i}$ then we need do the same to the corresponding label in $\hL^\ast_{3,i}$. Formally, this can be achieved via a second embdding for each $i$, applying again Lemma~\ref{lem:embed}.

                As in the previous Lemma~\ref{lem:sdiared-sound}, we now repeat this construction $n$ times, so that we obtain our desired $n$-unfolding of $\sqset S'$.
\end{proof}

\UnfoldingLemma*
\begin{proof}
  The set $\sqset S'_0 =\set{\sq G_0(\fm F)}$ is trivially $\fm F$-unfoldable. By the previous lemmas, this property is preserved by all the steps that are used to construct the sequence $\sqset S_0,\sqset S_1,\ldots$.
\end{proof}
  
\Provable*

\begin{proof}
  By the previous lemma, it follows that  the axiomatic set $\sqset S_k$ that caused the algorithm to terminate is $\fm F$-unfoldable. In particular, we have a derivation in $\labISfd$ with conclusion $\sq G_0(\fm F)$ where all premises are 1-unfoldings of elements of  $\sqset S_k$, which are all axiomatic. Therefore we can use instances of the $\idr$ and $\lef\BOT$ rules to give a complete proof of $\sq G_0(\fm F)$ in~$\labISfd$. By Proposition~\ref{prop:daggeradm}, there is also a proof of $\sq G_0(\fm F)$ in~$\labISfp$. This we can append with rules $\Rref$ and $\Lref$ at the bottom to get the proof of $\SEQ\labels rF$ in~$\labISfp$. By Corollary~\ref{cor:adm} and Theorem~\ref{thm:labIS4}, we have that  $\fm F$ is a theorem of $\ISfour$.
\end{proof}

\subsection{Proofs from Section \ref{sec:termination}}

\sizelabel*

\begin{proof}
  Only subformulas of $\fm F$ can occur in a sequent. Furthermore, the
  position of each subformula occurrence in $\fm F$ determines if such a subformula
  can occur on the left or on the right of the $\SEQ$. 
  Moreover, the algorithm introduces in a sequent only labelled formulas which do not already occur in the sequent (Def.~\ref{def:sat-sqset}).   
  Hence, we can have at most $n$
  formula occurrences at a given label.
  Also note that
  $\lb x\lbeq\lb y$ if{f} $\lb x$ and $\lb y$ contain the same set of
  subformulas of $\fm F$. Hence, there are $2^n$ different equivalence
  classes for~$\lbeq$ on labels
  which can occur in $\sqset S_i$.
\end{proof}

\sizecluster*

\begin{proof}
	We shall show that, 
	by construction, all labels in a cluster are different with respect to~$\lbeq$. 
  There are two ways of introducing non-singleton clusters: (i) 
  via unhappy $\lef\DIA$-formulas (Option~\ref{def:dia-sat:one} of Definition~\ref{def:dia-sat}): 
  if there are equivalent labels in such a cluster, the cluster would have been created 
  at an earlier step of the algorithm. 
  (ii) via unhappy R-triangle or U-triangle loops (Definitions~\ref{def:triangle},~\ref{def:u-triangle} and~\ref{def:loop-sat}): if at creation the chain from $\lb s$ to $\lb t$ contains duplicates, then they are collapsed immediately afterwards, as they are also valid choices of $\lb s$ and $\lb t$, fulfilling the
  same unhappy triangle conditions. Hence, when $\Loopred$ terminates, in all new clusters all labels are different with respect to~$\lbeq$. This limits the size of clusters to $2^n$, and we have at most  $2^{2^n}-1$ different equivalence classes 
  of non-empty clusters which can  occur in $\sqset S_i$.
\end{proof}

\lengthbranch*

\begin{figure}
		\begin{tikzpicture}[thick, every node/.style={scale=1.5},font = {\Large}]

		\tikzstyle{node}=[circle,fill=black,inner sep=1.5pt]
		\tikzstyle{nonode}=[inner sep=0pt]
		\tikzstyle{access}=[blue]
		\tikzstyle{future}=[dashed,->]
		
		\node[] (M0) at (-1.5,0) {$M_0$};
		\node[] (M1) at (-1.5,2) {$M_1$};
		\node[] (Mj-1) at (-1.5,5) {$M_{j-1}$};
		\node[] (Mj) at (-1.5,7) {$M_j$};
		\node[] (Mi) at (-1.5,10) {$M_i$};
		\node[] (Mk-1) at (-1.5,13) {$M_{k-1}$};
		\node[] (Mk) at (-1.5,15) {$M_k$};
		\node[font = {\Large}] (M) at (-1.5,18) {$M$};
		
		\node[] (a0) at (0,0) [node] {};
		\node[] (b0) at (2,0) [node] {};
		\node[] (a1) at (0,2) [node] {};
		\node[] (b1) at (2,2) [node] {};
		\node[] (c1) at (4,2) [node] {};
		\node[] (aj-1) at (0,5) [node] {};
		\node[] (bj-1) at (7,5) [node] {};
		\node[] (aj) at (0,7) [node] {};
		\node[] (bj) at (7,7) [node] {};
		\node[] (cj) at (9,7) [node] {};
		\node[] (ai) at (0,10) [node] {};
		\node[] (bi) at (7,10) [node] {};
		\node[] (ci) at (9,10) [node] {};
		\node[] (di) at (12,10) [node] {};
		\node[] (ak-1) at (0,13) [node] {};
		\node[] (bk-1) at (14,13) [node] {};
		\node[] (ak) at (0,15) [node] {};
		\node[] (bk) at (14,15) [node] {};
		\node[] (ck) at (16,15) [node] {};
		\node[] (a) at (0,18) [node] {};
		\node[] (b) at (16,18) [node] {};
		
		\draw [decorate, decoration = {brace,raise=5pt,amplitude=5pt}, thick] (b0) --  (a0) node[pos=.5,below=10pt]{$\bar M_0$};
		\draw [decorate, decoration = {brace,raise=5pt,amplitude=5pt}, thick] (c1) --  (b1) node[pos=.5,below=10pt]{$\bar M_1$};
		\draw [decorate, decoration = {brace,raise=5pt,amplitude=5pt}, thick] (cj) --  (bj) node[pos=.5,below=10pt]{$\bar M_j$};
		\draw [decorate, decoration = {brace,raise=5pt,amplitude=5pt}, thick] (ci) --  (bi) node[pos=.5,below=10pt]{$M_j^i$};
		\draw [decorate, decoration = {brace,raise=5pt,amplitude=5pt}, thick] (di) --  (ci) node[pos=.5,below=10pt]{$\bar M_{i/j}$};
		\draw [decorate, decoration = {brace,raise=5pt,amplitude=5pt}, thick] (ck) --  (bk) node[pos=.5,below=10pt]{$\bar M_k$};

		\draw[access] (a0) -- (b0);
		\draw[access] (a1) -- (c1);
		\draw[access] (aj-1) -- (bj-1);
		\draw[access] (aj) -- (cj);
		\draw[access] (ai) -- (di);
		\draw[access] (ak-1) -- (bk-1);
		\draw[access] (ak) -- (ck);
		\draw[access] (a) -- (b);

		\draw[future] (a0) -- (a);
		\draw[future] (b0) -- (2,18);
		\draw[future] (c1) -- (4,18);
		\draw[future] (bj-1) -- (7,18);
		\draw[future] (cj) -- (9,18);
		\draw[future] (di) -- (12,18);
		\draw[future] (bk-1) -- (14,18);
		\draw[future] (ck) -- (b);
	\end{tikzpicture}
	\caption{A branch and its pasts}
	\label{fig:termination}
\end{figure}
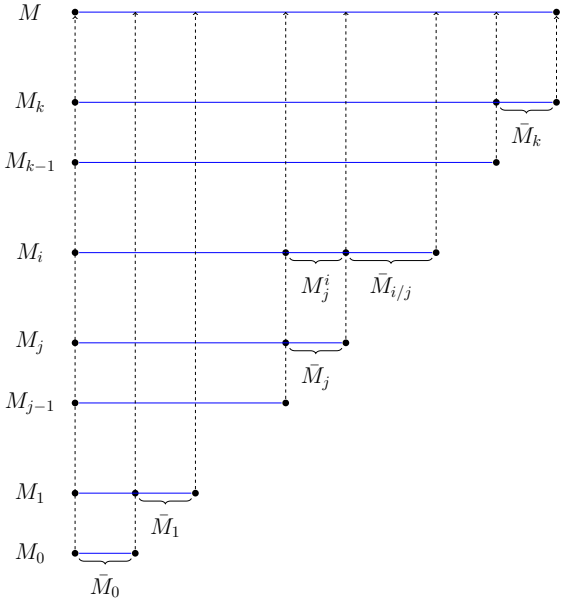

\begin{proof}
A branch $M'$ is a \emph{past} of a branch $M$ in a sequent $\sq G$, if for all $\lb x'\in M'$ there is a $\lb x\in M$ with $\futsq {x'}x$. In this case we write $M'\sle G M$.
  For a branch $M$ we write $\wip M$ for the labels in $M$ that have a
  past, and $\nop M$ for the ones that do not have a past. Then
  $M=\wip M\cup \nop M$ and $\wip M\cap \nop M=\varnothing$. We now
  consider a branch $M$ in a topmost layer $L$ and all its
  pasts, i.e., branches $M'$ with $M'\sle G M$.

  Let $M_0,M_1,\ldots,M_k$ be all the pasts of $M$ for which $\nop M_i\neq\varnothing$, and such that $M_i\sle G M_j$ whenever $0\le i\le j \le k$ (see Fig.~\ref{fig:termination}).
  We have $\sizeof{\nop M_i}\le 2^n+1$ for all $i$ because every label in~$\nop M_i$ 
   is created by Option~\ref{def:dia-sat:two} of Definition~\ref{def:dia-sat}, in correspondence to an unhappy $\lef\DIA$-formula (except for the first label in $\nop M_i$, which could be created by layer lifting in correspondence to an unhappy $\rig\BOX$ formula (Construction~\ref{def:liftingII}), or could be the initial label created in Step~0 of the algorithm). But in any case, after at most $2^n$ such steps, we encounter a label which is equivalent to a previous one (see also the proof of Lemma~\ref{lemma:sat}.\ref{lemma:sat:term}).

  Now let $\allfu Mji=\set{\lb y\in M_i\mid\exists \lb x\in\nop
    M_j.\,\futsq xy}$ be the set of futures of $\nop M_j$ in $M_i$
  for some $j<i$ (see Fig.~\ref{fig:termination}).  Because of the duplication of clusters in the
  layer lifting (Construction~\ref{def:lifting}), we can have
  $\sizeof{\allfu Mji}>\sizeof{\nop M_j}$. We can restrict the size of
  $\allfu Mji$ because there is only a limited amount of times a
  cluster can be duplicated before 
  a U-triangle loops is encountered.  
  Then, 
  the length of the branch from $\lb s$ and $\lb t$ before and
  after the repetition of the cluster is again bound by $2\cdot (2^n+1)$, because
  otherwise the U-triangle loop would be unhappy. (We have to take $2^n+1$ twice because the suricata label is not allowed to occur between $\lb s$ and $\lb t$.)
  Since the size of a
  cluster is bound by $2^n$, and the number of clusters in $\nop M_j$
  is bound by $2^n$, the size of $\allfu Mji$ is bound by $2\cdot 2\cdot
  2^n\cdot 2^n\cdot 2^n=2^{3n+2}$. 
  (This is a very naive
  argument, and the number could be restricted to a much smaller
  one. However, this paper is already technical enough, and we are
  here only interested in the fact that such a bound exists.)

  \def\kmax{k_{\mathrm{max}}}
  Let us now put a bound to $k$. Very naively, we see an R-triangle loop after $\kmax= 2^{2^n}$ (as we need to repeat the cluster $\lb{C_1}$, see Definition~\ref{def:triangle} and Lemma~\ref{lem:size-cluster}). Let $M_j$ be the branch
 that sees a repetition of $M_i$. Define $\nopin Mij$ (for $j<i$) to be the set of labels in $M_i$ that do not have
 a past in $M_j$ (see Fig.~\ref{fig:termination}). Then the R-triangle loop is unhappy if $\sizeof{\nopin Mij}\ge 2\cdot 2^n=2^{n+1}$ (same argument as above). Since $2^{3n+2}>2^{n+1}$, we have that $\sizeof M\le\kmax\cdot\sizeof{\allfu Mji}=2^n\cdot 2^{2^n}\cdot2^{3n+2}=2^{2^n+4n+2}$.

 Note that as soon we observe an unhappy triangle loop, the size of the newly created cluster is \hbox{$\sizeof{\lbc C}\le 2^n$}.
\end{proof}

\TerminationAlg*

\begin{proof}
	By the previous lemma, the size of a branch in a layer is bounded by
	the formula $\fm F$ in the end sequent. Hence the number of possible
	layers that can occur in a sequent in a set $\sqset S_i$ that is
	visited by the algorithm is also bounded. This puts a limit to the
	height of the tree of layers in the sequents, as eventually there
	will be a simulation by a previous layer.%
\end{proof}

\end{document}